\documentclass{ws-procs975x65}
\usepackage{epsfig,amssymb,psfrag,amsmath}

\def\1{\hbox{{1}\kern-.25em\hbox{l}}}
\newcommand{\bit}[1]{\mbox{\boldmath$#1$}}
\newcommand{\ft}[2]{{\textstyle\frac{#1}{#2}}}
\def \dcan {{d}}
\def \e  {\mathop{\rm e}\nolimits}
\def \be  {\begin{equation}}
\def \ee  {\end{equation}}
\def \ba  {\begin{eqnarray}}
\def \ea  {\end{eqnarray}}

\newcommand{\as}{\ifmmode\alpha_{\rm s}\else{$\alpha_{\rm s}$}\fi}
\newcommand\re[1]{(\ref{#1})}

\def \tr {\mbox{tr}}
\newcommand \vev [1] {\langle{#1}\rangle}

\newcommand\lr[1]{{\left({#1}\right)}}
\def \CO {{\mathcal O}}
\def\beqn{\begin{eqnarray}}
\def\eeqn{\end{eqnarray}}

\def \labeltest #1 {\label{#1}}
\def\beq{\begin{equation}}
\def\eeq{\end{equation}}

\renewcommand{\theequation}{\thesection.\arabic{equation}}
\begin{document}

\title{Integrability in QCD and Beyond}

\author{A.~V. Belitsky}

\address{Department of Physics and Astronomy, Arizona State University \\
         Tempe, AZ 85287-1504, USA \\and\\
         Department of Physics, University of Maryland at College Park \\
         College Park, MD 20742-4111, USA }

\author{V.~M. Braun}

\address{Institut f{\"u}r Theoretische Physik, Universit{\"a}t Regensburg \\
         D-93040 Regensburg, Germany }

\author{A.~S. Gorsky}

\address{Institute of Theoretical and Experimental Physics \\
         B.\ Cheremushkinskaya 25, Moscow, 117259 Russia}

\author{G.~P. Korchemsky}

\address{Laboratoire de Physique Th\'eorique\,\footnote{\,\uppercase{U}nit\'e
         mixte de recherche du \uppercase{CNRS} (\uppercase{UMR} 8627).},
         Universit\'e Paris Sud \\
         91405 Orsay C\'edex, France}

\maketitle


\abstracts{ Yang--Mills theories in four space-time dimensions possess a hidden
symmetry which does not exhibit itself as a symmetry of classical Lagrangians, but
is only revealed on the quantum level. It turns out that the effective Yang--Mills
dynamics in several important limits is described by completely integrable
systems that prove to be related to the celebrated Heisenberg spin chain and its
generalizations. In this review, we explain the general phenomenon of complete
integrability and its realization in several different situations. As a prime
example, we consider in some detail the scale dependence of composite (Wilson)
operators in QCD and super-Yang--Mills (SYM) theories. High-energy (Regge)
behavior of scattering amplitudes in QCD is also discussed and provides one with
another realization  of the same phenomenon that differs, however, from the first
example in essential details. As a third example, we address the low-energy
effective action in a $\mathcal{N}=2$ SYM theory which, contrary to the previous
two cases, corresponds to a classical integrable model. Finally, we include a
short overview of recent attempts to use gauge/string duality in order to relate
integrability of Yang--Mills dynamics with the hidden symmetry of a string theory
on a curved background. }

\newpage
\tableofcontents
\newpage
{\sl
We had the privilege
and pleasure of knowing Ian Kogan as a close friend. In our informal chats it was clear
that, whatever he was doing, there always was a single central problem that kept
his mind constantly busy -- the problem of confinement. All other projects
he pursued -- ranging from the Quantum Hall Effect to strings and quantum gravity --
he considered from this general perspective. He viewed the issue of integrability as
a unifying concept which could potentially explain many unsolved problems of
strongly coupled theories and was very enthusiastic about this new development
concerning four-dimensional gauge theories. If it were not for his untimely death, he
would definitely have had a profound impact on the whole field with his usual passion
for new concepts and trends.
}
\section{Introduction}
\setcounter{equation}{0}

Understanding the dynamics of gauge theories at strong coupling is one of the
outstanding problems in quantum field theory. It has been anticipated, since the
classic work by Wilson~\cite{Wilson}, that the Yang--Mills dynamics at strong
coupling can be reformulated as an effective theory of Faraday lines (gauge field
flux) which, in its turn, can be reinterpreted as a string-like theory. A natural
framework for discussing the stringy representation is offered by the multicolor
limit. In this limit, one can express observables as a sum over surfaces (strong
coupling expansion \textsl{\`a la} Wilson) but explicit realization of this
program in gauge theories is still lacking due to absence of an operational
gauge/string correspondence. It is expected that in the multicolor limit,
the path integral over all gauge field configurations will be dominated by a
saddle point -- the so-called ``master field''. Polyakov has suggested
\cite{Polyakov77} that the master field can be found as a solution to a
two-dimensional sigma model defined on a nontrivial background (Lobachevski
space-time). He also demonstrated that such models possess an infinite set of
conserved currents and, therefore, can be solved exactly. It has long been
speculated that similar nontrivial algebraic structures and, in particular,
integrability can be revealed in four-dimensional quantum gauge theories as well.
In the last few years several plausible indications that this actually occurs
have appeared and several intriguing connections have been established between
quantum gauge field theories and integrable lattice spin models. Recent
progress has been in several directions, which we describe here.

At first glance, the connection between four-dimensional gauge theories and
integrable systems may seem surprising. Indeed, in the latter case one is
dealing with quantum-mechanical systems, which have a finite number of
degrees of freedom and the same number of conserved charges. For such systems
the degrees of the freedom are just phase-space variables involved in the
Hamiltonian and the ``evolution time'' has a literal physical meaning. In
contrast, Yang--Mills theories in four dimensions are complex systems
with infinitely many degrees of freedom which are not integrable \textsl{per
se}. Complete integrability emerges as a unique feature of \textsl{effective}
Yang--Mills dynamics in various limits. The relevant degrees of freedom,
Hamiltonians and ``evolution times'' within the Yang--Mills theory are
different in different limits and their identification is not a priori
obvious. In this {work} we will elaborate on three examples which illustrate
the general phenomenon of hidden integrability.
They are\\[-5mm]
\begin{itemize}
\item  Scale dependence of composite (Wilson) operators in QCD and
super-Yang--Mills (SYM) theory
\item  High-energy (Regge) behavior of scattering amplitudes in QCD
\item Low-energy effective action in ${\mathcal N}=2$ SYM theory
\end{itemize}
In all three cases, the dynamics described by an integrable model
corresponds to the scale dependence of corresponding observables with the
``evolution time'' being identified as the logarithm of the relevant energy
scale.
In the first two cases one is dealing with quantum integrable models,
while the ${\mathcal N}=2$ low-energy action corresponds to a classical
model. The integrable systems that emerge in this context turn out to be
related to the celebrated Heisenberg spin magnet and its offspring. This
model was introduced by Heisenberg in 1926 and describes a one-dimensional
chain of atoms with an exchange interaction,
\begin{equation}
\label{XXX}
\mathbb{H}_{s=1/2}
=
- \sum_{n=1}^L \left(\mathbf{S}_n \cdot \mathbf{S}_{n+1} - \frac14 \right)
\end{equation}
where $\mathbf{S}_n = (S_n^x, S_n^y, S_n^z)$ is the spin$-1/2$ operator of
the $n-$th atom in the chain of length $L$ and periodic boundary conditions
are implied: $\mathbf{S}_{L+1}=\mathbf{S}_1$. The model (\ref{XXX}) is
completely integrable and its eigenspectrum was found in 1931 by Bethe
by an innovative method which is widely known nowadays as the Bethe Ansatz.
Much later, it was understood \cite{KRS81,TTF83} that the original Heisenberg
model (\ref{XXX}) can be generalized to arbitrary spins while preserving
complete integrability. The Hamiltonian of a completely integrable lattice
model describing a chain of interacting spin$-s$ operators was found to be
\cite{TTF83}
\begin{equation}
\label{XXX-gen}
\mathbb{H}_{s}
=
\sum_{n=1}^L H(J_{n,n+1})
\, , \qquad
J_{n,n+1} (J_{n,n+1} + 1) = (\mathbf{S}_n + \mathbf{S}_{n+1})^2
\, .
\end{equation}
Here the operator $J_{n,n+1}$ is related to the sum of two spins in the
neighboring sites, $\mathbf{S}_n^2 = s(s+1)$, and $H(x)$ is the following
harmonic sum
\begin{equation}
{H}(x) = \sum_{l=x}^{2s - 1} \frac1{l + 1} =  \psi(2 s + 1) - \psi(x + 1),
\label{XXX-gen1}
\end{equation}
where $\psi(x)=d\ln\Gamma(x)/dx$ is the Euler $\psi-$function. For $s=1/2$, the
two-particle spin takes the values $J_{n,n+1}=0$ and $J_{n,n+1}=1$. In that case,
$H(0)=1$ and $H(1)=0$ so that the two-particle Hamiltonian $H(J_{n,n+1})$ is
given by a projector onto $J_{n,n+1}=0$ subspace, $H(J_{n,n+1}) = \frac14 -
\mathbf{S}_n \cdot \mathbf{S}_{n+1}$, in agreement with (\ref{XXX}).

Approximately at the same time as the model (\ref{XXX-gen}) was formulated, QCD
calculations of the anomalous dimensions of twist-two Wilson operators
\cite{GW73} and high-energy asymptotics of scattering amplitudes \cite{BFKL} led
to expressions involving the very same combination $[\psi(J) - \psi(1)]$ with $J$
being the Lorentz and conformal $SL(2)$ spin respectively. In both situations,
the appearance of the $\psi$-functions is a generic feature related to the
existence of massless vector fields (gluons). For almost two decades, this
similarity remained unnoticed, mainly because of a lack of interaction between
the two communities. Matching the QCD expressions with (\ref{XXX-gen}), one
discovers the hidden integrability properties of gauge
theories~\cite{Lipatov94,FK95,Korchemsky95,BDM98}. Approximately at the same time
hidden integrability has been found in the $\mathcal{N}=2$ supersymmetric
Yang-Mills theory~\cite{n=2}. In the three cases mentioned above, the
identification goes as follows.

{}In the so-called  Bjorken kinematic limit for ``hard''
scattering processes, involving a large momentum transfer to a hadronic system,
the short-distance perturbative QCD dynamics can often be separated (factorized)
from the large-distance interactions and described through a set of gauge-invariant, local composite operators built from fundamental fields and covariant
derivatives. These (Wilson) operators mix under renormalization and their scale
dependence is governed by the renormalization group (RG) or Callan--Symanzik
equation
\begin{equation}
\label{RG}
\mu\frac{d}{d\mu} \mathcal{O}_N (x) = \sum_K \gamma_{NK}(g) \mathcal{O}_K (x)
\, ,
\end{equation}
where $\gamma_{NK}$ is the mixing matrix given by a series in the running
coupling constant $g = g (\mu^2)$. The size of the mixing matrix is constrained
by the symmetries and depends on the operators under consideration. The matrix
$\gamma_{NK}$ can be interpreted as a Hamiltonian acting in the space of
operators that get mixed via the RG flow \cite{BFLK85} with logarithm of the RG
scale $\tau=\ln\mu$ playing the r\^ole of  the ``evolution time''. In this way,
the evolution equation (\ref{RG}) takes the form of a Schr\"odinger equation. It
turns out that for a certain subclass of operators, and to one-loop accuracy, the
corresponding Hamiltonian can be identified as that of the open and/or closed
Heisenberg magnet with the spins being the generators of the $SL(2,\mathbb{R})$
group \cite{BDM98,Belitsky00,DKM00}. The number of sites in the spin chain is
given by the number of fundamental fields involved in the composite operators and
the value of the spin at each site is fixed by the $SL(2,\mathbb{R})$
representation to which the corresponding field belongs. It is different for
quarks and gluons. The emergence of the $SL(2,\mathbb{R})$ group as a symmetry group
of the spin chain is not accidental, since this group is just a reduction of the
four-dimensional conformal group $SO(2,4)$ for field operators ``living'' on the
light-cone \cite{Makeenko:bh,BKM03}. Integrability allows one to apply the Bethe
Ansatz to reconstruct the spectrum of the anomalous dimensions
\cite{Korchemsky96,Belitsky00,DKM00}. Work in this direction
\cite{BDKM99,Belitsky99a,Belitsky99b,BKM00,BKM01} has led to an almost complete
understanding of the spectrum of anomalous dimensions of twist-three operators in
QCD which are important for phenomenology. It has to be mentioned that, as a
rule, QCD evolution equations only become integrable in the large $N_c$ limit.
There is no chance that a quantum $SU(N_c)$ theory would turn out to be integrable for
any $N_c$ because the phenomena it describes are too complicated -- nuclear forces
being one of these. Indeed, in QCD it was found that corrections that break
integrability depend on the quantum numbers of the operators (states) and the
structure of such corrections are such that they lead to creation of ``mass
gaps'' in the spectrum of anomalous dimensions which can be interpreted as
creation of bound states in the corresponding quantum mechanical
model~\cite{BDKM99}.

Although historically the phenomenon of integrability was first discovered
in QCD for operators with maximum helicity, it is, in fact, a general hidden
symmetry of all Yang--Mills theories, which is not manifest an the classical level
and is enhanced in its supersymmetric extensions \cite{BS03}. In a supersymmetric
Yang--Mills theory, the mixing matrix in (\ref{RG}) gets modified due to the
presence of additional fields. Their contribution preserves the QCD-type
integrability and further augments it to an increasingly growing class of
operators as one goes from pure Yang--Mills ($\mathcal{N}=0$) to the maximally-supersymmetric $\mathcal{N}=4$ SYM theory \cite{BDKM04}. In particular, in
the $\mathcal{N}=4$ SYM theory, integrability holds in the sector of scalar
operators \cite{mz,BeiKriSta03}. This theory involves three complex scalars
and the one-loop mixing matrix for the scalar operators of the type ${\rm tr}\,
\{\Phi_1^{J_1}(0)\Phi_2^{J_2}(0)\Phi_3^{J_3}(0)\}$ can be identified as the
Heisenberg $SO(6)$ spin chain with $J_1+J_2+J_3$ sites. The $SO(6)$ group
is nothing but the $R-$symmetry group of the model with six real scalars
belonging to its fundamental representation. As a natural generalization of
these two integrable structures, discovered independently, it was found that
the one-loop renormalization of operators involving gauge, fermions and scalar
fields in $\mathcal{N} = 4$ SYM is described by a spin chain with $SU(2,2|4)$
group representations on each site \cite{BS03}.

As a second example, we consider the high-energy (Regge) asymptotics of
scattering amplitudes. The scattering amplitude $\mathcal{A}(s,t)$ in QCD
in the large-$N_c$ limit (here $s,t$ are Mandelstam variables) can be written
in the Regge limit $s \gg -t$ as
\begin{equation}
\mathcal{A}(s,t) = \sum_{N=2}^\infty \mathcal{A}_N(s,t)\,,\qquad
\mathcal{A}_N(s,t)\sim s^{\lambda E_N}\,,
\end{equation}
where $\lambda = g^2 N_c$ is the 't Hooft coupling constant and the positive
integer $N$ can be thought of as the number of (reggeized) gluons exchanged in
the $t-$channel. The partial amplitudes $\mathcal{A}_N(s,t)$ have a power-like
energy dependence which is governed by the numbers $E_N$. It turns out
that the $E_N$ coincide with the ground state energies of a completely integrable
spin chain model~\cite{Lipatov94,FK95,Korchemsky95}. The length of the
spin chain $N$ equals the number of the reggeized gluons involved, the
``evolution time'' is given by the total rapidity $\tau=\ln s$ and the
relevant degrees of freedom are the two-dimensional transverse coordinates of
the reggeized gluons in the scattering plane. This model can be identified as
a homogeneous Heisenberg magnet (\ref{XXX-gen}) with the spin operators
$\mathbf{S}_n$ being the generators of a (infinite-dimensional) unitary
representation of the $SL(2,\mathbb{C})$ group. This model can be solved by
the Bethe Ansatz and this allows one to calculate the spectrum of $E_N$
\cite{KKM02,DKKM02}.

We should stress that integrability of reggeon interactions is not specific for
QCD. Since the dominant contribution to the scattering amplitudes comes from
$t-$channel exchange of particles with maximal spins (gluons), it is not altered
(in the leading logarithmic  approximation) by the presence of additional fields
in SYM theories. In other words, the phenomenon is tied
to the gauge sector of a SYM theory.

Finally, the third example of integrability is provided by the Seiberg--Witten
solution for the low-energy effective action of the $\mathcal{N}=2$ SYM theory
\cite{SW1}. This solution is described by a Riemann surface bundled over the
space of order parameters characterizing the vacuum state of the $\mathcal{N}=2$
SYM theory. This surface fixes both the low-energy effective action depending on
the massless low-energy modes $u_k=\langle {\rm tr} \, \phi^k \rangle$ and the
spectrum of stable BPS states in the theory. The Seiberg--Witten solution admits
an elegant interpretation in the context of classical integrable models
\cite{n=2,Gorsky:1996hs,Gorsky:1997jq}. The particular pattern of the integrable
model depends on the matter content in the $\mathcal{N}=2$ theory, involving examples
of Toda and Calogero systems, as well as different types of spin chains. The number
of degrees of freedom in the underlying integrable system is equal to $N_c-1$
for gauge group $SU(N_c)$. As in previous examples, the crucial point is the
identification of the ``time'' variable. In the present case, the time is related
to the fundamental scale of the $\mathcal{N}=2$ SYM theory, $\tau = \ln
\Lambda_{\scriptscriptstyle\rm QCD}$, which arises through the dimensional-transmutation phenomenon. The gauge-invariant order parameters $u_k$ turn out to
be higher Hamiltonians of the integrable model, while the BPS spectrum coincides
with the action variables.

The phenomenon of integrability has two different aspects. From the point of view
of phenomenology, it allows one to apply the powerful Quantum Inverse Scattering
Method~\cite{QISM} to solve problems which cannot be treated by standard
techniques. Recent applications  include the calculation of the spectrum of the
multi-reggeon compound states responsible for the power-law rise of the
scattering amplitude as a function of the energy and the  exact solution of the evolution equations for
multi-particle distribution amplitudes in QCD. On the theory side, integrability
leaves a lot of questions to be answered. Currently, it has the status of an ``experimental
observation'' and its origin within the Yang--Mills dynamics remains unclear. It
is natural to ask whether integrability survives at higher loops, what the
meaning of the higher integrals of motion on the Yang--Mills side is and what the
impact of non-perturbative effects is. It is unlikely that the answers can be found
in the framework of perturbation theory. The string/gauge duality provides an
alternative approach. It offers a unifying picture for the gauge theory both at
weak and strong coupling regimes and allows one to relate integrability of the
Yang--Mills theory with the hidden symmetry of a (non)critical string theory
propagating in the curved background~\cite{bena,dolan2,alday,Polyakov04}.%
\footnote{\,Similar nonlocal Yangian type symmetries were previously discussed
in the context of high-energy asymptotics in QCD in Ref.~\cite{Mikhailov97}.}

Presently the gauge/string duality is rather firmly established only for the
maximally-supersymmetric gauge theory \cite{maldacena}. According to this conjecture,
at large $N_c$, $\mathcal{N} = 4$ SYM is equivalent to a string theory with $AdS_5
\times S_5$ target space and the flux of the four-form field inducing the number
of colors. It is assumed that the radii of $AdS_5$ and $S_5$
coincide and the tension of the string is proportional to $(g^2 N_c)^{1/2}$.
At strong coupling, the tension is large and the string can be treated
semi-classically. On the other hand, the  weak coupling regime in the field theory
is mapped into the quantum regime in the string sigma model. Unfortunately there
exists no satisfactory quantum treatment of the string in the $AdS_5\times S_5$
background. Therefore, it is necessary to select proper objects both on the gauge
theory and string sides to make their comparison explicit.

The anomalous dimensions of the gauge theory operators appear to be very
convenient objects for such a comparison. The dilatation operator in the gauge
theory can be identified with the Hamiltonian on the stringy side so that
calculation of the anomalous dimension of the operator is equivalent to the
calculation of the energy of the corresponding string configuration
\cite{bmn,gkp}. {Lacking the whole quantum spectrum of the string, the analysis
can be performed only in the special,  exactly solvable subsectors of the string
theory, or for selected operators in the gauge theory which can be treated
semi-classically on the string side.} The exactly-solvable example deals with a
string moving in the pp-background, which corresponds to the so-called BMN
operators \cite{bmn}, involving scalars, in the $\mathcal{N}=4$ SYM theory. This is
the only example of a gauge theory operator whose anomalous dimensions are
known at all values of the gauge coupling constant. A more general class of
operators with large quantum number with respect to the $R-$symmetry group
allow a semi-classical stringy treatment and the comparison of the stringy
calculation with the perturbative expansion in Yang--Mills theory.

The integrability arises along these lines in two different
ways. In the strong coupling regime, we treat the string classically and a
finite-dimensional integrable system of the Neumann type emerges if one
assumes a simple ansatz for the string motion \cite{arutyunov}. On the other
hand, one can start with the one-loop calculation at weak coupling and consider
the thermodynamical limit of the spin chain. In this limit, in the coherent
state basis, the string sigma model becomes manifest, with each field
incorporated into the composite operator playing the r\^ole of a single
string bit \cite{Kruczenski03}.

The mapping of the integrable system to the stringy picture behind the low-energy
effective actions is slightly different. In this case, the spectral curve of the
integrable system is part of the background since the M5 brane providing the
world-volume for the four-dimensional theory under consideration is wrapped
around it \cite{vafa}. The states of the string wrapped around the spectral curve
amount to the spectrum of BPS states in the gauge theory which is a counterpart of
the gauge/string correspondence in this case. The first step
towards the derivation of the stringy picture for the Regge case based on the
semiclassical limit of the $SL(2,\mathbb{C})$ spin chains was done in
\cite{GKK02}.

The presentation is organized as follows. In Sect.~2, we explain the phenomenon
of hidden integrability for the scale dependence of local operators, alias the
renormalization group (RG) dilatation operator. We  consider three-quark
operators in some detail, and using this example illustrate the main steps that lead
to the spin-chain interpretation. This includes the treatment of RG equations as
a Hamiltonian problem, finding an explicitly $SL(2)$-covariant representation
and identification of the conserved charges.
{The main approaches used in the
quantum integrability framework are described, including the method of
Separated Variables, the method of the Baxter $Q-$operator and
semiclassical solution to the Baxter equation in terms of Riemann surfaces.}
 We will
demonstrate that integrability in QCD is restricted to operators built of fundamental
fields with maximal helicity (and covariant derivatives), whereas  the ``interaction''
between quarks with opposite helicity breaks integrability and leads to creation of
mass gaps in the spectrum. We also give a short summary of the applications to
quark-antiquark-gluon operators, in which case open spin chains arise.

In Sect.~3, we consider the extension of these results to the case of SYM theories
using the non-covariant light-cone superspace formalism due to
Mandelstam~\cite{Mandelstam83} and Brink
et al~\cite{BLN83}. {We give a short review of this formalism and present the
results of the one-loop calculation of the dilatation operator in
the $\mathcal{N}=2$ and $\mathcal{N}=1$ SUSY theories.}
We explain how the super spin chains emerge in the dilatation
operator, which we map to a $SL(2|{\mathcal{N}})$ Heisenberg (super)spin chain.

Section 4 is devoted to the high-energy asymptotics of scattering amplitudes. We
explain both the differences and the similarities of this problem with the case of
local-operator renormalization and give a short summary of the existing results.
In this way, we identify the compound states of reggeized gluons in multicolor
QCD as ground states of the quantum spin chain with $SL(2,\mathbb{C})$ group.
We present the exact solution to the eigenproblem for this integrable model
based on the method of the Baxter $Q-$operator and discuss the properties of the
energy spectrum within the semiclassical approach.

Section 5 contains a discussion of the low-energy effective actions in SYM
while in Sect.~6 we give an overview of the relation between the spectrum of the
anomalous dimensions of {composite operators in gauge theory }
and energies of the string in some background
{within} the string/gauge duality. In particular, the operators with large quantum numbers will
be described semi-classically on the stringy side. The mapping between integrable
spin chains and strings will be briefly outlined. The final Sect.~7 contains
a summary and concluding remarks.


\section{Light-cone dominated processes in QCD}
\setcounter{equation}{0}

For a practitioner, QCD as a theory of strong interactions has
been mainly of use in scattering processes at high energies. In this case, world
lines of quarks and gluons participating in the scattering event are close
to the light cone. It is therefore not surprising that the separation of
transverse and longitudinal coordinates with respect to the scattering
plane proves to be essential. Thus, {before going into details}
let us introduce some notation.

Let $n$ and $\bar n$ be two independent light-like vectors
\begin{equation}\label{nbarn}
n^2 = \bar n^2 =0
\, ,
\qquad n \cdot \bar n = 1
\, .
\end{equation}
For an arbitrary four-vector, $A_\mu$ we define
\begin{equation}
\label{dot}
A_+ \equiv A_\mu n^\mu
\, , \qquad
A_- \equiv A_\mu \bar n^\mu
\, ,
\end{equation}
and the metric tensor in the directions orthogonal to the light-cone
\begin{equation}\label{gperp}
g_{\mu\nu}^\perp = g_{\mu\nu} - n_\mu \bar n_\nu - n_\nu \bar n_\mu
\, .
\end{equation}
We will also use the notation $A_\perp$ for a generic transverse projection
and $A_\perp^\mu = (0, \bit{A}_\perp, 0)$ for a vector that only has
transverse components. For example,
\begin{equation}
\label{sudakov}
x^\mu = x_- n^\mu + x_+ \bar n^\mu + x^\mu_\perp
\, , \qquad
x^\mu_\perp \equiv g^{\mu\nu}_\perp x_\nu
\end{equation}
and, therefore, $x^2 = 2 x_+ x_- - \bit{x}_\perp^2$ with $\bit{x}_\perp^2
= - x_\perp \cdot x_\perp$.

In the parton model, hadron states are described by a bunch of partons,
all moving in the same light-like direction, say $\bar n_\mu$, and  bound
in a hadron wave function. If, in addition to high energy, the physical
process of interest also involves a large momentum transfer, then in a rather
generic situation the transverse structure of the parton system is not resolved
by the interaction and appears to be irrelevant. For such processes, which we
refer to as light-cone dominated, the nonperturbative hadron structure is encoded
in suitable matrix elements of gauge-invariant nonlocal operators built of quark,
antiquark and gluon fields located on the same light-ray (hence the name light-ray
operators \cite{AZ78}) and connected by light-ray ordered gauge links (Wilson
lines). For example,
$$
\bar \psi (z_1 n) {\not\!n} \, {\rm P}
\exp \left\{i g \int_{z_2}^{z_1} d u\, A_+ (un)\right\}
\psi(z_2 n)\,,
$$
where $\psi (x)$ is a quark field and $z_i$'s are real numbers $z_i \equiv
x_{i-}$. In what follows we will not show the gauge links in order not to
complicate the formulae.
However, they are always implied.%
\footnote{\,We hope that using the same notation $\psi$ for the quark field
and the Euler $\psi$-function will not create confusion.}

As always in a field theory, taking the asymptotic limit (here, by approaching
the light-cone) induces extra divergences in addition to conventional
ultraviolet ones. Both have to be renormalized. The nonlocal light-ray operator
can be understood as the generating function of renormalized local operators, e.g.,
\begin{equation}
\bar \psi (-z n) {\not\!n} \, \psi(z n)
=
\sum_{N} \frac{(2z)^N}{N!}
\bar \psi(0) {\not\!n} \stackrel{\leftrightarrow}{D}{\!}^{N}_+ \psi (0)
\, ,
\label{example}
\end{equation}
where $\stackrel{\leftrightarrow}{D}{\!\!}_+=
n \cdot\!\! \stackrel{\rightarrow}{D}- n \cdot\!\! \stackrel{\leftarrow}{D}$, $D_\mu$
is the covariant derivative. Note that the contraction of covariant derivatives
with the light-like vector picks up the contribution of the maximum Lorentz spin,
and all arising local operators have the same twist = dimension -- spin. The
operators in (\ref{example}) can mix with similar operators containing
total derivatives. However, the corresponding mixing matrix is triangular, so
that the anomalous dimensions are not affected and can be calculated by
taking the matrix elements over states with equal momenta, in which case
operators with total derivatives do not contribute. For a given $N$,
therefore, one is left with a single anomalous dimension $\gamma_N$, and
the corresponding expression has been known since the pioneering work in \cite{GW73}.
By now it is known to three-loop accuracy. Taking into account the mixing with gluons
makes $\gamma_N$ a $2\times 2$ matrix, but does not add complications of principle.

Going over to operators built from three and more fields, the situation becomes
considerably more involved. In the bulk of this section we will consider a
particular example of the light-ray operators built out of three quark fields of
definite helicity $\psi^{\uparrow (\downarrow)}=\frac12 \left (1 \pm \gamma_5
\right) \psi$ at a light-like separation
\begin{eqnarray}
B^{3/2}_{\alpha\beta\gamma}(z_1,z_2,z_3)
&=&
\varepsilon^{ijk}(\!\not\!n \psi_{i}^\uparrow)_\alpha(z_1n)
(\!\not\!n \psi_{j}^\uparrow)_\beta(z_2n)
(\!\not\!n \psi_{k}^\uparrow)_\gamma(z_3n),
\labeltest{B3/2}
\\
B^{1/2}_{\alpha\beta\gamma}(z_1,z_2,z_3) &=&
\varepsilon^{ijk}(\!\not\!n \psi_{i}^\uparrow)_\alpha(z_1n)
(\!\not\!n \psi_{j}^\downarrow)_\beta(z_2n)
(\!\not\!n \psi_{k}^\uparrow)_\gamma(z_3n),
\labeltest{B1/2}
\end{eqnarray}
where $\alpha,\beta,\gamma$ are spinor and $i,j,k$ color indices and the
superscript refers to the total helicity $\lambda=3/2$ or $\lambda=1/2$.
The factors $\!\not\!n$ project onto the $\psi_+$
components  of the four-component Dirac spinor fields
\begin{equation}
\psi = \psi_+ + \psi_-
\equiv \ft12
{\not\!\bar{n}} {\not\! n} \psi + \ft12 {\not\! n} {\not\!\bar{n}} \psi
\, ,
\label{goodpsi}
\end{equation}
cf.\ Eq.~(\ref{example}). We will tacitly assume that the three quarks have
different flavor. Identity of the quarks does not influence renormalization
but rather introduces certain selection rules which pick up eigenstates with
particular symmetries. The operators (\ref{B3/2}) and (\ref{B1/2}) appear in the
QCD theory of hard exclusive processes. They are used to define the longitudinal
momentum fraction distributions of quarks inside the $\Delta$-resonance and
the proton, respectively.

Similarly to (\ref{example}) the renormalized three-quark light-ray operators
are defined through their Taylor expansion
\begin{equation}
B(z_1,z_2,z_3)
=
\sum_N \sum_{k_1 + k_2 + k_3 = N}
\frac{z_1^{k_1}}{k_1!}
\frac{z_2^{k_2}}{k_2!}
\frac{z_3^{k_3}}{k_3!}
\, D^{k_1}_+ \psi(0) \, D^{k_2}_+ \psi(0) \, D^{k_3}_+ \psi(0)
\, .
\labeltest{Tailor}
\end{equation}
Local three-quark operators having the same total number of derivatives all mix
together but the ones containing total derivatives can be discarded by taking
forward matrix elements. One is then left with a nontrivial mixing matrix of
size $N + 1$ for given $N$. Eigenvalues of this matrix $\gamma_{N,q}$ now have
two indices: $N$, which refers to the total number of derivatives, and $q$, which
enumerates the anomalous dimensions (e.g., from below) for a given $N$. In the
case of $B^{3/2}$, we will be able to identify $q$ with a conserved charge.

The rest of this section is organized as follows. After a short
exposition of the conformal symmetry in Sect.~2.1, we develop in Sect.~2.2 an
approach to operator renormalization in which this symmetry becomes manifest. In
this form the relation to spin-chain Hamiltonians also becomes apparent and this
connection is elaborated in detail in Sect.~2.3. In Sect.~2.4, we use complete
integrability to construct a semiclassical expansion of the anomalous dimensions
$\gamma^{3/2}_{N,q}$  in powers of $1/N$. In Sect.~2.5, the difference between
$B^{3/2}$ and $B^{1/2}$ is discussed and we find that breaking of integrability
in the case of $B^{1/2}$ produces a mass gap in the spectrum of anomalous dimensions
by a mechanism that can be interpreted as binding of quarks into scalar diquarks.
Sect.~2.6 contains a short summary of other applications, in particular, the case
of quark-antiquark-gluon operators that are related to open spin chains. In Sect.~2.7,
we introduce the general notion of quasi-partonic operators which will be important
for the discussion of supersymmetric extensions. Finally, in Sect.~2.8 we comment
on the conformal symmetry breaking in QCD at higher orders.

\subsection{Conformal symmetry and the collinear subgroup}

Among the general coordinate transformations of the four-dimensional
Minkowski space that conserve the interval $ds^2 = g_{\mu\nu}(x) dx^\mu
dx^\nu$, there are transformations that change only the scale of the
metric:
\begin{equation}
\label{scale1}
g'_{\mu\nu}(x') = \omega(x) g_{\mu\nu}(x)
\,
\end{equation}
and, consequently, preserve angles and leave the light-cone invariant.
All transformations belonging to this subclass form, by definition, the
conformal group. It is obvious that conformal transformations correspond
to a generalization of the usual Poincar\'e group, since the Minkowski
metric is not changed by translations and Lorentz rotations. Examples of
specific conformal transformations are dilatations (global scale
transformations) and inversion
\begin{equation}
\label{scale2}
x^\mu \to x^{\prime\mu} = \lambda x^\mu
\, , \qquad
x^\mu \to x^{\prime\mu} = \frac{x^\mu}{x^2}
\, ,
\end{equation}
with real $\lambda$.  Another important example is the so-called
special conformal transformation
\begin{equation}
\label{special}
x^\mu \to x^{\prime\mu}
=
\frac{x^\mu + a^\mu x^2}{1+2 a\cdot x + a^2 x^2}
\, ,
\end{equation}
which corresponds to the sequence of inversion, translation by an arbitrary
constant vector $a_\mu$ and a further inversion with the same origin. The
full conformal algebra in 4 dimensions includes fifteen generators:
$\mathbf{P}_\mu$ (4 translations), $\mathbf{M}_{\mu\nu}$ (6 Lorentz rotations),
$\mathbf{D}$~~\,(dilatation) and $\mathbf{K}_\mu$ (4 special conformal
transformations) and is a generalization of the well-known 10-parameter
Lie algebra of the Poincar\'e group generated by $\mathbf{P}_\mu$ and
$\mathbf{M}_{\mu\nu}$.

For fields ``living'' on the light-ray,
only a few conformal transformations are relevant.
One of them is a special case of the conformal
transformation in Eq.\ (\ref{special}), with $a_\mu$ being a light-like vector
$a_\mu = a \bar n_\mu$. One finds
\begin{equation}\label{lc1}
x_- \to x^\prime_- = \frac{x_-}{1+ 2 a\, x_-}
\end{equation}
so that these transformations map a light-ray in the $x_-$-direction into
itself. Together with the translations and dilatations in the same
direction, $x_- \to x_- + c$ and $x_- \to \lambda x_-$, they form a so-called
collinear subgroup of the full conformal group. This subgroup is nothing but the familiar $SL(2,{\mathbb R})$ group of projective transformations on
a line:
\begin{eqnarray}
\label{project}
z \to z' &=& \frac{\alpha z+\beta}{\gamma z + \delta}
\, , \quad
\alpha\delta - \beta\gamma = 1
\, ,
\nonumber\\
\Phi(z) \to \Phi'(z) &=& (\gamma z +\delta )^{-2j}
\Phi\left(\frac{\alpha z+\beta }{\gamma z+\delta}\right)
\, ,
\end{eqnarray}
where $\Phi$ is a generic quantum field (quark or gluon), $\alpha,\beta,\gamma,\delta$
are real numbers and $j$ is the conformal spin
\begin{equation}
\label{cspin}
j = (d + s)/2 \, .
\end{equation}
The parameter $d$ is called the scaling dimension and it specifies the field
transformation under the dilatations
\begin{equation}
\delta_D \Phi(x) \equiv  i [\Phi(x), \mathbf{D}]
= - \left( x \cdot \partial + \dcan \right) \Phi(x)
\, .
\label{dila}
\end{equation}
In a free theory (i.e., at the classical level) the scaling dimension coincides
with the canonical dimension $d^{\rm can}$, which is fixed by the requirement
that the action of the theory is dimensionless. In the quantum theory $d \neq
d^{\rm can}$, in general, and the difference is called the anomalous dimension.
In turn, $s$ is the spin projection of $\Phi$ on the ``+'' direction:
\begin{equation}
\label{spin}
\Sigma_{-+} \Phi(z) = s \,\Phi(z)
\, .
\end{equation}
Here we assume that the field $\Phi$ is chosen to be an eigenstate of the spin
operator $\Sigma_{-+}$. In the general case, one should use suitable projection
operators to separate different spin components.

The collinear conformal transformations are generated by the four generators
$\mathbf{P}_{+}$, $\mathbf{M}_{-+}$, $\mathbf{D}$ and $\mathbf{K}_{-}$ which
form a subalgebra of the full conformal algebra. In order to bring the
commutation relations to the standard form, it is convenient to introduce the
following linear combinations: $\mathbf{L}_+ = \mathbf{L}_1 + i \mathbf{L}_2
= - i \mathbf{P}_+$, $\mathbf{L}_- = \mathbf{L}_1 - i \mathbf{L}_2  = ({i}/{2})
\mathbf{K}_-$ and $\mathbf{L}_0 = ({i}/{2}) (\mathbf{D}+\mathbf{M}_{-+})$.
The action of the generators on primary quantum fields can  be traded for the algebra of
differential operators acting on the field coordinates on the light cone
\begin{eqnarray}
\label{SL2R2}
{}[\mathbf{L}_+,\Phi(z)] &=& - \partial_z \Phi(z)
\equiv L_+ \Phi(z)
\, , \nonumber\\
{}[\mathbf{L}_-,\Phi(z)] &=&  \left(z^2\partial_z+2 j z \right)\Phi(z)
\equiv L_- \Phi(z)
\, , \nonumber\\
{}[\mathbf{L}_0,\Phi(z)] &=&  \left(z\partial_z+ j\right)\Phi(z)
\equiv L_0 \Phi(z)
\, ,
\end{eqnarray}
where $\partial_z = \partial/\partial z$.
The differential operators $L_i$ satisfy the
familiar $SL(2)$ commutation relations:
\begin{equation}\label{SL2R}
{}[L_0, L_\mp] = \pm L_\mp \, , \quad  [L_-, L_+] = 2 L_0
\, .
\end{equation}
The remaining generator $\mathbf{E} = ({i}/{2})(\mathbf{D} - \mathbf{M}_{-+})$
counts the so-called collinear twist of the field $\Phi$: $[\mathbf{E},\Phi(z)]
= \frac12(d - s)\Phi(z)$. It commutes with all $\mathbf{L}_i$ and is not relevant
for most of our discussions.

A local field operator $\Phi(z)$ that has fixed spin projection (\ref{spin}) on
the light-cone is an eigenstate of the quadratic Casimir operator
\begin{equation}
\label{casim1}
\sum_{i=0,1,2} [\mathbf{L}_i,[\mathbf{L}_i,\Phi(z)]]
= j (j - 1) \Phi(z) = L^2 \Phi(z)
\end{equation}
with the operator $L^2$ defined as
\begin{equation}\label{casimir}
{L}^2 = \sum_{i = 0, 1, 2} L_i^2 = ( L_0 )^2 - L_0+ L_- L_+
\, , \qquad
[L^2 , L_i] = 0
\, .
\end{equation}
Eqs.~(\ref{project}) and (\ref{casim1}) imply that the field $\Phi(z)$ is
transformed under the projective transformations according to a representation
of the $SL(2,\mathbb{R})$ group specified by the parameter $j$; hence the
rationale of referring  to it as the {conformal spin} of the field.
We will see below that we are dealing with infinite-dimensional representations
of the collinear conformal group. In general, different spin components of the
fields have different conformal spin. For example, for the ``good'' components
of the quark field, Eq.~(\ref{goodpsi}), $s=+1/2$ and the conformal spin takes
the value $j_q=1$; for the transverse component of the gauge strength tensor
$n^\mu F_{\mu\perp}$ the conformal spin equals $j_g=3/2$ and for the scalar field
one has $j_s=1/2$.

Besides the collinear subgroup just described, one can consider another subgroup
corresponding to transformations of the two-dimensional transverse plane
$x_\perp^\mu = (0,x_1,x_2,0)$, introduced in (\ref{sudakov}). This ``transverse''
subgroup is relevant for the Regge kinematics and will be considered in the
corresponding section.

\subsection{Hamiltonian approach to operator renormalization}

In quantum theory, composite operators mix with each other under the
renormalization group flow in the cut-off parameter. This flow is driven
by the dilatation operator,
\begin{equation}
{}[ \mathcal{O} , \mathbf{D}]
= i
\left(
d^{\rm can}
-
\mu \frac{\partial}{\partial \mu}
-
\beta (g) \frac{\partial}{\partial g}
\right)
\mathcal{O}
\, ,
\end{equation}
where $d_{\rm can}$ is the canonical dimension of the composite operator
$\mathcal{O}$. This equality is understood as an operator insertion into
a Green function and is known as the Callan--Symanzik equation. Conformal
symmetry of the QCD Lagrangian (for massless quarks) leads to important
constraints on the above operator mixing. Indeed, the one-loop renormalization
is governed by the infinite parts of the one-loop Feynman diagrams which have the
symmetry of the classical Lagrangian. It follows that the QCD operators belonging
to different representations of the (collinear) conformal group cannot mix under
renormalization at leading-logarithmic accuracy. This simplification
proves to be crucial, e.g., for the QCD description of hadron form factors
at very large momentum transfers. The aim of this section is, first, to
explain this symmetry of the QCD renormalization group equations using a simple
example and, second, to re-interpret these equations as a quantum-mechanical
Hamiltonian problem.

To make the conformal symmetry explicit, it is convenient to work with
light-ray operators (\ref{B3/2}), (\ref{B1/2}) directly, rather than decompose
them in local operators. The renormalization group equations for light-ray
operators can be written as
\cite{AZ78,BB89,MRGDH94}
\begin{equation}
\left\{
\mu\,\frac{\partial}{\partial \mu}
+
\beta(g)\,\frac{\partial}{\partial g}
\right\}
B
=
{\mathbb H}\cdot B
\, ,
\labeltest{RG1}
\end{equation}
where ${\mathbb H}$ is a certain integral operator corresponding, to the
one-loop accuracy, to contributions  of the Feynman diagrams shown in
Fig.~\ref{figure1}.

\setlength{\unitlength}{0.7mm}
\begin{figure}[t]
\vspace{5.3cm}
\hspace*{-2cm}
\begin{picture}(120,200)(0,1)
\mbox{\epsfxsize16.0cm\epsffile{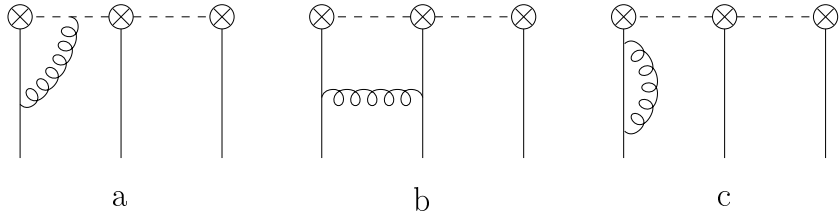}}
\end{picture}
\vspace*{-15.0cm}
\caption{\labeltest{QCD96fig1}
\small Examples of a ``vertex'' correction (a), ``exchange'' diagram (b) and
self-energy insertion (c) contributing to the renormalization of
three-quark operators in Feynman gauge. Path-ordered gauge factors
are shown by the dashed lines. The set of all diagrams includes
possible permutations.
}
\label{figure1}
\end{figure}

To simplify notations we factor out the QCD coupling constant, the color
factors and trivial contributions of the self-energy insertions:\,\footnote{\,The
color factors of the diagrams shown in Figs.~\ref{figure1}a and b can be
calculated using the identity $\varepsilon_{i'j'k} t^a_{i'i} t^a_{j'j} = -
\varepsilon_{ijk}(1 + 1/N_c)/2$ with $t^a$ being the generators of the $SU(N_c)$
in the quark representation.}
\begin{equation}
{\mathbb H} = \frac{g^2}{8\pi^2}\Big[(1 + 1/N_c){\mathcal H} + 3C_F/2\Big],
\end{equation}
where $C_F=(N_c^2-1)/(2N_c)$ is the Casimir operator in the fundamental
representation of $SU(N_c)$. In the Feynman gauge, the gluon exchange
diagram in Fig.~\ref{figure1}b vanishes unless the participating quarks
have opposite helicity. The renormalization of the $\lambda=3/2$ operator
$B_{3/2}$ (\ref{B3/2}) is therefore determined by the vertex correction in
Fig.~\ref{figure1}a alone. By explicit calculation one finds~\cite{LB79,Peskin79}
\begin{equation}
{\mathcal H}_{3/2}
=
{\mathcal H}_{12}^{v} + {\mathcal H}_{23}^{v} + {\mathcal H}_{13}^{v}
\, ,
\labeltest{H32}
\end{equation}
where ${\mathcal H}_{ik}^{v}$ are the two-particle kernels involving the
$i$-th and $k$-th quarks, for example,
\begin{eqnarray}
{\mathcal H}_{12}^{v}\,B(z_i)
=
- \int_{0}^{1}\frac{d\alpha}{\alpha}
\Big\{
\!&{\bar\alpha}&\! \left [ B(z_{12}^\alpha,z_2,z_3) - B(z_1,z_2,z_3)\right]
\nonumber\\
+
\!&{\bar\alpha}&\!
\left [
B(z_1,z_{21}^\alpha,z_3) - B(z_1,z_2,z_3)
\right ]
\!
\Big\}
\, ,
\labeltest{H32-part}
\end{eqnarray}
with $\bar \alpha \equiv 1-\alpha$ and $z_{ik}^\alpha\equiv z_{i} \bar\alpha
+ z_{k}\alpha$.

In the case of $B_{1/2}$, the vertex correction remains the same, but one has
to add the contributions of gluon exchange between the quarks with opposite
helicity. One obtains
\begin{equation}
{\mathcal H}_{1/2}
=
{\mathcal H}_{3/2} - {\mathcal H}_{12}^e - {\mathcal H}_{23}^e
\, ,
\labeltest{H12}
\end{equation}
where we assume that the first and the third quark have the same helicity,
as in Eq.~(\ref{B1/2}). The kernels $H^e_{ik}$ act on $i$-th and $k$-th
arguments of the nonlocal operators only, and can be written in the form
\begin{equation}
{\mathcal H}_{12}^{e}\,B(z_i)
=
\int \!{\mathcal D}\alpha\,
B(z_{12}^{\alpha_{1}},z_{21}^{\alpha_{2}},z_3)
\, ,
\labeltest{H12-part}
\end{equation}
with the integration measure ${\mathcal D}\alpha$ defined as
\begin{equation}
\int\! {\mathcal D} \alpha
\equiv
\int_0^1\! d\alpha_1\,d\alpha_2\,d\alpha_3\,
\delta (1-\alpha_1-\alpha_2-\alpha_3)\,.
\labeltest{Dx}
\end{equation}

The expected conformal invariance of the evolution equation for baryonic
operators implies that the two-particle kernels ${\mathcal H}_{ik}$ commute
with the generators of the ${\it SL}(2)$ transformations $L_\alpha$ defined
in ({\ref{SL2R2}). To show this, consider the following expression that
generalizes both  (\ref{H32-part}) and (\ref{H12-part}):
\begin{equation}
{\mathcal H}_{12} B(z_1,z_2,z_3)
=
\int {\mathcal D}\alpha\, \omega(\alpha_1,\alpha_2)
B (z_1-\alpha_1 z_{12},z_2+\alpha_2 z_{12},z_3),
\labeltest{ansatz}
\end{equation}
where $z_{12}=z_1-z_2$. The integration measure is defined in (\ref{Dx}).
This operator has a simple meaning: acting on the three-particle nonlocal
operator $B(z_1,z_2,z_3)$ it displaces the quarks with the coordinates
$z_1$ and $z_2$ on the light-cone in the direction of each other. The
vertex correction in (\ref{H32-part}) is obtained with $\omega^v
(\alpha_1,\alpha_2) = \delta\left(\frac{\alpha_1\alpha_2}{\bar\alpha_1
\bar\alpha_2}\right)$ and the exchange contribution in (\ref{H32-part})
corresponds to $\omega^e(\alpha_1,\alpha_2) =1$.

It is easy to see that, for this ansatz, $[{\mathcal H}_{12},L_+] =
[{\mathcal H}_{12},L_0]=0$ for an arbitrary function
$\omega(\alpha_1,\alpha_2)$, whereas the condition
$[{\mathcal H}_{12},L_-]=0$ leads to the following constraint:
\begin{equation}
\left(
\frac{\partial}{\partial \alpha_1} \alpha_1\bar\alpha_1
+
2 \alpha_1\,j_1\right) \omega(\alpha_1,\alpha_2)
=
\left(
\frac{\partial}{\partial \alpha_2} \alpha_2\bar\alpha_2
+
2\alpha_2\,j_2
\right)
\omega(\alpha_1,\alpha_2)
\, ,
\end{equation}
where $\bar\alpha=1-\alpha$. Its general solution has the form
\begin{equation}
\omega(\alpha_1,\alpha_2) = \bar\alpha_1^{2j_1-2}
\bar\alpha_2^{2j_2-2} \varphi\left(\frac{\alpha_1\alpha_2}
{\bar\alpha_1\bar\alpha_2}\right),
\label{funny}
\end{equation}
with an arbitrary $\varphi$. Note that the conformal spins $j_k=1$ for
all the three quark fields entering (\ref{B3/2}) and (\ref{B1/2}) so that
the prefactor $\bar\alpha_1^{2j_1-2}\bar\alpha_2^{2j_2-2}$ in (\ref{funny})
disappears and the conformal symmetry requires that the kernel
$\omega(\alpha_1,\alpha_2)$ is an arbitrary function of the ratio
$\frac{\alpha_1\alpha_2}{\bar\alpha_1\bar\alpha_2}$ which is indeed
the case for both expressions (\ref{H32-part}) and (\ref{H12-part}).

Once conformal symmetry of the two-particle kernels is established,
the group theory tells us that ${\mathcal H}_{ik}$ may only depend on
the corresponding two-particle Casimir operators $L^2_{ik}$. To find
the functional form of this dependence one has to compare their action
on a suitable basis of trial functions. For definiteness, let us find
${\mathcal H}_{12}$ as a function of $L^2_{12}$. To this end, it is
enough to compare their action on the homogeneous polynomials of two
variables $z_1$ and $z_2$:
$$
B(z_1,z_2,z_3) \longrightarrow b_n(z_1,z_2)
$$
which we choose to be eigenfunctions of the operator
\begin{equation}
L_{12}^2=-\partial_1\partial_2z_{12}^2; \qquad z_{12}\equiv z_1-z_2\,.
\label{Cas2}
\end{equation}
It is easy to see that the polynomials defined thus form an infinite-dimensional
representation of the $SL(2)$ group on which the operators $L_+ \equiv L_{1,+}
+ L_{2,+}$ and $L_-\equiv L_{1,-}+L_{2,-}$ act as the lowering  and raising
operators respectively. It is thus sufficient to consider only the functions
(polynomials) annihilated by $L_+$, or equivalently, the highest weight of the
representation, since all other eigenfunctions of $L_{12}^2$ can then be
obtained by a repeated application of $L_-$. Since $L_+ = - (\partial_1 +
\partial_2)$, the latter condition is simply translation invariance, which
leaves one with
\begin{equation}
b(z_1,z_2) = (z_1-z_2)^n \equiv z_{12}^n
\, , \quad
n = 0, 1, 2, \ldots
\label{bz}
\end{equation}
An explicit calculation gives
\begin{eqnarray}
L_{12}^2 z_{12}^n &=& (n+2)(n+1)z_{12}^n,
\nonumber\\
{\mathcal H}_{12}^{v}z_{12}^n &=& 2[\psi(n+2)-\psi(2)] z_{12}^n,
\nonumber\\
{\mathcal H}_{12}^{e}z_{12}^n &=& 1/[(n+2)(n+1)] z_{12}^n,
\labeltest{H-L}
\end{eqnarray}
where $\psi(x)=d\ln\Gamma(x)/{dx}$ is the Euler $\psi$-function.
To cast (\ref{H-L}) in an operator form, define $J_{12}$
as a formal solution of the operator relation
\begin{equation}
L_{12}^2 = J_{12}(J_{12}-1).
\labeltest{def:J12}
\end{equation}
The eigenvalues of $J_{12}$ equal $j_{12} = n+2$ and specify the
possible values of the sum of two $j=1$ conformal spins of quarks
in the $(12)$-pair, cf.~(\ref{casimir}). Then
\begin{eqnarray}
{\mathcal H}_{12}^{v} &=& 2[\psi(J_{12})-\psi(2)]
\, , \nonumber\\
{\mathcal H}_{12}^{e} &=& 1/[J_{12}(J_{12}-1)] = 1/L^2_{12}
\, .
\labeltest{H-SL2}
\end{eqnarray}
Substituting the representation  (\ref{H-SL2}) into (\ref{H32}) and
(\ref{H12}) one obtains a Schr\"odinger equation for three particles
with the coordinates $z_1$, $z_2$ and $z_3$ on the light-ray line.
The `Hamiltonians' ${\mathcal H}_{3/2}$ and ${\mathcal H}_{1/2}$
entering this equation for different baryon states have a pairwise
structure and are expressed in terms of the corresponding two-particle
Casimir operators (\ref{H-SL2}).

The precise form of the Schr\"odinger equation and the meaning of the
corresponding ``wave function'' depend on the representation that is
used for the $SL(2)$ generators. Looking for polynomial solutions in
light-cone coordinates $\Phi(z_1,z_2,z_3)$ as in (\ref{bz}) is one
possibility:
\begin{equation}
{\mathcal H}\cdot \Phi_{N,q}  =  {\mathcal E}_{N,q} \Phi_{N,q}
\, .
\labeltest{Sch3}
\end{equation}
Here $N$ refers to the degree of the polynomial alias the total number
of derivatives, and $q$ enumerates the energy levels.

Another option is to go over to local operators. A generic local operator
with $N$ derivatives can be written as the sum of monomials entering the
expansion (\ref{Tailor}) with arbitrary coefficients
\begin{equation}
{\mathcal O}=\sum_{k_1+k_2+k_3=N}c_{k_1,k_2,k_3}
\, D^{k_1}_+ \psi(0) \, D^{k_2}_+ \psi(0) \, D^{k_3}_+ \psi(0)
\, ,
\labeltest{operF}
\end{equation}
and can be represented by a polynomial in three variables
\begin{equation}
{\Psi(x_1,x_2,x_3)}
=
\sum_{k_1+k_2+k_3=N} c_{k_1k_2k_3} x_1^{k_1} x_2^{k_2} x_3^{k_3}
\, .
\labeltest{coefF}
\end{equation}
Note that $\Psi(x_i)$ serves as a projector, separating out the contribution of
the local operator $O_\Psi$ to the nonlocal operator $B(z_i)$, which can be
made explicit by writing
\begin{equation}
{\mathcal O}_\Psi=\Psi(\partial_1,\partial_2,\partial_3)B(z_1,z_2,z_3)|_{z_i=0}.
\labeltest{PsiOp}
\end{equation}
Therefore, one can call $\Psi(x_i)$ the coefficient function of a local
operator. Since a local operator is completely determined by its
coefficient function, diagonalization of the mixing matrix for operators
can be reformulated as diagonalization of the mixing matrix for the
coefficient functions. Requiring that ${\mathcal O}_\Psi$ (\ref{PsiOp})
is multiplicatively renormalized, one ends up with a matrix equation in
the space of homogeneous polynomials of degree $N$ of three variables
\begin{equation}
{\mathcal H}\cdot \Psi_{N,q}  =  {\mathcal E}_{N,q} \Psi_{N,q}
\, ,
\labeltest{Sch1}
\end{equation}
where the Hamiltonian ${\mathcal H}$ is given by the same expression,
Eq.\ (\ref{H-SL2}), in terms of the two-particle Casimir operators,
but the $SL(2)$ generators have to be taken in a different, adjoint
representation (see \cite{BDKM99,BKM01} for details). The eigenvalues
of ${\mathcal H}$ in (\ref{Sch3}) and (\ref{Sch1}) are of course the
same; they correspond to the anomalous dimensions
\begin{equation}
\gamma_{N,q}
\equiv
\frac{\alpha_s}{2 \pi}
\Big[
(1 + 1/N_c)\,{\mathcal E}_{N,q} + 3/2\, C_F
\Big]
\, .
\label{anomal}
\end{equation}
The wave functions in (\ref{Sch3}) and (\ref{Sch1}) are related by a certain
integral transformation, see \cite{BDKM99}. The advantage of using $\Phi(z_i)$
is that is this representation the highest weight vector condition $L_+
\Phi(z_i)=0$ reduces to the requirement of the translation invariance of
$\Phi(z_i)=0$. On the contrary, the corresponding equation for the operator
coefficient functions is more complicated (see also \cite{Ohrndorf82}).

The Hamiltonian in (\ref{Sch1}) is hermitian with respect to the
conformal scalar product \cite{BKM03}. This implies that the anomalous
dimensions take real quantized values and the corresponding eigenfunctions
are mutually orthogonal with the weight function $x_1x_2x_3$
\begin{equation}
\int {\mathcal D}x\, x_1 x_2 x_3 \,\Psi_{N,q}(x_i)\Psi_{N,q'}(x_i)
\sim \delta_{q,q'}
\,.
\labeltest{aaa}
\end{equation}
The construction that we have described was first suggested in \cite{BFLK85}.
It is general and can be used in all other cases.

\subsection{Complete integrability and non-compact Heisenberg magnets}

The Hamiltonian ${\mathcal H}_{3/2}$ possesses an additional ``hidden''
symmetry. One can construct an integral of motion (conserved charge) that
commutes with ${\mathcal H}_{3/2}$ and with the $SL(2)$ generators:
\begin{eqnarray}
{\mathcal Q} &=& \frac{i}{2}[L_{12}^2,L_{23}^2]
=
i^3 \partial_1\partial_2\partial_3 z_{12} z_{23} z_{31}
\, , \nonumber\\
&&
[{\mathcal Q}, L_\alpha] = [{\mathcal Q}, {\mathcal H}_{3/2}] = 0
\, ,
\labeltest{Q3}
\end{eqnarray}
where the explicit expression in the first line refers to the representation
of the generators in Eq.~(\ref{SL2R2}). The existence of a non-trivial integral
of motion makes the corresponding Schr\"odinger equation (\ref{Sch3})
completely integrable. Indeed, Eq.~\re{Q3} tells us that ${\mathcal H}_{3/2}$,
${\mathcal Q}$, the total three-quark conformal spin $L^2 = \sum_{i=0,1,2}
(L_{1,i}+L_{2,i}+L_{3,i})^2$ and its projection $L_0 = L_{1,0}+L_{2,0}+L_{3,0}$
can be diagonalized simultaneously. Since there are only three independent
degrees of freedom, this implies that the Hamiltonian ${\mathcal H}_{3/2}$
is a (complicated) function of the conserved charges. As a reward, one can
replace the Schr\"odinger equation by a simpler problem of diagonalization
of the conserved charge ${\mathcal Q}$. The eigenvalues of ${\mathcal Q}$ provide
one with a ``hidden'' quantum number that can be used to parameterize the
spectrum of the Hamiltonian, ${\mathcal E}_{3/2}={\mathcal E}_{3/2}(N,q)$.%
\footnote{\,For the local conformal operator with the coefficient function
$\Psi(x_i)$ that satisfies the highest weight condition $L_+\Psi(x_i)=0$
one finds $L_0\Psi(x_i) = J \Psi(x_i)$ and  $L^2\Psi(x_i) = J(J-1)\Psi(x_i)$.
Here  $J=N+3j$ is
the total conformal spin with $j=1$ being the conformal spin of
the quark field. Hence the dependence of the energy eigenvalues on the
conformal spin (and spin projection) can be traded for the dependence on $N$,
the total number of derivatives.}

It turns out that ${\mathcal H}_{3/2}$ can be viewed as the Hamiltonian of
the non-compact $SL(2,\mathbb{R})$ Heisenberg spin magnet. As a hint,  notice
that the expression for the Hamiltonian ${\mathcal H}_{3/2}$ as the sum of
the Euler $\psi-$functions, Eqs.~\re{H32} and \re{H-SL2}, resembles the
definition of generalized $SU(2)$ Heisenberg magnet, Eqs.~\re{XXX-gen} and
\re{XXX-gen1}. The important difference between the two expressions is,
however,  that the two-particle ``spin'' $J_{12}$ in \re{H-SL2} takes an
infinite set of quantized values $J_{12}=n+2$ with $n\ge 0$, while for the
same operator in \re{XXX-gen} one has $0 \le J_{12} \le 2s$. In other words,
in these two cases we deal with the \textsl{infinite}-dimensional representations
of the $SL(2,\mathbb{R})$ group and \textsl{finite}-dimensional representation of
the $SU(2)$ group, respectively.

To establish the exact correspondence, we can think of three quark operators
inside $B(z_1,z_2,z_3)$ as three sites of the lattice and reinterpret the
generators of the collinear $SL(2,\mathbb{R})$ subgroup, $L_0,L_+$ and $L_-$,
acting on $k-$th quark field as defining the spin operators in $k-$th site
\begin{equation}
S_k^+=z_k^2\partial_k - 2sz_k
\, , \qquad
S_k^-=-\partial_k
\, , \qquad
S_k^3=z_k\partial_k -s
\, , \label{spin1}
\end{equation}
with $S^\pm=S^1\pm iS^2$ and spin $s$ related to the conformal spin of the
quark field $s=-j=-1$.

The spin operators \re{spin1} are the generators of a unitary representation
of the $SL(2,\mathbb{R})$ group of the discrete series. The definition of
the integrable spin chain is based on the existence of a fundamental
operator, the $R-$matrix, $R_{km}(u)$, which acts on the coordinates in
sites $k$ and $m$, depends on an arbitrary complex parameter $u$ and
satisfies the Yang--Baxter equation \cite{QISM}
\begin{equation}
R_{1 2}(u) R_{1 3}(v) R_{2 3}(u-v) = R_{2 3}(u-v) R_{1 3}(v) R_{1 2}(u)\,,
\label{YB}
\end{equation}
with $u$ and $v$ being arbitrary complex spectral parameters. For
infinite-dimensional representations of the $SL(2,\mathbb{R})$ group of
the discrete series, the solution to Eq.~(\ref{YB}) is given by \cite{KRS81}
\begin{equation}
R_{km}(u)
=
(-1)^{J_{km}}\frac{\Gamma(J_{km} - iu)}{\Gamma(J_{km} + iu)}
\, ,
\label{R1}
\end{equation}
where $J_{km}$ is the two-particle spin operator, cf.\ Eq.~(\ref{def:J12}).
The Hamiltonian of the $SL(2,\mathbb{R})$ Heisenberg magnet of length $L$
is defined in terms of the $R-$matrix as \cite{TTF83}
\begin{equation}
\mathcal{H}_{\scriptscriptstyle\rm XXX}
=
\sum_{n=1}^L H_{n,n+1}
\, , \qquad
H_{n,n+1}=-i\frac{d}{d u} \ln R_{n,n+1}(u)\bigg|_{u=0}
\, ,
\label{X}
\end{equation}
with $H_{L,L+1}\equiv H_{L,1}$. It is easy to verify that for $L=3$ the
Hamiltonian \re{X} coincides with $\mathcal{H}_{3/2}$ up to an irrelevant
additive constant.

Complete integrability of the Hamiltonian $\mathcal{H}_{3/2}$ allows one
to solve the Schr\"odinger equation by applying the Bethe Ansatz. The exact
expression for the energy spectrum is given by
\begin{equation}
{\mathcal E}_{N,q} = i \frac{d}{d u}\ln \frac{Q(u + ij)}{Q(u - ij)}\bigg|_{u=0}
\, ,
\label{E-Q}
\end{equation}
where the function $Q(u)$ is the eigenvalue of the Baxter operator for the
$SL(2,\mathbb{R})$ magnet~\cite{Derkachov99}. It satisfies the finite-difference
Baxter equation
\begin{equation}
t(u) Q(u) = (u+ij)^3 Q(u+i) + (u-ij)^3 Q(u-i)
\label{Baxter-eq}
\end{equation}
with the transfer matrix
\begin{equation}
t(u) = 2u^3 - (N + 3j)(N + 3j - 1) u + q
\label{transfer}
\end{equation}
and $j=1$ being the conformal spin of the quark field, under the additional
condition that $Q(u)$ is polynomial in the spectral parameter. These conditions
determine $Q(u)$ up to an overall normalization and allow one to establish the
quantization conditions for the charge $q$. The polynomial $Q(u)$ can be
parameterized by its roots
\begin{equation}
Q(u)=\prod_{j=1}^N (u-\lambda_j)
\, ,
\label{Bethe}
\end{equation}
with non-negative $N$ related to the total $SL(2,\mathbb{R})$ spin of the
magnet $J=N+3j$. Substituting this expression into the Baxter equation
\re{Baxter-eq}, one finds that the roots $\lambda_j$ satisfy the Bethe
equations for spin $s=-1$. The fact that the spin is negative leads to
a number of drastic differences compared to conventional ``compact''
$SU(2)$ magnets. In particular, the Bethe roots take real values only and the
number of solutions is infinite~\cite{Korchemsky95}.

Solving the Baxter equation \re{Baxter-eq} for $N=0,1,\ldots$ one can
reconstruct the spectrum of the Hamiltonian $\mathcal{H}_{3/2}$ and, as
a consequence, determine the exact spectrum of the anomalous dimensions
of the maximal helicity baryon operators, see Fig.~\ref{figure2}.

\psfrag{a}[cc][cc]{$a$}
\psfrag{b}[cc][cc]{$b$}
\psfrag{N}[cc][cc]{$\mbox{\small N}$}
\psfrag{q}[cc][cc]{$q$}
\begin{figure}[t]
\centerline{\epsfysize14.0cm\epsffile{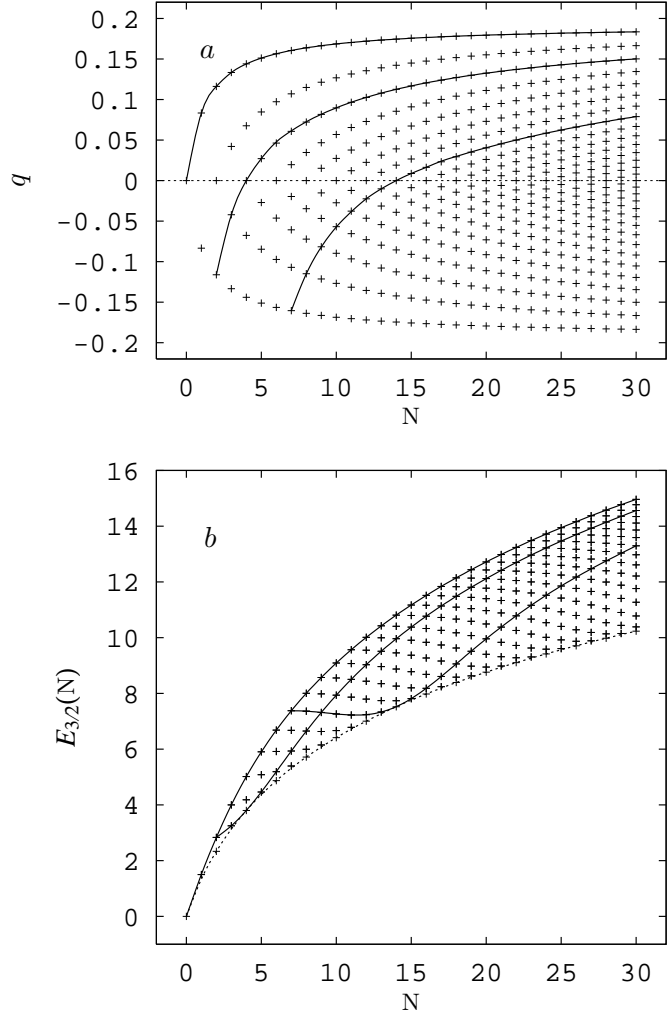}}
\caption[]{ The spectrum of eigenvalues for the conserved charge, $q$ in ($a$)
and for the helicity-3/2 Hamiltonian, $\mathcal{E}_{N,q}$ in ($b$). }
\label{figure2}
\end{figure}
We would like to stress that the function $Q(u)$ has profound physical meaning:
It defines the eigenstates of the $SL(2,\mathbb{R})$ magnet in the representation
of the Separated Variables~\cite{Sklyanin}. Due to complete integrability of the
Hamiltonian $\mathcal{H}_{3/2}$, its eigenfunctions $\Psi_{N,q} (z_1,z_2,z_3)$
have to diagonalize the conserved charge $\mathcal{Q}$ defined in \re{Q3}. In
general, the solutions to the differential equation
$\mathcal{Q}\Psi_{N,q}(z_1,z_2,z_3)=q\Psi_{N,q}(z_1,z_2,z_3)$ are complicated
functions of the light-cone variables $z_k$. Obviously, the spectrum of
$\mathcal{H}_{3/2}$ and $\mathcal{Q}$ does not depend on the representation in
which these operators are defined. The main idea behind the method of Separated
Variables is to perform a unitary transformation on the Hilbert space of
$\mathcal{H}_{3/2}$
\begin{equation}
\mathcal{H}_{3/2} {\rightarrow} \mathcal{H}_{3/2}^{({\rm SoV})}
=
U \mathcal{H}_{3/2} U^\dagger
\, , \qquad
\Psi_{N,q} \rightarrow \Psi_{N,q}^{({\rm SoV})} = U \Psi_{N,q}
\label{SoV-general}
\end{equation}
in order to go over to a representation in which the eigenfunction
$\Psi_{N,q}^{({\rm SoV})}$ is factorized into a product of functions
depending on a single variable. Denoting the coordinates in the SoV
representation as
\begin{equation}
(z_1,z_2,z_3)\quad \stackrel{\rm SoV}{\longrightarrow} \quad (z_0,x_1,x_2)
\end{equation}
with $z_0$ being the center-of-mass coordinate of the system of three quarks
on the light-cone and $x_{1,2}$ being new collective coordinates, one has
\begin{equation}
\Psi_{N,q}^{({\rm SoV})} = {\rm e}^{i p z_0} Q(x_1) Q(x_2)\,.
\label{Sov-wave-function}
\end{equation}
The explicit form of the unitary transformation to the Separated Variables,
Eqs.~\re{SoV-general} -- \re{Sov-wave-function}, was found in \cite{DKM2002}.
Here $p$ is the total (light-cone) momentum of three quarks $\sum_{k=1}^3
i\partial_{z_k} \Psi_{N,q}(z_1,z_2,z_3) = p \Psi_{N,q}(z_1,z_2,z_3)$, or
equivalently the ``$-$'' component of the total momentum of the
$SL(2,\mathbb{R})$ spin chain. The function $Q(x)$ defines the wave function
in the separated variables, Eq.~\re{Sov-wave-function}.
In this way the Schr\"odinger equation for $\Psi_{N,q}$ is translated into
the Baxter relation for $Q(x)$,
Eq.~\re{Baxter-eq}. We conclude that the Baxter equation encodes the
 complete information about the system --
its solutions $Q(x)$ govern the energy
spectrum, Eq.~\re{E-Q}, and define the wave functions in the representation of
Separated Variables, Eq.~\re{Sov-wave-function}.

\subsection{Semiclassical expansion}

The spectrum of the anomalous dimensions shown in Fig.~\ref{figure2}
exhibits remarkable regularity. One can reveal its origin by making
use of the interpretation of $Q(u)$ as the wave function in the separated
variable. Assuming a  parametrization
$Q(x) = \exp(\frac{i}{\hbar}S(x))$, one can solve the Baxter equation
\re{Baxter-eq} by making a semiclassical expansion for
$\hbar\to 0$. Along this line of reasoning, the r\^ole of the Planck
constant is played by the (inverse) total conformal spin $1/J =
1/(N + 3j)$ for $N \gg 1$.

To be more precise, we introduce a small parameter $\eta$
\begin{equation}
\eta = [J(J - 1)]^{-1/2}
\, , \qquad J = N + 3j
\label{eta}
\end{equation}
and replace $u=x/\eta$ as the parameter in Eq.~\re{Baxter-eq}. Introducing
the notation $f(x)$ for the rescaled $Q$-function
\begin{equation}
f(x)=\eta^J (x/\eta)^3 Q(x/\eta)\,, \qquad \widehat q = q\eta^3\,.
\label{hat}
\end{equation}
the Baxter equation \re{Baxter-eq} takes the form
\begin{equation}
f(x+i\eta)+f(x-i\eta)=x^{-3} t(x) f(x)\,, \label{B-f}
\end{equation}
where in the expression \re{transfer} for the transfer matrix $t(x)$, the
parameter  $q$ is replaced by $\widehat q$ defined in \re{hat}. We now
look for the solutions of the rescaled Baxter equation in the form of
the WKB expansion
\begin{equation}
f(x)=\exp\lr{\frac{i}{\eta}S_0(x)+ i S_1(x) + ...}\,, \label{exp}
\end{equation}
where each term is assumed to be uniformly bounded and the expansion is assumed to
be convergent. Substituting the ansatz \re{exp} in Eq.~(\ref{B-f}) and
expanding both sides to order $\eta$, one obtains \cite{Korchemsky96}
\begin{equation}
2\cosh S_0'(x)=t(x)/x^3\,,\qquad S_1'(x)=\frac{i}2 S_0''(x)\coth S_0'(x)\,,
\label{S0}
\end{equation}
where a prime denotes a derivative with respect to $x$. To this accuracy,
the solution of the Baxter equation \re{B-f} is given by
\begin{equation}
f(x)=\frac1{\left[\sinh S_0'(x)\right]^{1/2}} \exp\lr{\frac{i}{\eta}\int^x d
S_0(x)+ \CO(\eta)}\,. \label{f}
\end{equation}
As usual, the leading term $S_0(x)$ defines the classical action function
of the underlying quantum system. Introducing the notation
\begin{equation}
\omega(x)=\exp(S_0'(x))= \exp(p(x))
\label{S0-1}
\end{equation}
one can rewrite the first equation in \re{S0} as
\begin{equation}
\omega+\frac1{\omega}=x^{-3} t(x)\,.
\label{spe}
\end{equation}
This equation establishes the relation between the classical momentum $p$ and the
coordinate $x$ for ``equal energy levels'' defined by the charge $q$. Taking into
account that $\omega$ is real on classical trajectories, we obtain the conditions
that define the classically allowed regions of motion for quarks as
\begin{equation}
t^2(x_j) \ge 4 x_j^{6} \,, \qquad (j=1,2)\,,
\label{band}
\end{equation}
where the subscript refers to the $j-$th allowed band.

Next, we continue the relation \re{spe} into the complex domain, define
$y=x^N\lr{\omega-\frac1{\omega}}$ and rewrite \re{spe} as~\cite{Korchemsky96}
\begin{equation}
\Gamma: \qquad y^2=t^2(x)-4 x^{6}=(-x+\widehat q)(4 x^3 - x + \widehat q)
\, ,
\label{GamN}
\end{equation}
where $t(x)= 2x^3 - x + \widehat q$\,\, is a polynomial depending on the
charge $\widehat q$. The curve \re{GamN} determines an elliptic Riemann
surface $\Gamma$ equipped with a meromorphic differential $dS_0=p(x)dx$
defined in \re{S0}. Integrating this differential along some path that
terminates at $x$ one obtains the ``action'' function $\int^x dS_0 =
\int^x p(x) dx$. In general, for arbitrary values of the conserved charge
$q$, this function is single-valued on the Riemann surface but double-valued
on the complex $x-$plane. We recall that the same ``action'' function
defines the wave function in the separated coordinates, Eq.~\re{f}, which
has to be a single-valued function on the real $x-$axis. To satisfy this
requirement, the charge $q$ has to satisfy certain quantization conditions.
In the semiclassical approach, they take the form of standard Bohr--Sommerfeld
quantization conditions imposed on the solution in Eq.~\re{f}
\begin{equation}
a_k\equiv \oint_{\alpha_k} dS_0 = 2\pi \eta \left(n_k+\frac12\right) + O(\eta^2)
\label{B-S}
\end{equation}
with $k=1,2$ and $n_k$ being integer, such that $n_1+n_2=N$. Here the integration
goes over the closed contours $\alpha_k$ on the Riemann surface encircling the
intervals of the classical motion on the real $x-$axis, Eq.~\re{band}. Solving
\re{B-S}, one finds the spectrum of quantized charges
\begin{equation}
q(N,n_1) =\frac{\eta^{-3}}{\sqrt{27}}\left[1-3 \eta \left(n_1+\frac12\right)+\eta^2
\left(2n_1^2+2n_1-\frac{13}{24}\right)+\ldots \right]\,,
\label{WKB-q}
\end{equation}
with $\eta$ defined in \re{eta}. The integer $n_1$ parameterizes different
``trajectories'' as shown in Fig.~\ref{figure2}. Note that all non-zero
eigenvalues of $\mathcal{Q}$ come in pairs: If $q$ is an eigenvalue,
then $-q$ is also an eigenvalue. As a consequence of the symmetry of
quantization conditions in the separated coordinates, the trajectories
$q (N,n_1)$ have the reflection symmetry $q (N,n_1) = - q (N,N-n_1)$,
which maps positive values of $q$ on the $n_1$-th trajectory into the
negative $q$ on the $n_2 = (N - n_1)$-th trajectory.

Finally, the energy eigenvalues $\mathcal{E}_{N,q}$ can be calculated
as a function of the spin $N$ and the charge $q$ with the help of
Eq.~(\ref{E-Q}). At the semiclassical level, one obtains
\begin{equation}
\mathcal{E}_{N,q}
=
2 \ln 2 - 6 + 6\gamma_{\rm E} + 2 \, \Re{\rm e}\sum_{k=1}^3
\psi(1+i\delta_k/\eta^3) + \mathcal{O}(\eta^6)
\label{WKB-E}
\end{equation}
where $\delta_k$ are the roots of the transfer matrix $t_3 (\delta_k)
= 2\delta_k^3-\delta_k+\widehat q=0$. Combined together,
Eqs.~\re{WKB-q} and \re{WKB-E} provide a good quantitative description
of the spectrum of the anomalous dimensions for helicity$-3/2$ baryon
operators. Note that $\mathcal{E}_{N,q}= \mathcal{E}_{N,-q}$ so that
the anomalous dimensions are double degenerate, apart from the lowest
eigenvalues for even $N$ which correspond to $q=0$. The origin of this
degeneracy can be traced to the permutation symmetry between the three
quarks \cite{BDKM99}.

\subsection{Breakdown of integrability}

There is no reason to expect that QCD is an integrable theory, simply because of
the complexity of phenomena that it is supposed to describe. This
complexity is certainly related to breaking of the conformal symmetry,
but even so breaking of integrability in cases where the conformal
symmetry is preserved may lead to important clues. The renormalization
of the three-quark operator $B^{1/2}$ provides one with a very
instructive example of what can happen. As mentioned above, the
operators $B^{3/2}$ (integrable) and $B^{1/2}$ (non-integrable) are
related to quark distributions at small transverse separations inside
the $\Delta$-isobar and the nucleon, respectively, and it is natural
to speculate that the difference in light-cone dynamics in these two
hadrons is related to the loss of integrability in the latter case.
Let us study this difference in some detail.

The scale dependence of  $B^{1/2}$ is driven by the Hamiltonian
${\mathcal H}_{1/2}$ defined in (\ref{H12}), which differs from
${\mathcal H}_{3/2}$ by the two terms corresponding to gluon exchange
between quarks of opposite chirality (see Fig.~\ref{figure1}):
\begin{equation}
{\mathcal H}_{1/2} = {\mathcal H}_{3/2} + {V}\,,\qquad
{V}= -\left(\frac1{L_{12}^2}+\frac1{L_{23}^2}\right).
\labeltest{V}
\end{equation}
To visualize the effects of this additional interaction it proves to be
convenient to consider a somewhat more general Hamiltonian
\begin{equation}
{\mathcal H}(\epsilon) = {\mathcal H}_{3/2} + \epsilon {V}\,,\qquad
\labeltest{Heps}
\end{equation}
with $\epsilon$ being a new coupling constant. ${\mathcal H}(\epsilon=0)$
reproduces ${\mathcal H}_{3/2}$ whereas ${\mathcal H}(\epsilon=1)$ coincides
with the Hamiltonian ${\mathcal H}_{1/2}$. Thus, the spectra of
${\mathcal H}_{3/2}$ and ${\mathcal H}_{1/2}$ are related to each other
through the flow of the energy levels of ${\mathcal H}(\epsilon)$ from
$\epsilon=0$ to $\epsilon=1$ (see Fig.~\ref{figure7}).
\begin{figure}[t]
\centerline{
\epsfig{file=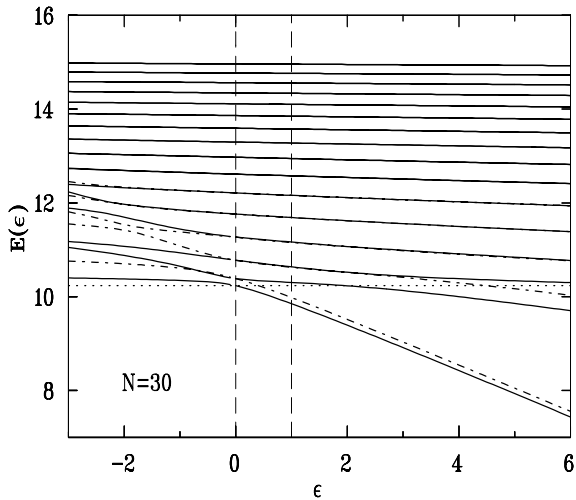, height=7.5cm, width=6.5cm, angle=90}
}
\caption[]{\small The flow of energy eigenvalues for the Hamiltonian
${\mathcal H}(\epsilon)$ for $N=30$ (see text).
The solid and the dash-dotted curves show
the parity-even and parity-odd levels, respectively.
The two vertical dashed lines
indicate ${\mathcal H}_{3/2}\equiv{\mathcal H}(\epsilon=0)$ and
${\mathcal H}_{1/2}\equiv{\mathcal H}(\epsilon=1)$, respectively. The horizontal
dotted line shows position of the ``ground state'' that corresponds to
$q=0$.
}
\label{figure7}
\end{figure}
Note that for  $\epsilon\neq 0$, the cyclic permutation symmetry
between the three quarks is also lost, but ${\mathcal H}(\epsilon)$ is still
invariant under permutations of the first and the third quarks;
$[{\mathcal H}(\epsilon),{\mathcal P}_{13}]=0$. As a result, the
degeneracy between parity-odd and parity-even eigenstates is lifted
and, in fact, the flows of levels with different parity are completely
independent from one another.

The spectra in Fig.~\ref{figure7} exhibit the following characteristic
features:
\begin{enumerate}
\item[--] In the upper part of the spectrum the effect of
$\epsilon-$proportional terms on the spectrum of the `unperturbed'
Hamiltonian ${\mathcal H}(\epsilon=0)={\mathcal H}_{3/2}$ is very
mild. While at $\epsilon=0$ the energy levels are double degenerate,
their splitting at $\epsilon\neq 0$ remains (exponentially, as one
can show \cite{BDKM99}) small for large $N$.
\item[--] For $\epsilon>0$, the two lowest levels are decoupled from
the rest of the spectrum and fall off with $\epsilon$ almost linearly.
In the large$-N$ limit they become separated from the rest of the spectrum
by a finite ``mass gap''.
\end{enumerate}
This structure suggests that the difference $\epsilon {V} =
{\mathcal H}(\epsilon) - {\mathcal H}(\epsilon=0)$ can be considered
as a perturbation for most of the levels, but not for the few lowest
ones (for large $N$). Moreover, for $\epsilon > 0$, which corresponds to
attractive interaction, we seemingly witness creation of a certain
bound state.

To formalize the argument, one has to evaluate the matrix elements
of $V$ between the eigenstates of ${\mathcal H}_{3/2}$ and
compare them with the energy splittings for the ``unperturbed''
Hamiltonian. By explicit calculation one confirms that the
perturbation theory in ${V}$ is justified for large $N$ for the
upper part of the spectrum, while several lowest energy eigenstates
are affected strongly and have to be re-diagonalized. It turns out
that the matrix elements ${\langle \Psi_q|{V} |\Psi_q\rangle}$ for
low-lying energy eigenstates can be computed in terms of eigenvalues
of the cyclic permutation operator
\begin{equation}
{\mathcal P} \Psi_q(x_1,x_2,x_3)
=
\Psi_q(x_3,x_1,x_2) = e^{-i\phi_q} \Psi_q(x_1,x_2,x_3)
\, ,
\end{equation}
where the phases $\phi_q$ take quantized values $ \phi_q =
0,\frac{2\pi}{3}, \frac{4\pi}{3}$ (since ${\mathcal P}^3=1$).
One obtains \cite{BDKM99}
\begin{equation}
\langle \Psi_{q'}|V|\Psi_{q} \rangle
=
-\frac{\pi^2}9 \frac{1}{\ln N}\cos(\phi_q-\phi_{q'})
\labeltest{Mr12}
\end{equation}
It is easy to see that the possible values of $\cos(\phi_q\pm\phi_{q'})$
are $1$ and $-1/2$ depending on whether  the phases $\phi_q$ and $\phi_{q'}$
coincide. Introducing an integer $k = 0,\pm 1, \pm2,\ldots$ to enumerate
quantized values of $q$ starting from the ones with the lowest absolute
value, we get
\begin{equation}
\langle \Psi_{q'}|\epsilon V| \Psi_q \rangle =
-g
\Lambda_{k'k}\,,\qquad
\Lambda_{k'k}=\left(
\begin{array}{rrrrr}
1&-\frac12&-\frac12&1&\ldots\\
-\frac12&1&-\frac12&-\frac12&\ldots\\
-\frac12&-\frac12&1&-\frac12&\ldots\\
1&-\frac12&-\frac12&1&\ldots\\
\vdots&\vdots&\vdots&\vdots&\ddots
\end{array}
\right),
\labeltest{potential}
\end{equation}
where the dependence on $\epsilon$ and $\eta$ is absorbed
in the `effective coupling' $ g= \frac{\epsilon\pi^2}{9\ln N} $.
Notice the $\sim 1/\ln N$ falloff for large $N$, which is slower
than the level splitting for ${\mathcal H}_{3/2}$, $\delta E_{3/2}
\sim 1/(\ln N)^2$, confirming our expectation that $\sim \ln N$
lowest levels have to be re-diagonalized. Neglecting, for a moment,
the  splitting between unperturbed levels up to some number
$k_{\rm max}\sim \ln N$, the true eigenstates would coincide with
those of $\Lambda_{kk'}$ (of  the size $k_{\rm max}$), and the
corresponding eigenvalues would define the energies. Remarkably,
the spectrum of $\Lambda_{kk'}$ is extremely simple. One can easily
convince oneself that $\Lambda_{kk'}$ has two and only two nonzero
eigenvalues which are both equal to $k_{\rm max}$. Remembering that
$g \sim 1/\ln N$ and $k_{\rm max} \sim \ln N $, this implies that
two energy levels will get shifted by the finite amount $g k_{\rm max}
= \CO(\ln^0 N )$ while all the other ones remain exactly degenerate
to this accuracy.  Since (\ref{potential}) was derived up to
corrections of order $\sim 1/\ln^2 N$, this implies that the true
energy shift for all levels apart from the lowest two is, at most,
$\sim 1/\ln^2 N$. This explains the creation of the mass gap observed
in Fig.~\ref{figure7}.

The size of the gap can be calculated for small $\epsilon$ by
constructing an effective Hamiltonian for the low-lying eigenstates
\cite{BDKM99}. The idea of the solution is to interpret the integer
$k$ as a discrete momentum variable. The corresponding wave functions
in configuration space correspond to Bloch--Floquet waves and the
resulting Schr\"odinger equation turns out to be a generalization
of the famous Kroning--Penney model of a single particle in a periodic
$\delta$-function potential. The result for the mass gap reads
\begin{equation}
 \Delta (\epsilon) = {\mathcal E}_{\rm bound}-{E}_0
     =-\epsilon^2\frac{\pi^4}{72\zeta(3)} +{\mathcal O}(\epsilon^3)\,.
\label{gap1}
\end{equation}
Extension of this approach to the physically interesting case
$\epsilon \sim 1$ is nontrivial. However, the mass gap can
easily be calculated in the opposite limit $\epsilon \gg 1$, and
the result for $\epsilon =1$ found by matching the two expansions
using Pad\'e approximants: $\Delta_{1/2} \equiv \Delta(\epsilon =1)
= - (0.32 \pm 0.02)$ \cite{BDKM99}.

Creation of the mass gap implies that the wave function of the
ground state is drastically modified. {}For the integrable case,
the eigenfunction corresponding to the lowest energy and hence $q=0$
can be found exactly. The physical interpretation of this result is
most transparent in coordinate space.  Up to contribution of operators
with total derivatives, one can represent the `ground state' of three
quarks with the same helicity in a concise form as the nonlocal
light-cone operator
\begin{equation}
\label{Blow32}
B_{3/2}^{(q=0)}(z_1,z_2,z_3)=\frac12 \sum_{a,b=1,2,3
\atop
a\neq b} \!\varepsilon^{ijk}\!\int_0^1\!dv\, \!\not\!n \psi_{i}^\uparrow (z_a)
\!\not\!n \psi_{j}^\uparrow(vz_a+(1-v)z_b)
\!\not\!n \psi_{k}^\uparrow (z_b)\,.
\end{equation}
The Taylor expansion of (\ref{Blow32}) at short distances, $z_{12},
z_{32} \to 0$, generates the series of local multiplicatively
renormalizable three-quark operators with the lowest anomalous
dimension for each even $N$. Note the  integration in Eq.\
(\ref{Blow32}) with unit weight over the position of the quark
in the middle that goes in between  the light-cone positions
of the other two quarks, up to permutations.  If renormalization
of the operator is interpreted as interaction, this unit weight
can in turn be interpreted as the statement that the quark in the
middle is effectively `free': In the `ground state' with the lowest
``energy'', the interaction of the quark in the middle with its
right and left neighbors exactly compensate each other.

{}For the $B^{1/2}$, case the ground wave function  changes
dramatically and in the limit of large $N$ becomes, in the
same sense as (\ref{Blow32}),
\begin{eqnarray}
B^\pm(\alpha_1,\alpha_2,\alpha_3)&=&
\varepsilon^{ijk}\left[ (\!\not\!n \psi_{i}^\uparrow
\!\not\!n \psi_{j}^\downarrow)(\alpha_1)
\!\not\!n \psi_{k}^\uparrow (\alpha_3)\delta(\alpha_2-\alpha_1)\right.
\nonumber\\
&&{}\left.
\hspace*{0.5cm}\pm
\!\not\!n \psi_{i}^\uparrow(\alpha_1)
(\!\not\!n \psi_{j}^\downarrow)
\!\not\!n \psi_{k}^\uparrow)(\alpha_3)\delta(\alpha_2-\alpha_3)\right],
\label{Blow12}
\end{eqnarray}
where the two solutions correspond to states with different parity (cf.\
Fig.~\ref{figure7}). Formation of the mass gap in the spectrum of anomalous
dimensions is, therefore, naturally interpreted as due to binding of the
quarks with opposite helicity into scalar diquarks.

Note that while the expression (\ref{Blow32}) for the eigenfunction is
exact, the result (\ref{Blow12}) is only valid in the asymptotic
$\ln N\to\infty$ limit. In the coordinate space picture, the restriction
to large $N$ is translated to the condition that the light-cone separation
between the same helicity quarks is very large to allow for the formation
of a diquark. In other words, the diquark itself has a large size. In the
momentum space, the result means that at sufficiently large normalization
scale $\mu^2$ the quark carrying a very large momentum fraction is more
often with the same helicity as of the parent baryon. This observation
seems to be in qualitative agreement with phenomenological models of baryon
distribution amplitudes.

\subsection{Open spin chains}

There are other multiparticle operators which represent phenomenological interest
and which were studied along the same line as the three-quark operators. For the
twist-three sector, in addition to three-quark operators that we have addressed
above one considers quark-antiquark-gluon and three-gluon operators. They arise
in particular in connection with the structure function $g_2(x,Q^2)$ measured in
polarized deep-inelastic scattering. This is attracting increasing interest as
it provides a direct measurement of three-parton correlations in the proton. The
scale-dependence of $g_2(x,Q^2)$ can be traced to the renormalization-group
equations for the quark-antiquark-gluon operator~\cite{SV81}
\begin{eqnarray}\label{Spm}
S^\pm (z_1,z_2,z_3)
=
\frac12\bar \psi(z_1)
{}[ ig\widetilde F_{\perp +}(z_2) \pm g F_{\perp +}(z_2)\gamma_5]
{\not\!n}
\psi(z_3)
\, ,
\end{eqnarray}
where $\widetilde F_{\mu \nu}=\frac12\varepsilon_{\mu\nu\rho\lambda}
F^{a,\rho\lambda}t^a$, $F_{\mu \nu}= F_{\mu\nu}^a t^a$. In the flavor
singlet sector, one also has to include the three-gluon operator
\begin{eqnarray}
\label{Odual}
\widetilde O (z_1,z_2,z_3)
=
\frac{i}{2} g f^{abc}
F^a_{+\lambda}(z_1) \widetilde F^b_{+\perp}(z_2) F^c_{+\lambda}(z_3)
\, ,
\end{eqnarray}
which mixes with the C-even part of the flavor singlet
quark-antiquark-gluon operator operator
(\ref{Spm}).

To begin with, consider the flavor non-singlet sector. In this case the
three-gluon operator drops out and the evolution of $S^\pm$ simplifies
dramatically in the limit of large number of colors, $N_c\to\infty$. The
corresponding Hamiltonian reads, e.g., for $S^+$ \cite{BFLK85,BDM98}
\begin{equation}
\label{qGq-1}
\mathbb{H}_{S^+}
=
\frac{\alpha_s}{2 \pi} N_c
\Big[ V_{qG}(J_{12})+ U_{Gq}(J_{23}) \Big]
+
{\mathcal O}\left(1/{N_c}\right)
\, ,
\end{equation}
where
\begin{eqnarray}\label{qGq-2}
V_{qG}(J) &=& \psi \left( J + \ft{3}{2} \right)
+
\psi \left( J - \ft{3}{2} \right) - 2 \psi (1) - \ft{3}{4}
\, ,
\nonumber\\
U_{Gq}(J) &=&
\psi \left( J + \ft{1}{2} \right)
+
\psi \left( J - \ft{1}{2} \right)
-
2 \psi (1) - \ft{3}{4}
\, ,
\end{eqnarray}
and it turns out to be equivalent to the Hamiltonian of the completely
integrable \textsl{open} Heisenberg magnet \cite{BDM98}. The corresponding
conserved charge is equal to
\begin{equation}\label{qGq-charge}
{\mathcal Q}_{S^+} = \{{ L}_{12}^2,{ L}_{23}^2\}
-
\frac12{ L}_{12}^2 - \frac92{ L}_{23}^2\,,
\end{equation}
where $\{,\}$ stands for an anticommutator. Properties of the Hamiltonian
\re{qGq-1} were
studied in detail in Refs.\ \cite{BDM98,Belitsky99a,DKM00,Belitsky00} and the
corrections $\mathcal{O}(1/N_c^2)$ to the spectrum were calculated in Ref.\
\cite{BKM00}. The most interesting result is that the lowest energy, albeit the
lowest anomalous dimension in the spectrum of quark-gluon operators can be
exactly found, and the corresponding eigenfunction coincides to leading
logarithmic accuracy with the structure function $g_2$ itself. Thus, a simple
evolution equation arises~\cite{ABH91}.

The study of the flavor singlet sector is much more challenging \cite{BKM01}.
Ignoring quarks for a moment, the evolution kernel for the three-gluon
operators (\ref{Odual}) {can be mapped to}
the Hamiltonian of \textsl{closed}
Heisenberg magnet with three ``gluonic'' sites of the spin $j_g=3/2$, whose
integrability is ``softly'' broken by additional terms \cite{Belitsky00}.%
\footnote{\,The exactly integrable chain arises for maximal-helicity
gluonic operators
\cite{Belitsky00}.}
This case was studied using the same techniques as the
three-quark operators, although the algebra becomes more complicated because
of the higher spins involved. The most interesting part appears to be the
mixing between quark-gluon and three-gluon operators which for large $N$
can be interpreted as describing the interaction between open and closed
Heisenberg magnets. It turns out that this mixing has rather peculiar
features, which we cannot discuss in detail in this review, but the outcome
is that the mixing can to a large extent be reduced to a few participating
levels. Identifying important degrees of freedom (at least for large $N$)
one can write down an approximate two-channel evolution equation for the
structure function $g_2$ in terms of transverse spin densities \cite{BKM01}.

\subsection{Quasipartonic and non-quasipartonic operators}
\label{QPOvsNONQPO}

The operators we have discussed above all fall into the same class of
the so-called quasipartonic operators, introduced in Ref.~\cite{BFLK85}.
Since an offspring of this formalism will be {useful in the
discussion of supersymmetric extensions of QCD, let us elaborate
upon it.} The approach heavily relies on the light-cone
gauge, $A_+ = 0$. In this gauge, only physical degrees of freedom
propagate, i.e., ``good'' components of the quark field $\psi_+$ and
circularly polarized gluons $\bit{A}^\mu_\perp$. Thus the identification
of the elementary fields entering the Lagrangian with partonic states is
straightforward -- hence the name quasipartonic operators. Introducing
the helicity operator as%
\,\footnote{\,Do not confuse helicity $h$ with the light-cone spin, discussed above.}
\begin{equation}
\label{Helicity}
h \equiv
\bar{\bit{e}}^i_\perp {\bit{e}}^j_\perp \Sigma^{ij}
\, ,
\qquad
\bit{e}^i_\perp = \left( \bar{\bit{e}}^i_\perp \right)^\ast
= \ft{1}{\sqrt{2}} (1, i)
\, ,
\end{equation}
one decomposes all fields in components with a fixed helicity. In particular,
the left $\psi_+^{\downarrow}$ and right $\psi_+^{\uparrow}$ quarks,
together with (anti)holomorphic components of the two-component
propagating gauge potential $A \equiv \ft{1}{\sqrt{2}} (A^x + i A^y)$
and $\bar A \equiv A^\ast$ are eigenfunctions of the helicity operator
corresponding to
\begin{equation}
\begin{array}{cc}
h \psi^\uparrow_+ (x) = \ft12 \psi^\uparrow_+ (x)
\, , \qquad
&
h \psi^\downarrow_+ (x) = - \ft12 \psi^\downarrow_+ (x)
\, , \\
h \bar A (x) = \bar A (x)
\, , \qquad
&
h A (x) = - A (x)
\, .
\end{array}
\end{equation}
Let us notice in passing that the ``good'' quark fields
of definite helicity have a single nonzero complex component, which
suggests that one can entirely get rid of the spinor indices. A general
quasipartonic operator is {built} from the above parton fields
\begin{equation}
\mathbb{O} (z_1 , \dots)
=
\prod_i A (z_i)
\prod_j \bar A (z_j)
\prod_k \psi^\uparrow_+ (z_k)
\prod_l \psi^\downarrow_+ (z_l)
\, .
\end{equation}
Its covariant analogue can be found {expressing} the (anti)holomorphic gauge
potentials in terms of the field strength tensor,
\begin{eqnarray}
A
&=&
\ft{1}{\sqrt{2}} \partial_+^{- 1} (F^{+ x} - i \widetilde F^{+ x})
=
\ft{i}{\sqrt{2}} \partial_+^{- 1} (F^{+ y} - i \widetilde F^{+ y})
\, , \nonumber\\
\bar A
&=&
\ft{1}{\sqrt{2}} \partial_+^{- 1} (F^{+ x} + i \widetilde F^{+ x})
=
- \ft{i}{\sqrt{2}} \partial_+^{- 1} (F^{+ y} + i \widetilde F^{+ y})
\, .
\end{eqnarray}

{Contrary to ``good'' operators,
the evolution of operators containing ``bad'' components of
field operators does not have, in general, a simple pairwise pattern of the
renormalization group mixing even at leading order in the coupling
constant.} A few exceptions do exist, however, for example in the case of the
two-quark sector with Dirac structures
\begin{equation}
\bar\psi (z_1 n)
(1, \gamma^\mu_\perp,  \gamma^\mu_\perp \gamma^5)
\psi (z_2 n)
\, .
\end{equation}
After separation of twist-two and three components from these operators,
one finds that, in the large-$N_c$ limit, both of them evolve autonomously
with different pairwise Hamiltonians. The twist-two part is well known.
The eigenvalues of the twist-three Hamiltonian coincide with the lowest
anomalous dimensions for quark-antiquark-gluon operators introduced
in Sect.~2.6 ( see \cite{ABH91,BBKT96,Belitsky97a,Belitsky97b}).

The situation with more generic non-quasipartonic operators is an open
issue. One of the major complications is that, in distinction with
quasipartonic operators, the construction of the irreducible basis of
non-quasipartonic operators is a nontrivial task. Up to now there exists
no constructive procedure for the classification of non-quasipartonic
operators of arbitrary high twist. The representation theory used in the
analysis of usual quasipartonic operators is not very handy since it does
not allow one to get rid of the redundancy of the operators in the basis.
The redundancy in the basis, in turn,  leads to complications in finding
the eigensystem for the mixing matrix. Currently it can be solved only on
a case-by-case basis. Finding a systematic procedure is an open question
for further research.

A particular case of certain non-quasipartonic local multi-gluon higher-dimension
operators of the type
\begin{equation}
\label{LocalGluonOper}
\prod_{j = 1}^L F_{\mu_j \nu_j} (0)
\end{equation}
in pure gluodynamics have been addressed recently and found to possess integrable
structures corresponding to the compact spin-one Heisenberg magnet
\cite{Ferretti:2004ba}.%
\footnote{\,Other operators exist but do not affect the anomalous dimensions matrix as was
demonstrated in \cite{Gracey02} for $L=3,4$ by explicit calculation.}
 Without loss of generality we can rephrase their analysis
entirely in Euclidean space. The strength tensor can be decomposed into
irreducible components
\begin{equation}
F_{\mu\nu} = \eta^A_{\mu\nu} F^A_{+} + \bar\eta^A_{\mu\nu} F^A_-
\end{equation}
with the help of 't Hooft symbols, $O (4) \sim SU (2) \otimes SU(2)$.
The selfdual and anti-selfdual components transform as $(1, 0)$ and
$(0, 1)$, respectively. The part of the RG Hamiltonian responsible for
eigenvalues of autonomous components does not change the number of
fields in the local gluonic operator (\ref{LocalGluonOper}). By matching
the coefficient of different irreducible components, extracted by means
of projectors $\mathbb{P}^P_{(j_1, j_2)}$ for spin-$j$ and parity $P$,
to the available one-loop calculations for gluonic operators up to dimension eight,
the pairwise Hamiltonian was found to be  \cite{Ferretti:2004ba}
\begin{equation}
\mathbb{H}_{12}
= 7
\left( \mathbb{P}_{(2,0)} + \mathbb{P}_{(0,2)} \right)
+
\mathbb{P}_{(1,0)} + \mathbb{P}_{(0,1)}
- 11
\left( \mathbb{P}_{(0,0)}^+ + \mathbb{P}_{(0,0)}^- \right)
+ 3
\mathbb{P}_{(1,1)}^-
\, .
\end{equation}
The projection on the selfdual operators, i.e., built from products
of $F^A_+$, reduces the above Hamiltonian to
\begin{equation}
\mathbb{H}_{12}^{\rm sd}
= 7
\mathbb{P}_{(2,0)}
+
\mathbb{P}_{(1,0)}
- 11
\mathbb{P}_{(0,0)}
\, ,
\end{equation}
where the projection operators extract maximal-spin, antisymmetric and trace
components and have the following obvious representation
\begin{eqnarray*}
\mathbb{P}_{(2,0)}
F^A_+ F^B_+
&=&
\ft12
\left(
F^A_+ F^B_+ + F^A_+ F^B_+ - \ft23 \delta^{AB} F^C_+ F^C_+
\right)
\, , \\
\mathbb{P}_{(1,0)}
F^A_+ F^B_+
&=&
\ft12
\left(
F^A_+ F^B_+ - F^A_+ F^B_+
\right)
\, , \\
\mathbb{P}_{(2,0)}
F^A_+ F^B_+
&=&
\ft13
\delta^{AB} F^C_+ F^C_+
\, .
\end{eqnarray*}
They can be easily related to the permutation $P F^A_+ F^B_+ = F^B_+
F^A_+$, trace $K F^A_+ F^B_+ = \delta^{AB} F^C_+ F^C_+$ and identity
$I F^A_+ F^B_+ = F^A_+ F^B_+$ operators,
$$
\mathbb{P}_{(2,0)}
=
\ft12 \left( I + P \right) - \ft13 K
\, , \quad
\mathbb{P}_{(1,0)}
=
\ft12 \left( I - P \right)
\, , \quad
\mathbb{P}_{(0,0)}
=
\ft13 K
\, .
$$
Then, the pairwise Hamiltonian can be brought to the form
\begin{equation}
\mathbb{H}_{12}^{\rm sd}
= 4 I_{12} + 3 P_{12} - 6 K_{12}
= 7 + 3 \bit{s}_1 \cdot \bit{s}_2 \left( 1 - \bit{s}_1 \cdot \bit{s}_2 \right)
\, ,
\end{equation}
where in the second equality we have used the representation in terms of spin-one
$SU (2)$ generators. This is a Hamiltonian of an exactly solvable spin-one
Heisenberg magnet which can be diagonalized by means of the Bethe Ansatz
\cite{KRS81,TTF83,Zamolodchikov80}. It is interesting that the ground state of
the corresponding spin-one chain is antiferromagnetic contrary to the
ferromagnetic situation for the spin chains governing the renormalization of the
scalar operators in $\mathcal{N}=4$ SYM theory.

\subsection{Conformal symmetry breaking}

Nothing is known at present about whether integrable structures survive in
QCD beyond leading logarithms, in which case conformal symmetry
is broken by the trace anomaly of the energy-momentum tensor. The
constraints imposed by conformal symmetry, and, in particular, by
conformal Ward identities, appear to be nontrivial and can be used
to advance perturbative calculations to higher orders. The idea has
been to establish the following schematic structure of the perturbative
series for a generic quantity ${\mathcal Q}$
\begin{eqnarray}
\label{DefForConLim}
{\mathcal Q} = {\mathcal Q}^{\rm con} + \frac{\beta(g)}{g} \Delta{\mathcal Q}\,
,\quad\mbox{where}\quad
 \Delta{\mathcal Q} = \mbox{power\ series in\ } \alpha_s\,.
\end{eqnarray}
Here ${\mathcal Q}^{\rm con}$ is the result in the formal conformal limit,
obtained by setting the $\beta$-function to zero by hand. It has the full
symmetry of a conformally-invariant theory. The extra term $\Delta{\mathcal Q}$
can be perturbatively evaluated as a power series in $\alpha_s$
and vanishes to leading order (LO). Note that the evaluation of the leading
${\mathcal O}(\alpha_s)$ contribution to $\Delta{\mathcal Q}$ requires
little effort, since it is sufficient to calculate the $N_f$ proportional
terms in $\beta_0=11N_c/3 - 2N_f/3$ via quark bubble insertions.

The possibility of the separation of the conformally-symmetric and
$\beta$-proportional contributions is by no means trivial. The reason is that
although the dilatation anomaly to ${\CO(\alpha_s)}$ accuracy receives
contributions only from UV-divergent parts of relevant Feynman diagrams and is
diagonal for conformal operators, the special conformal anomaly (to the same
accuracy) is affected by the finite parts of the same (one-loop) diagrams and is not
diagonal in a general, ${\rm MS}$-like, renormalization scheme
\cite{Muller:1993hg}. The nondiagonal entries in the special conformal anomaly
can, however, be removed and the special conformal covariance (neglecting the
$\beta$-function) restored by performing a finite renormalization
\cite{Muller:1993hg}. {This} transformation defines the conformal subtraction
scheme in which  the separation in (\ref{DefForConLim}) becomes meaningful.
Whether the remnants of integrability survive in {this} procedure, remains to be
seen.

\section{Dilatation operators in supersymmetric gauge theories}
\setcounter{equation}{0}

During the last year, a lot of new and interesting results have been
obtained on the integrability properties of the $\mathcal{N}=4$ SYM theory.
These studies are important and provide hints on the long standing quest to
find the correspondence between (supersymmetric) Yang--Mills theories
at strong coupling and noncritical strings propagating on the curved
background. Our aim in this section is to establish a connection
between these new developments and the integrability of RG equations
in QCD discussed in the previous section. In particular, the key question
is to distinguish between phenomena that are already present in
{ pure Yang--Mills theory}
(and only get enhanced by the supersymmetry), and ones that
present genuinely new features of SYM gauge theories.

In order to trace this connection, we introduce a superfield
formalism in which all symmetries of the theory become manifest and
calculations can be performed in a unified manner for different
numbers of supercharges $\mathcal{N}=0,1,2,4$.
The maximally-supersymmetric $\mathcal{N}=4$ SYM theory is a finite,
four-dimensional conformal field theory~\cite{Mandelstam83,BLN83,SW81,HST84},
while the $\mathcal{N}=0$ theory corresponds to pure gluodynamics.
Unfortunately, no covariant superfield formulation for
$\mathcal{N}=4$ exists, which makes this case somewhat special.
Because of this, we employ a noncovariant, light-cone
superspace formalism due to Mandelstam~\cite{Mandelstam83} and
Brink~{\it et al.}~\cite{BLN83}. This technique was used, e.g.,
to prove the finiteness of the $\mathcal{N}=4$ SYM theory but is
not very popular nowadays. Hence we begin this section
by reviewing this beautiful framework.

We will find it useful to introduce multi-particle single trace operators
built from light-cone superfields at ``super-light-cone'' separations
in the superspace $Z_k=(z_k,\theta^A_k)$ where $z_k$ are the usual ``even''
light-cone coordinates and $\theta^A_k$ are a set of Grassmanian ``odd'' coordinates. The one-loop dilatation
operator for such operators can be found in terms of simple
field displacements on the super-light-cone, similarly to the QCD case.
Expanding such operators in ``even'' and ``odd'' coordinates, we will
be able to reproduce the integrability properties of QCD operators
with multiple derivatives {discussed in the previous section} \cite{BDM98},
on one side, and composite scalar operators,
considered in \cite{mz} on the other side, which are related to $SL(2,\mathbb{R})$
and $SO(6)$ spin chains, respectively. Hence, these two results, that were found independently,
appear to be just two sides of the same coin.
 We will then comment on the
extension of these results to other operators.

\subsection{SYM theories in light-cone superspace}

All super-Yang--Mills theories in four dimensions can be obtained through
dimensional reduction from a higher-dimensional ${\mathcal N} = 1$ gauge
theory~\cite{Sohnius85}
\begin{equation}
\label{S-general}
S
=
\int d^D x \,
\left\{
- \ft{1}{4} F^a_{MN} F^{MN}_a
+
\ft{i}{2} \,
\bar{\psi}^a {\Gamma}_M
\left( \partial^M \delta_{ab} + g f^{acb} A^M_c \right)
{\psi}^b
\right\}
\, ,
\end{equation}
describing the interaction of the $D$-dimensional gauge field $A^M$ with the
$2^{[D/2]}$-component Majorana--Weyl fermion $\psi$, both belonging to the adjoint
representation of the gauge $SU(N_c)$ group. In particular, the ${\mathcal N} =
2$ SYM theory can be obtained from the $D = 6$ action,
\begin{eqnarray}
\label{N2Lagrangian}
{\mathcal L}_{{\mathcal N} = 2} = {\rm tr} \, \bigg\{ \!\!\!&-& \ft12 F_{\mu\nu}
F^{\mu\nu} + 2 i \bar\lambda_{\dot\alpha A} \sigma^{\dot\alpha \beta}_\mu
{\mathcal D}^\mu \lambda^A_\beta + 2 \left( {\mathcal D}_\mu \bar \phi \right)
\left( {\mathcal D}^\mu \phi \right)
\\
&-& g^2 [\phi , \bar\phi]^2 - \sqrt{2} g \varepsilon_{AB} \lambda^{\alpha A}
[\bar\phi, \lambda_\alpha^B] + \sqrt{2} g \varepsilon^{AB}
\bar\lambda_{\dot\alpha A}
[\phi , \bar\lambda^{\dot\alpha}_B] \bigg\} \, , \nonumber
\end{eqnarray}
(with ${\scriptstyle A,B}=1,2$), and the ${\mathcal N = 4}$ SYM theory follows
from from $D = 10$ action,
\begin{eqnarray}
{\mathcal L}_{{\mathcal N} = 4} = {\rm tr} \,
\bigg\{
\!\!\!&-& \ft12
F_{\mu\nu} F^{\mu\nu}
+
\ft12
\left( {\mathcal D}_\mu \phi^{AB} \right)
\left( {\mathcal D}^\mu \bar\phi_{AB} \right)
+
\ft{1}8 g^2
[\phi^{AB}, \phi^{CD}] [\bar\phi_{AB}, \bar\phi_{CD}]
\nonumber\\
&& \hspace*{-1cm}{}+
2 i \bar\lambda_{\dot\alpha A}
\sigma^{\dot\alpha \beta}_\mu {\mathcal D}^\mu
\lambda^A_\beta
-
\sqrt{2} g
\lambda^{\alpha A}
[\bar\phi_{AB}, \lambda_\alpha^B]
+
\sqrt{2} g
\bar\lambda_{\dot\alpha A}
[\phi^{AB}, \bar\lambda^{\dot\alpha}_B] \bigg\} ,
\end{eqnarray}
(with ${\scriptstyle A,B}=1,...,4$). Obviously, the first term on the r.h.s.\ of
Eq.~(\ref{N2Lagrangian}) gives the Lagrangian of a pure gauge, $\mathcal{N}=0$,
theory, while the sum of the first two terms defines (for ${\scriptstyle A}=1$)
the ${\mathcal N} = 1$ SYM theory in the Weyl form. Thus, augmentation of
supersymmetry increases the number of fundamental fields in the theory, which has
profound consequences. The ${\mathcal N} = 0, 1$ SYM theories possess a
non-vanishing beta-function which receives contributions at all orders of the
perturbative expansion in the coupling. In ${\mathcal N} = 2$ theory the
beta-function gets a contribution at one-loop only, while for ${\mathcal N} = 4$
the beta function vanishes to all orders. Hence the ${\mathcal N} = 4$ theory
remains superconformally-invariant at the quantum
level~\cite{Mandelstam83,BLN83,SW81,HST84}.

In order to construct a SYM theory on the light-cone, one starts with the
component form of the action and fixes the light-cone gauge $A_+(x) = 0$. In this
gauge, one sacrifices four-dimensional Lorentz covariance in favor of a unified
description of $\mathcal{N}-$extended SYM theories in superspace. The main steps
in designing this formalism~\cite{Mandelstam83,BLN83,CJ85} consist of (i) making
use of the decomposition of the elementary fields into propagating and
non-propagating modes in light-cone time $x_+$~\footnote{\,I.e., those that
have or do not have the kinetic term with light-cone time derivative
$\partial_-$.}, in the same way as indicated above in the context of QCD, (ii)
integrating non-dynamical fields out in the functional integral, (iii) assembling
the propagating fields into superfields, and finally (iv) casting the light-cone
component action into the superspace form.

The first two steps can be done for a generic $D$-dimensional ${\mathcal N} = 1$
Lagrangian without specifying the value of $D$. One finds that in the light-cone
gauge $A_+(x)=0$, the action \re{S-general} involves two non-propagating
fields~\cite{Mandelstam83,BLN83,CJ85}: one bosonic $S^a$ and one fermionic
$\chi^a$
\ba
S^a &=& \partial_+ A_-^a - \partial_+^{- 1} \bit{\mathcal D}^{ab}_i \partial_+
\bit{A}_i^b - \ft{i}{2} g f^{bac}
\partial_+^{- 1} \bar{\psi}^b_+ {\Gamma}_+ {\psi}_+^c
\,,
\nonumber \\[3mm]
\chi^a &=& {\psi}^a_- - \ft{1}{2} \partial_+^{- 1} {\Gamma}_+ \bit{\Gamma}^i
\bit{\mathcal D}_i^{ab} {\psi}^b_+\,,
\ea
which can be set to zero. Here $\bit{\mathcal D}^i_{ab} = - \delta_{ab}
\partial^i_{\bit{\scriptstyle x}} + g f_{acb} \bit{A}^i_c$ is the transverse
covariant derivative. Solving $S^a = \chi^a = 0$, one obtains the expressions
for $A^a_-$ and ${\psi}^a_-$ in terms of the propagating fields and substitutes
them back into the SYM action. As the next step, one performs the reduction of
the resulting action to four dimensions, giving rise to various SYM theories
on the light-cone. Their explicit form for $\mathcal{N}> 1$ is very lengthy and
is not of interest {\it per se}, but rather as an intermediate result
on the way to achieving the superspace form.

We go directly to step (iii) and decompose all propagating, ``physical'' fields
into definite helicity components. In the case of $\mathcal{N}=4$ SYM,\footnote{\,As
we will see in a moment, the SYM light-cone theories with less supersymmetry can
be obtained from the $\mathcal{N}=4$ theory through the truncation procedure.}
they include (anti)holomorphic, helicity-$\pm 1$ fields, $A=(A_1+iA_2)/\sqrt{2}$
and $\bar A=(A_1-iA_2)/\sqrt{2}$, built from two-dimensional transverse
components of the gauge field, $A_\perp(x)=(A_1,A_2)$, complex scalar fields
$\phi^{AB}$ of helicity $0$ and helicity $\pm 1/2$ components of Majorana--Weyl
fermions, $\lambda_\alpha^A$ and $\bar\lambda^{\dot\alpha}_A$. An important
property of the light-cone formalism, which makes it advantageous over the
covariant one, is that the latter fields have only one non-vanishing component
$\lambda_1^A \equiv \sqrt[4]{2} \lambda^A$ and $\bar\lambda^{\dot{2}}_A \equiv i
\, \sqrt[4]{2} \bar\lambda_A$. As a consequence, one can describe helicity-$\pm
1/2$ fermions by Grassmann-valued complex fields without any Lorentz index.
Introducing the fermionic coordinates $\theta^A$ possessing the helicity $h
\theta^A = - \ft12 \theta^A$ and their conjugates $\bar\theta_A$ with $h
\bar\theta_A = \ft12 \bar\theta_A$, we can assemble the above fields into a
single, complex, $\mathcal{N}=4$ superfield~\cite{BLN83}
\begin{eqnarray}
\hspace{-3mm}\Phi (x, \theta^A, \bar\theta_A) &=& {\rm e}^{\frac12 \bar\theta_A \theta^A \,
\partial_+} \bigg\{
\partial_+^{-1}A(x)
+ \theta^A \partial_+^{-1}\bar\lambda_A (x) + \frac{i}{2!} \theta^A \theta^B \bar
\phi_{AB} (x)
\nonumber\\
\hspace{-3mm}&-&\!\frac{1}{3!} \varepsilon_{ABCD} \theta^A \theta^B \theta^C \lambda^D
(x)\! - \frac{1}{4!} \varepsilon_{ABCD} \theta^A \theta^B \theta^C \theta^D
\partial_+ \bar{A} (x) \bigg\} , \quad \ \label{N=4-field}
\end{eqnarray}
which satisfies the chirality condition
\begin{equation}
\bar{D}^B
\Phi (x, \theta^A, \bar\theta_A) = \left(
\partial_{\bar\theta_B} - \ft{1}{2}\theta^B \partial_+
\right) \Phi (x, \theta^A, \bar\theta_A) = 0 \, .
\end{equation}
It embraces all particle helicities, from $-1$ to $1$ with half-integer step,
and, therefore, $\Phi (x, \theta^A, \bar\theta_A)$ describes a CPT-self-conjugate
supermultiplet. Another advantage of the light-cone formalism is that the full
supersymmetry algebra is reduced on the light-cone to a subalgebra which acts
linearly on the ``physical'' fields and closes off-shell. In particular, in
the $\mathcal{N}=4$ SYM theory, the supersymmetric transformations on the
light-cone read as
\begin{eqnarray}
\label{N=4susy}
&& \delta A = \xi^A \bar\lambda_A ,
\\
&& \delta \lambda^A = - \left( \partial^+ \bar{A} \right) \xi^A - i \left(
\partial^+ \phi^{AB} \right) \bar\xi_B ,
, \nonumber \\
&& \delta \phi^{AB} = i \left( \xi^{[A} \lambda^{B]} - \varepsilon^{ABCD}
\bar\xi_C
\bar\lambda_D \right) ,
\nonumber
\end{eqnarray}
where ${\scriptstyle [A, B] \equiv AB - BA}$.

Gauge theories on the light-cone with less or no supersymmetry can be deduced
from the maximally supersymmetric $\mathcal{N}=4$ theory by removing ``unwanted''
physical fields. In the superfield formulation this amounts to a truncation of
the $\mathcal{N}=4$ superfield, or equivalently, reduction of the number of
fermionic directions in the superspace~\cite{BT83,Smith85}. For instance, the
${\mathcal N} = 2$ chiral superfield is
\begin{equation}
\Phi^{(2)} = \Phi^{(4)}|_{\theta^3 = \theta^4 = 0}={\rm e}^{\frac12 \bar\theta_A
\theta^A \,
\partial_+} \bigg\{
\partial_+^{-1}A(x)
+ \theta^A \partial_+^{-1}\bar\lambda_A (x) + i \theta^1 \theta^2 \bar \phi(x)
\bigg\}
\label{N=2-field}
\end{equation}
with the identification $\phi^{AB} = \varepsilon^{AB} \phi$ and ${\scriptstyle
A,B}=1,2$. To get the ${\mathcal N} = 1$ superfields one removes three odd
coordinates $\theta^2=\theta^3 = \theta^4 = 0$, and, finally for ${\mathcal N} =
0$ all $\theta$'s in \re{N=4-field} have to be set to zero.  Notice that under
this procedure the truncated $\mathcal{N}=2$, $\mathcal{N}=1$ and $\mathcal{N}=0$
theories involve only half of the fields described by the $\mathcal{N}-$extended
SYM theory and the other half of the needed particle content arises from the
complex conjugated superfields $\bar\Phi \equiv \Phi^\ast$. {}For ${\mathcal N}
= 4$ the two superfields $\Phi$ and $\bar\Phi$ are related to each other through
the reality condition~\cite{Mandelstam83,BLN83}
\begin{equation}
\Phi (x, \theta^A , \bar\theta_A)
=
- \partial_+^{-2} \bar{D}^1 \bar{D}^2 \bar{D}^3 \bar{D}^4
\bar\Phi (x, \theta^A, \bar\theta_A)
\, ,
\label{N=4-bar-field}
\end{equation}
but they are independent for $\mathcal{N}\le 2$.

The final step consists of constructing the action for the SYM theory in terms of
the light-cone superfields. One starts with the $\mathcal{N}=4$ theory, which
involves a single superfield \re{N=4-field}. The action consists of the kinetic
term $\Phi \Phi$ as well as the cubic $\Phi \Phi \Phi$ and quartic couplings
$\Phi \Phi \Phi \Phi$. Since light-cone supersymmetry is realized linearly,
each of the above three terms is separately invariant under the supersymmetric
transformations \re{N=4susy}. Thus one can individually construct superspace
invariants for each number of superfields and match them with the light-cone
action in the component form to fix the overall normalization. This
yields~\cite{BLN83}
\begin{eqnarray}
S_{\mathcal{\scriptscriptstyle N}=4}
&=&\!
\int\! d^4 x\, d^4\theta\, d^4 \bar\theta\,
\Big\{
\ft12 \bar{\Phi}^a \Box \partial_+^{- 2} {\Phi}^a
-
\ft23 g f^{abc}
\lr{
\partial_+^{- 1} \bar{\Phi}^a{\Phi}^b\bar\partial{\Phi}^c
+
\partial_+^{- 1} {\Phi}^a\bar{\Phi}^b\partial\bar{\Phi}^c
}
\nonumber
\\
&-&
\ft12 g^2 f^{abc}f^{ade}
\left(
\partial_+^{- 1} ({\Phi}^b\partial_+{\Phi}^c)
\partial_+^{- 1} (\bar{\Phi}^d\partial_+\bar{\Phi}^e)
+ \ft12{\Phi}^b\bar{\Phi}^c{\Phi}^d\bar{\Phi}^e \right) \Big\} \, ,
\label{N=4SYM}
\end{eqnarray}
with $\bar{\Phi}$ given by \re{N=4-bar-field}. The SYM theories with less
supersymmetry, ${\mathcal N} \le 2$, can be derived from \re{N=4SYM} through the
truncation procedure. For $\mathcal{N}=0,1,2$ the resulting light-cone action has
the universal form~\cite{BT83}
\begin{eqnarray}
S_{\mathcal{\scriptscriptstyle N}\le 2} &=& \int\! d^4 x \,
d^{\mathcal{\scriptscriptstyle N}} \theta \, d^{\mathcal{\scriptscriptstyle N}}
\bar\theta \bigg\{ \sigma_{\mathcal N}
\bar{\Phi}^a {\Box} \partial_+^{2 - {\mathcal{\scriptscriptstyle N}}} {\Phi}^a
\\
&-& 2 g f^{abc} \sigma_{\mathcal N}\left(
\partial_+ {\Phi}^a
\partial_+^{2 - {\mathcal{\scriptscriptstyle N}}} \bar{\Phi}^b \bar\partial {\Phi}^c
+ (- 1)^{\mathcal N}
\partial_+ \bar{\Phi}^a
\partial_+^{2 - {\mathcal{\scriptscriptstyle N}}} {\Phi}^b \partial \bar{\Phi}^c
\right)
\nonumber \\
&-& (- 1)^{{\mathcal{\scriptscriptstyle N}}}
2 g^2 f^{abc}f^{ade} \partial_+^{- 1}
\left(
\partial_+ {\Phi}^b
\bar{\bit{D}}^{\mathcal{\scriptscriptstyle N}}
\partial_+^{2 - {\mathcal{\scriptscriptstyle N}}} \bar{\Phi}^c
\right)
\partial_+^{- 1}
\left(
\partial_+\bar {\Phi}^d
{\bit{D}}_{\mathcal{\scriptscriptstyle N}}
\partial_+^{2 - {\mathcal{\scriptscriptstyle N}}} {\Phi}^e
\right)
\bigg\}
, \nonumber
\label{allNSYM}
\end{eqnarray}
where $\sigma_{\mathcal N}=(- 1)^{\frac{{\mathcal N} ({\mathcal N} + 1)}{2}}$,
${\bit{D}}_{\mathcal{\scriptscriptstyle N}} \equiv  D_{1} \dots
D_{{\mathcal{\scriptscriptstyle N}}}$ and analogously for
$\bar{\bit{D}}^{\mathcal{\scriptscriptstyle N}}$. Here, in contradistinction with
Eq.~(\ref{N=4SYM}), the superfields $\Phi$ and $\bar\Phi$ are independent on
each other.\\[-7mm]~

\subsection{Superconformal symmetry on the light-cone}

Our main objects of interest will be multiparticle single-trace operators built
from light-cone superfields
\begin{equation}
\label{WilsonOperators}
\mathbb{O}(Z_1,\ldots,Z_L) = {\rm tr} \, \{ \Phi (Z_1) \Phi (Z_2) \cdots \Phi
(Z_L) \} \,,
\end{equation}
where $\Phi(Z) \equiv \Phi^a(Z) t^a$ is a matrix valued superfield in the
fundamental representation of the gauge group and $Z$ denotes its position in
the superspace with four even coordinates, $x_\mu$, and $2\mathcal{N}$ odd
coordinates, $\theta^A$ and $\bar\theta_A$ with $A=1,\ldots, \mathcal{N}$. In
addition, we choose all superfields to be located along the light-cone
direction in the four-dimensional Minkowski space  defined by the light-like
vector $n_\mu$ (with $n^2=0$) defined as in Eq.~(\ref{nbarn}), so that
$n \cdot A = A_+ =0$, cf. (\ref{dot}). Similarly to the QCD case, the positions
of the superfields on the light-cone are parameterized by real numbers
$x_\mu = z n_\mu$. One can further simplify the notations by taking into account
that a $\bar\theta-$dependence of the chiral superfields can be absorbed into
the redefinition of $z-$variables
\begin{equation}
\Phi (zn_\mu,\theta^A,\bar\theta_A)
=
\Phi \left((z+\ft12\theta\cdot\bar\theta)n_\mu,\theta^A,0\right)
\, .
\end{equation}
Redefining $z + \ft12 \theta\cdot\bar\theta \to z$, we shall specify the
position of the superfield in the superspace as $Z_k=(z_k,\theta_k^A)$ so
that $\Phi(Z_k) \equiv \Phi(z_k n_\mu,\theta_k^A,0)$.

The single-trace operators \re{WilsonOperators} represent a natural
generalization of nonlocal light-ray operators in QCD, cf. Eq.~\re{B3/2}. To
obtain the latter it is sufficient to expand $\mathbb{O}(Z_1,\ldots,Z_L)$ in
powers of odd variables $\theta^{A_1}_1\ldots \theta^{A_L}_L$. As in QCD,
nonlocal operators \re{WilsonOperators} serve as generating functions for Wilson
operators with the maximal Lorentz spin and minimal twist equal to the number of
constituent fields $L$. A question arises as to whether this set of Wilson operators
is complete. In $\mathcal{N}=4$ SYM theory there is only one independent chiral
superfield $\Phi(Z)$ and, as a consequence, the operators \re{WilsonOperators}
generate \textsl{all} Wilson operators of twist$-L$ built from $L$
fundamental fields. For $\mathcal{N}\le 2$, the superfields $\Phi(Z)$ and
$\bar\Phi(Z)$ are independent of each other and, in addition to the operators in
\re{WilsonOperators}, one can introduce ``mixed'' operators built from both
superfields. This means that in the $\mathcal{N}=0,1$ and $2$ SYM theories, the
operators \re{WilsonOperators} only generate a certain subset of the existing
Wilson operators. The reason why we will concentrate on this subset is that it
inherits integrability properties found in QCD in the sector of maximal helicity
operators. Namely, as we shall argue below in this section, the one-loop
dilatation operator acting on the space of single-trace operators
\re{WilsonOperators} can be mapped in the multicolor limit into a Hamiltonian of
a completely integrable Heisenberg spin magnet with the symmetry group being a
collinear $SL(2|\mathcal{N})$ subgroup of the full superconformal
group~\cite{BDKM04}.

To explain this correspondence, we would like to recall the symmetries of the SYM
theories. At the classical level, a four-dimensional supersymmetric Yang--Mills
theory with $\mathcal{N}$ supercharges is invariant under the $SU(2,2 | {\mathcal
N})$ superconformal transformations~\cite{Sohnius85}. The corresponding algebra
contains both bosonic (even) and fermionic (odd) generators. The even generators
are the fifteen generators of the $SO(2,4)$ conformal algebra familiar from QCD
plus the chiral ${\bf R}$ charge and the $SU ({\mathcal N})$ internal ``flavor''
symmetry generators ${\bf T}_A{}^B$ (for ${\mathcal N} \ge 2$ only). The odd
generators are the generators of supersymmetric and superconformal
transformations, ${\bf Q}_{\alpha A},~ {\bf \bar Q}^{\dot\alpha A}$ and ${\bf
S}_\alpha^A,~{\bf \bar S}^{\dot\alpha}_A$, respectively. Since the superfields
entering \re{WilsonOperators} ``live'' on the light-cone, we expect, by analogy
with the QCD case, that the full superconformal group will be reduced to its
collinear subgroup. We know from Sect.~2.1 that in Yang--Mills theories with no
supersymmetry, the collinear subgroup is generated by three bosonic operators
$\mathbf{L}^{\pm, 0}$ which form the $SL(2,\mathbb{R})$ algebra. Supersymmetry
enlarges the $SL(2,\mathbb{R})$ subgroup by adding the following generators: the
chiral charge ${\mathbf R}$, $SU ({\mathcal N})$ charges ${\mathbf T}_A{}^B$,
transverse boosts ${\mathbf M}_{12}$ and, finally, the ``good'' components of the
fermionic charges $\mathbf{Q}_{A}$, $\mathbf{\bar Q}^{A}$, ${\mathbf S}^A$ and
$\mathbf{\bar S}_{A}$. Being combined together, these operators form the
$SL(2|\mathcal{N})$ algebra~\cite{BDKM04}.

To one-loop accuracy, the generators of the $SL(2|\mathcal{N})$ subgroup
can be realized as linear differential operators, $[\mathbf{G} , \Phi] =
G(\partial_z,\partial_\theta) \Phi$, acting on the light-cone coordinates of
the chiral superfield ${\Phi} (Z)$~\cite{DKK01,BDKM04}. The corresponding
expressions for the generators of the $SL(2;\mathbb{R})$ collinear subgroup
generalize similar expressions in the QCD case, Eq.~\re{SL2R2},
\begin{equation}
\label{sl2N-1}
 {L}^- = -\partial_z \, , \quad  {L}^+ = 2 j\, z + z^2\partial_z + z \left(
\theta\cdot \partial_\theta \right) \, , \quad   {L}^0 = j + z
\partial_z + \ft12\left( \theta\cdot \partial_\theta \right)
\, ,
\end{equation}
and the twist operator is ${E}= t$. The four odd generators can be represented by
the following differential operators
\begin{equation}
{W}{}^{A,-} = \theta^A \, \partial_z
\, , \ \
{W}{}^{A,+} = \theta^A [ 2j  +  z \partial_z +
\left( \theta\cdot \partial_\theta \right) ]
\, , \ \
{V}^-_{A} = \partial_{\theta^A}
\, , \ \
{V}^+_{A} = z\partial_{\theta^A} \, .
\end{equation}
Finally, the $SU(\mathcal{N})$ generators and linear combination of the abelian
$R-$charge and transverse boosts ${\mathbf M}_{12}$ are realized as
\begin{equation}
\label{sl2N-3}
{T}_B{}^A = \theta^A
\partial_{\theta^B} - \ft1{\mathcal{N}} \, \delta_B^A \left( \theta\cdot
\partial_\theta \right)\,,\qquad
{B} = - j - \ft12 \left( 1 - \ft{2}{{\mathcal N}} \right) \left( \theta\cdot
\partial_\theta \right)
 \, .
\end{equation}
In these expressions, $\partial_z \equiv \partial/\partial z$ and $\theta \cdot
\partial_\theta \equiv \theta^A \partial/\partial\theta^A$. The twist $t$ and
conformal spin $j$ are the parameters defining the representation. For the chiral
superfield $\Phi(Z)$ entering the Lagrangian of $\mathcal{N}$-extended SYM
theories, Eqs.~\re{N=4SYM} and \re{allNSYM}, one has $j = - \ft12$ and $t=1$.

The light-cone operator $\mathbb{O}(Z_1,\ldots,Z_L)$, Eq.~\re{WilsonOperators},
serves as the generating function of an infinite tower of Wilson operators with an
arbitrary number of derivatives, when expanded in the bosonic separations, and of
different field content, when expanded in fermionic coordinates. The most
efficient way to renormalize an infinite set of local Wilson operators is to
study the renormalization properties of the light-ray operators $\mathbb{O}(Z_1,
\ldots,Z_L)$ themselves. Superconformal invariance imposes severe restrictions on
the possible form of the dilatation operator in the $\mathcal{N}-$extended SYM
theory acting on the space of light-cone operators \re{WilsonOperators}. For the
operator $\mathbb{O}(Z_1,\ldots,Z_L)$, the generators of the $SL(2|\mathcal{N})$
collinear subgroup are given by the sum of single-particle generators,
Eqs.~\re{sl2N-1} -- \re{sl2N-3}, acting on the coordinates in the superspace
$Z_k=(z_k,\theta_k^A)$ with $k=1,\ldots,L$. In particular, the total twist of the
operator $\mathbb{O}(Z_1,\ldots,Z_L)$ is equal to the number of superfields $L$.
It is well known from QCD \cite{BFLK85} that to one-loop accuracy the RG
evolution of local quasipartonic operators preserves the number of particles. In
addition, the mixing of single trace operators with operators involving larger
number of traces is suppressed by powers of $1/N_c^2$. Taken together,
these two observations imply that to one-loop accuracy the set of light-cone
operators $\mathbb{O}(Z_1,\ldots,Z_L)$ is closed under renormalization in the
multicolor limit. The corresponding one-loop dilatation operator can be represented
in the light-cone superfield formalism  as a quantum mechanical Hamiltonian
$\mathbb{H}$ leading to the following RG equation
\begin{equation}
\left\{ \mu\,\frac{\partial}{\partial\mu} +
\beta(g)
\frac{\partial}{\partial g} + L \gamma (g) \right\}
\mathbb{O}^{\scriptscriptstyle\rm R} (Z_1,\ldots,Z_L) = - \frac{g^2
N_c}{8\pi^2}\left[ \mathbb{H} \cdot \mathbb{O}^{\scriptscriptstyle\rm R} \right]
(Z_1,\ldots,Z_L) .
\label{EQ}
\end{equation}
The renormalized operator $\mathbb{O}^{\scriptscriptstyle\rm R}$ generates
finite Green functions and is expressed in terms of the bare operator
(\ref{WilsonOperators}) with an infinite multiplicative counterterm. Here
$\gamma (g)$ is the anomalous dimension of the superfields in the light-like
gauge $A_+(x)=0$. Thanks to the Ward identity it is related to the beta-function
of the theory $\gamma(g) = \beta (g)/g$. The Hamiltonian $\mathbb{H}$ in
Eq.~\re{EQ} is an integral operator acting on the coordinates of the superfields
in the superspace. Superconformal invariance implies that the dilatation operator
$\mathbb{H}$ commutes with the generators of the $SL(2|\mathcal{N})$ supergroup
in  Eqs.~\re{sl2N-1}--\re{sl2N-3}.

\subsection{One-loop dilatation operator}

\begin{figure}[ht]
\centerline{\epsfxsize12.0cm\epsfbox{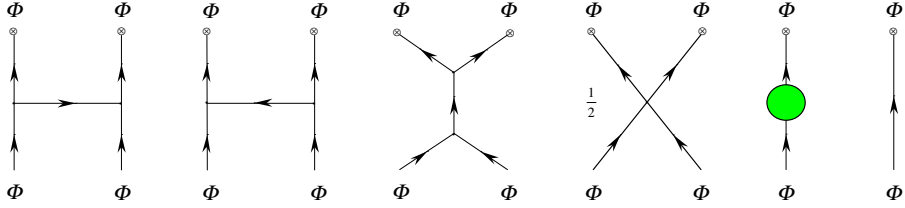}}
\caption{Feynman diagrams contributing to the one-loop
dilatation operator for the two chiral superfields on the light-cone.}
\label{supergraphs}
\end{figure}
To one-loop order, the dilatation operator $\mathbb{H}$ has a pairwise
structure. In addition, only the nearest-neighbor
interactions survive in the multicolor limit,
leading to
\begin{equation}
\label{H-general}
\mathbb{H} = \mathbb{H}_{12} + \mathbb{H}_{23} + \dots + \mathbb{H}_{L1} \, .
\end{equation}
The two-particle interaction kernel $\mathbb{H}_{k,k+1}$ receives contributions
from the Feynman diagrams shown in Fig.~\ref{supergraphs}. They involve cubic and
quartic interaction vertices which are uniquely defined by the light-cone actions
\re{N=4SYM} and \re{allNSYM}. As a consequence of superconformal symmetry the
kernel $\mathbb{H}_{k,k+1}$ has to commute with the generators of the
$SL(2|\mathcal{N})$ superalgebra acting on the superfields with labels $k$ and
$k+1$ and, therefore, it is a function of the two-particle quadratic Casimir
operator
\begin{equation}
\mathbb{H}_{k,k+1} = H(L^2_{k,k+1})\,,
\end{equation}
where a general expression for the quadratic Casimir in the $SL(2|\mathcal{N})$
algebra is~\cite{FSS96}
\begin{eqnarray}
\label{Casimir}
L^2
&=&
(L^0)^2 + L^+ L^-
\\
&+&
(\mathcal{N}-1) L^0 +\frac{\mathcal{N}}{\mathcal{N}-2}
B^2 - V^+_A W^{A,-} - W^+_A V^{A,-}-\ft12\, T^B{}_A T^A{}_B
\, . \nonumber
\end{eqnarray}
For $\mathcal{N}=0$, this expression coincides with Eq.~\re{casimir}. In that case,
the $SL(2)$ invariance allows one to obtain the general form of the two-particle
kernels in QCD, Eq.~\re{ansatz}.

We remind the reader that in the light-cone superspace formalism, the Yang--Mills
theory without supersymmetry can be obtained from the one with
$\mathcal{N}$ supercharges by putting the odd coordinates to zero, $\theta^A=0$.
For this reason the integral operator \re{ansatz} can be considered as the special
limit of a general two-particle kernel $\mathbb{H}_{k,k+1}$ for
$\theta_k^A=\theta_{k+1}^A=0$. Making use of the odd generators in the
$SL(2|\mathcal{N})$ algebra one can ``lift'' the operator \re{ansatz} into the
superspace and restore the dependence on the odd coordinates. In this way, one
finds a general form of the two-particle $SL(2| {\mathcal N})$ invariant
evolution kernel
\begin{eqnarray}
\label{GenericFormDilatation}
&&[ \mathbb{H}_{12} \cdot \mathbb{O} ] (Z_1, Z_2 , \dots , Z_L )
\\
&&\qquad = \int_0^1 d \alpha \int_0^{\bar \alpha} ( \bar{\alpha} \bar{\beta} )^{2
j - 2} \varphi \left( \frac{\alpha\beta}{\bar{\alpha}\bar{\beta}} \right)
\mathbb{O} (\bar\alpha Z_1 + \alpha Z_2, \beta Z_1 + \bar{\beta} Z_2 , \dots ,
Z_L) \, , \nonumber
\end{eqnarray}
where $j=-1/2$ is the conformal spin of the chiral superfield $\Phi(Z)$ entering
into the light-cone operator \re{WilsonOperators}. Here the kernel $\mathbb{H}_{12}$
acts on the coordinates of the superfields $\Phi(Z_1)$ and $\Phi(Z_2)$ and displaces
them simultaneously along bosonic and fermionic directions in the superspace $Z =
(z, \theta^A)$.

The evolution kernel \re{GenericFormDilatation} involves an arbitrary function
$\varphi$ of the harmonic ratios. Its form is not fixed by superconformal
invariance and is determined by the dynamics of the underlying SYM theory. The
one-loop calculation of super-Feynman graphs, displayed in Fig.\
\ref{supergraphs}, gives $\varphi (\xi) \sim \delta (\xi)$. This suggests that
the two-particle kernel looks like~\cite{BDKM04}
\begin{eqnarray}
\label{KernelDilatation}
\mathbb{V}_{12} \mathbb{O} (Z_1,...,Z_L) &=& \int_0^1
\frac{d\alpha}{\bar\alpha\alpha^2} \, \big\{ 2 \alpha^2\, \mathbb{O}
(Z_1,Z_2,...,Z_L)
\\
&-& \mathbb{O} (\alpha Z_1+\bar\alpha Z_{2},Z_2,...,Z_L) - \mathbb{O} (Z_1,\alpha
Z_2+\bar\alpha Z_{1},...,Z_L) \big\}
\, . \nonumber
\end{eqnarray}
This relation admits an obvious generalization to $\mathbb{V}_{k,k+1}$. However,
Eq.~\re{KernelDilatation} cannot be identified as the two-particle dilatation
operator since the integral is divergent for $\alpha\to 0$. The divergence arises
due to a negative value of conformal spin $j=-1/2$ of the light-cone superfield -- see
Eqs.~\re{N=4-field} and \re{N=2-field} -- which can be traced back to the fact that
the lowest component of the superfield is a \textsl{nonlocal} operator
$\partial_+^{-1} A(z)$. This means that expanding the nonlocal operator
$\mathbb{O} (Z_1,...,Z_L)$ around the origin $Z_1=...=Z_L=0$, in terms of local
operators one obtains both Wilson operators and ``spurious'' operators involving
nonlocal fields $\partial_+^{-1} A(0)$, $A_+(0)$ and $\partial_+^{-1}\bar\lambda(0)$. To
eliminate the latter it is sufficient to subtract the two first terms in
the expansion of  the superfield $\Phi(Z)$ around $Z=0$:
\begin{equation}
\Phi(Z) \to \Phi^{\scriptscriptstyle\rm W} (Z)= \Phi (Z) - \Phi (0) - Z \cdot
\partial_Z \Phi (0)\,,
\end{equation}
where the superscript indicates that the superfield $\Phi^{\scriptscriptstyle\rm
W} (Z)$ does not generate nonlocal fields. Let us define the operator $\Pi$ that
projects the nonlocal operator \re{WilsonOperators} onto the subspace of Wilson
operators
\begin{equation}
\mathbb{O}^{\scriptscriptstyle\rm W} (Z_1,...,Z_L) = \Pi \,\mathbb{O}(Z_1,...,Z_L)=
{\rm tr} \, \{ \Phi^{\scriptscriptstyle\rm W} (Z_1)  \cdots
\Phi^{\scriptscriptstyle\rm W} (Z_L) \}\,,
\end{equation}
so that $\Pi^2=\Pi$. Since Wilson operators mix among themselves and cannot mix
with spurious operators (but the opposite is possible!), one expects that
nonlocal operators $\mathbb{O}^{\scriptscriptstyle\rm W} (Z_1,...,Z_L)$ evolve
autonomously under the RG transformations. Indeed, decomposing the operators as
\begin{equation}
\mathbb{O} (Z_1,...,Z_L)=\mathbb{O}^{\scriptscriptstyle\rm W} (Z_1,...,Z_L) +
(1-\Pi)\mathbb{O}^{\scriptscriptstyle\rm spur} (Z_1,...,Z_L)
\end{equation}
and substituting this relation into \re{KernelDilatation}, one finds that the
operators $\mathbb{O}^{\scriptscriptstyle\rm W} (Z_1,...,Z_L)$ obey the same
relation \re{KernelDilatation} but with the two-particle kernel replaced
by~\cite{BDKM04}
\begin{equation}
\mathbb{V}_{12} \to \mathbb{H}_{12}^{\scriptscriptstyle\rm W} =  \Pi
\mathbb{V}_{12} \Pi\,.
\label{H-w}
\end{equation}
It is straightforward to verify that the integral operator
$\mathbb{H}_{12}^{\scriptscriptstyle\rm W}$ is well defined, that is, the
above-mentioned divergences for $\alpha\to 0$ do not appear in the expression
for $\mathbb{H}_{12}^{\scriptscriptstyle\rm W}\mathbb{O} (Z_1,...,Z_L)$. Thus,
as far as the renormalization of Wilson operators is concerned, the one-loop
dilatation operator is given in the multicolor limit by Eq.~\re{H-general} with
the two-particle kernel $\mathbb{H}_{k,k+1}^{\scriptscriptstyle\rm W}$ defined
in \re{H-w}.

The expression obtained for the dilatation operator does not exhibit an explicit
dependence on $\mathcal{N}$. This dependence enters entirely through the number
of odd directions in the superspace, $Z=(z,\theta^A)$. This means that for
nonlocal light-cone operators \re{WilsonOperators} the dilatation operator has
the same, universal form in $\mathcal{N}=0$, $\mathcal{N}=1$, $\mathcal{N}=2$ and
$\mathcal{N}=4$ SYM theories on the light-cone. The important difference between
these theories is that for $\mathcal{N}=4$ the operators \re{WilsonOperators}
generate all Wilson operators of the twist $L$, whereas for $\mathcal{N} \le 2$
they capture only a subsector of Wilson operators. To describe the remaining
operators, one has to consider single-trace operators built from both chiral and
anti-chiral superfields, $\Phi$ and $\bar\Phi$, respectively. Remarkably enough,
the dilatation operator for such mixed nonlocal operators can be deduced from the
dilatation operator (\ref{KernelDilatation}) in maximally supersymmetric
$\mathcal{N}=4$ theory via the truncation procedure described above.

\subsection{Dilatation operator as a $SL(2 | {\mathcal N})$ Heisenberg spin chain}

Let us consider the properties of the one-loop dilatation operator
(\ref{KernelDilatation}) in more detail and demonstrate how one can use the
obtained expressions to evaluate the anomalous dimensions of various Wilson
operators in $\mathcal{N}-$extended SYM theories.

To begin with, consider (\ref{KernelDilatation}) for $\mathcal{N}=0$. In this
case, the nonlocal operator \re{WilsonOperators} is reduced to the product of
holomorphic components of gauge fields $\partial_+^{-1}A(z)$. To avoid spurious
operators it is convenient to go over to the nonlocal light-cone operators
\begin{equation}
\label{N=0-Operators}
\mathbb{O}_g(z_1,\ldots,z_L)=(\partial_{z_1} \ldots \partial_{z_L})^2
\mathbb{O}(z_1,\ldots,z_L) = {\rm tr} \, \{ \partial_+ A(z_1)
 \cdots \partial_+ A(z_L) \} \,.
\end{equation}
Remember that we have adopted the light-like axial gauge $A_+(x)=0$. In the
covariant formulation the field $\partial_+ A(z)$ is replaced by $n^\mu (F_{\mu
\perp}(z)-i\tilde F_{\mu\perp}(z))$. Recall that the field $A(z)$ carries the
helicity $-1$ and, therefore, $\mathbb{O}_g(z_1,\ldots,z_L)$ is the generating
function for pure gluonic Wilson operators of maximal helicity $-L$.

Substituting Eq.~\re{N=0-Operators} in (\ref{KernelDilatation}) one gets
\begin{eqnarray}
&&[ \mathbb{H}_{12}^g \cdot \mathbb{O}^g ] (z_1, z_2 , \dots , z_L ) = \int_0^1
\frac{d \alpha}{\alpha} \Big\{ 2 \mathbb{O}^g (\bar\alpha z_1 + \alpha z_2, z_2 ,
\dots , z_L)
\label{H-gluon}
\\
&& \qquad -
\bar{\alpha}^2
\mathbb{O}^g (\bar\alpha z_1 + \alpha z_2, z_2 , \dots , z_L) -
\bar{\alpha}^2
\mathbb{O}^g (z_1 , \bar\alpha z_2 + \alpha z_1, \dots , z_L) \Big\} \, .
\nonumber
\end{eqnarray}
This result should be compared with the similar expression for the dilatation
operator for quarks with the aligned helicities, Eq.~\re{H32-part}. The only
difference is in the power of $\bar\alpha$ in front of the second and third terms.
In general, this power is determined by the conformal spin of the field,
$\sim \bar\alpha^{2j - 1}$. For the quark field, one has $j_q=1$, whereas for
the gauge field strength, $j_g=3/2$, and for scalar fields,  $j_s = 1/2$.

To diagonalize the dilatation operator, one has to repeat the same steps as
described in Sect.~2.2 in the case of aligned helicity quark operators. Namely,
one substitutes
$$
\mathbb{O}^g  (z_1, z_2 , \dots , z_L ) \to (z_1-z_2)^n \equiv z_{12}^n
$$
into \re{H-gluon} and obtains
\begin{equation}
\label{IntegrableForm}
\mathbb{H}_{12}^g \, z_{12}^n = 2[\psi(n+3)-\psi(1)]z_{12}^n\,.
\end{equation}
This relation defines the anomalous dimension of the conformal operator built
from two gauge fields of the same helicity
\begin{equation}
\mathcal{O}_g(0) = \partial_+ A(0) (i\partial_+)^n \partial_+ A(0) + \ldots\ .
\label{Og}
\end{equation}
Here ellipses stand for terms involving total derivatives. The conformal
$SL(2;\mathbb{R})$ invariance fixes them uniquely and allows one to write down a
closed expression in terms of Gegenbauer polynomials \cite{Makeenko:bh}.
The conformal spin of the
operator \re{Og} equals $J_{12} = n+ 2j_g = n+3$ and, therefore, the two-particle
dilatation operator \re{H-gluon} can be written in the $SL(2)$ invariant form
\begin{equation}
\mathbb{H}_{12}^g =2\big[\psi(J_{12})-\psi(1)\big]
\label{Hg=psi}
\end{equation}
with $J_{12}$ being the two-particle conformal spin. In contrast with the
quark operator, the corresponding $SL(2;\mathbb{R})$ representation has
$j_g=3/2$.

Combining together \re{Hg=psi} and \re{H-general}, we conclude that the dilatation
operator in the sector of maximal helicity gluon operators
can be identified as the Hamiltonian of the $SL(2;\mathbb{R})$
Heisenberg magnet. The length of the spin chain is equal to $L$ and the value of
the spin in each site is given by the conformal spin of the gauge strength
tensor $j_g=3/2$. Notice that the maximal-helicity gluonic operators evolve
autonomously and the corresponding one-loop mixing matrix is sensitive to the
gluonic sector of the gauge theory only. This implies that the one-loop dilatation
operator in the sector of maximal-helicity gluonic operators is the \textsl{same}
in all $\mathcal{N}-$extended SYM theories.

The Wilson operators built from six scalars $\phi^{AB}(x)$ with no derivatives
provide another limiting case which reflects the specifics of the $\mathcal{N}=4$ SYM theory.
It proves convenient to introduce their linear combinations~\cite{Sohnius85}
\begin{equation}
\phi_j(x) =  \ft1{2\sqrt{2}}\, \Sigma_j^{AB} \bar \phi_{AB} (x)
\end{equation}
where  $j = 1, \ldots, 6$ and $\Sigma_j^{AB}$ are the chiral blocks of Dirac
matrices in six-dimensional Euclidean space. Then, the composite scalar operators
look like
\begin{equation}
O_{j_1 j_2 \ldots j_L}
= {\rm tr}
\{ \phi_{j_1}(0) \phi_{j_2}(0) \ldots \phi_{j_L}(0) \}
\, .
\label{real-scalar}
\end{equation}
To one-loop order, these operators evolve autonomously under the RG
transformations and the corresponding dilatation operator has the form
\re{H-general} with the two-particle kernel given by a finite-dimensional matrix
\begin{equation}
\mathbb{H}_{12}\, \phi_{j_1}(0) \phi_{j_2}(0) = \sum_{j_1', j_2'} V_{j_1
j_2}^{j_1' j_2'}\,  \phi_{j_1'}(0) \phi_{j_2'}(0)\,.
\end{equation}
The scalar field can be
obtained from the $\mathcal{N}=4$ superfield \re{N=4-field} by means of a simple
projector
\begin{equation}
\phi_j(x) = \ft{i}{2\sqrt{2}} \Sigma_j^{AB}
\partial_{\theta^A}\partial_{\theta^B} \Phi (x, \theta^A , 0)\bigg|_{\theta^A = 0}\,.
\end{equation}
Using this projection, the mixing matrix $ V_{j_1 j_2}^{j_1' j_2'}$ can be
extracted from the general expression
for the $\mathcal{N}=4$ one-dilatation operator given in Eqs.~\re{H-w} and
\re{KernelDilatation}:
\begin{equation}
V_{j_1 j_2}^{j_1' j_2'}=- \ft18 \left( \Sigma_{j_1}^{AB} \partial_{\theta_1^A}
\partial_{\theta_1^B} \right) \left( \Sigma_{j_2}^{CD} \partial_{\theta_2^C}
\partial_{\theta_2^D} \right) \left[ \mathbb{H}_{12} \cdot \mathbb{O} \right]
(Z_1, Z_2) \bigg|_{Z_1 = Z_2 = 0}\,.
\end{equation}
The calculation of fermionic derivatives then yields
\begin{equation}
V_{j_1 j_2}^{j_1' j_2'} =\delta_{j_1}^{j_1'} \delta_{j_2}^{j_2'} + \ft12\,
\delta_{j_1j_2}\delta^{j_1'j_2'} - \delta_{j_1}^{j_2'} \delta_{j_2}^{j_1'} .
\end{equation}
This matrix acts on the isotopic indices of two neighboring scalar fields in
\re{real-scalar} and has the size $6^2\times 6^2$. Hence, the one-loop dilatation
operator for the scalar composite operators \re{real-scalar} is given by the
$6^L\times 6^L-$matrix. As in the previous case it has a hidden symmetry -- this
matrix can be mapped into the  Hamiltonian of the completely integrable
Heisenberg $SO(6)\sim SU(4)$ spin chain of length $L$~\cite{mz}.

We see that in the $\mathcal{N}=4$ SYM theory, the integrability phenomenon occurs
in two different sectors of Wilson operators. In both cases, one encounters the
Heisenberg spin chain of length $L$ but with different symmetry groups. For the
maximal helicity Wilson operators with arbitrary number of derivatives, the
$SL(2;\mathbb{R})$ symmetry group arises and it is related to conformal symmetry
on the light-cone. For scalar operators with no derivatives, the arising $SU(4)$
symmetry group is related to the $R-$symmetry; that is, rotations in the isotopic
space. From the point of view of the superspace, $Z=(z,\theta^A)$, the two
groups act separately on the bosonic and fermionic coordinates of the superfields
and as such they are subgroups of a bigger, $SL(2|\mathcal{N})$ group. Since the
$SL(2;\mathbb{R})$ and $SO(6)$ spin chains emerge as special limits of a general
expression for the one-loop dilatation operator  Eq.~\re{KernelDilatation}, a
question arises about integrability properties of the latter operator as a whole. Quite
naturally, the one-loop dilatation operator can be mapped into a Hamiltonian of the
Heisenberg spin magnet with the $SL(2|\mathcal{N})$ symmetry group. The latter is
defined in \re{H-general} with the two-particle kernels given by the familiar
expression~\cite{DKK01,BDKM04}
\begin{equation}
\mathbb{H}_{12}^{\scriptscriptstyle \rm W} = 2 \left[ \psi(J_{12})-\psi(1)
\right]\,,\qquad J_{12} (J_{12}-1) = L_{12}^2
\end{equation}
in which $J_{12}$ is the two-particle $SL(2|\mathcal{N})$ spin and the quadratic
Casimir was introduced in \re{Casimir}.

For $\mathcal{N}=4$ this (super) spin chain describes renormalization of
arbitrary Wilson operators in the underlying SYM theory built from ``good''
components of the fundamental fields. For $\mathcal{N}\le 2$, it covers only a special
subsector of Wilson operators built from fundamental fields entering into the
expression for the chiral superfield, Eq.~\re{N=2-field}. In particular, at
$\mathcal{N}=0$ this sector corresponds to maximal helicity gluonic operators.
For the remaining Wilson operators generated by mixed products of chiral and
antichiral superfields, the dilatation operator does not exhibit integrability
properties. As was illustrated in Sect.~2.5 in the case of non-maximal helicity
quark operators, the additional terms in the dilatation operator responsible for
breaking the integrability generate a mass gap in the spectrum of the anomalous
dimensions.

The findings of Ref.\ \cite{BS03} suggest  that in the $\mathcal{N}=4$ theory,
not only quasi-partonic operators, addressed above, but all operators
obey integrable renormalization group equations. Thus the spin chain discussed
above gets enhanced to the $SU (2, 2| 4)$ superchain. As we have emphasized in
Sect.~\ref{QPOvsNONQPO}, this is a very nontrivial statement in the QCD context.
It suggests that operators of different twists are intertwined in a very subtle
manner. The superconformal symmetry does indeed relate operators of different
twists. However, this is not enough to conclude that the anomalous dimensions of
non-quasipartonic operators are expressed via the  quasipartonic operators, since this
is not the case in the most straightforward supersymmmetric generalization of QCD,
the $\mathcal{N} = 1$ SYM. The presence of the elementary scalar fields in the
$\mathcal{N} = 4$ model does matter, so that multiple full\,\footnote{\,I.e., not
restricted to the light-cone as in Eq.\ (\ref{N=4susy}).} supersymmetric
transformations of the operator $\phi^i D_+^N \phi^i$, having the integrable form
of the anomalous dimension (\ref{IntegrableForm}) will yield combinations of
operators with different twists but having the same anomalous dimension. This is
not the case in QCD where anomalous dimensions of different twist operators are
believed to be dynamically independent. This issue deserves thorough further
study.

There have been several recent related developments that we cannot address
here in detail. In particular, the anomalous dimension matrices in different
autonomous sectors in $\mathcal{N} = 4$ SYM were studied in
Refs.~\cite{BMSZ03,BBMS04,luza,Freyhult04}, mostly in the different
subsectors of scalar operators. Another very important question is whether the
integrability of the dilatation operator persists at higher orders in the loop
expansion. Several positive demonstrations that this does happen have been given in
certain scalar subsectors \cite{Bei03,RyzTse04,Min04}. To extend these findings to
the full dilatation operator, the analysis of the $SL(2)$ subsector has to be
performed. The two-particle part of the corresponding dilatation operator was
calculated in Refs.\ \cite{KotLip} up to three-loop order.

Ultimately, one has to understand what symmetry of the quantum field theory is
responsible for the appearance of integrability
\cite{Roiban:2003dw,Berenstein:2004ys} and what part of it survives after decreasing
the amount of supersymmetry \cite{Wang:2003cu,DeWolfe:2004zt,DiVecchia:2004jw}, as
we explained for one particular case above.

\newcommand \Mybf[1] {\mbox{\boldmath$ {#1} $}}
\newcommand \mybf[1] {\mbox{\boldmath$ {\scriptstyle #1} $}}

\section{High Energy Scattering}
\setcounter{equation}{0}

In the previous section, we have considered the perturbative QCD dynamics of
partons constrained to small transverse separations, which can be achieved
by considering ``hard'' scattering processes in generalized Bjorken
kinematics. When the transverse distance becomes smaller than any other
scale in the process,  the corresponding dependence
can be treated by renormalization-group methods and disappears from
the dynamics. The problem becomes essentially one dimensional: the relevant
degrees of freedom correspond to partons ``living'' on the light-ray and
the interaction between them amounts to displacement along the light-ray,
with a certain weight-function. We were able to analyze this interaction
using a Hamiltonian formulation and found its hidden integrability property.
The scale dependence in this case corresponds to ``time'' evolution in
the quantum-mechanical picture.

The present section is devoted to perturbative QCD dynamics in the high-energy
limit. In this case, the invariant energy $s=(p_A+p_B)^2$ of colliding hadrons
$A$ and $B$ is considered to be the largest scale and, in the limit $s\to\infty$,
the energy dependence corresponds to a renormalization  group flow for a
dynamical system ``living'' in the two transverse dimensions orthogonal to the
scattering plane. The relevant degrees of freedom correspond to reggeized gluons
(see below) with assigned transverse coordinates, and the interaction between
them leads to a displacement in the transverse plane. Compared to the ``hard''
scattering situation, the longitudinal and transverse degrees of freedom
change places, and the dynamics becomes two dimensional in the transverse plane
instead of being one dimensional in a given light-ray direction as in the former
case. In what follows, we will sketch the Hamiltonian approach to high-energy
scattering and establish its hidden symmetry properties. The main result of
this analysis is that the system of interacting reggeized gluons can be mapped
into a \textsl{quantum} XXX Heisenberg magnet with the $SL(2,\mathbb{C})$
symmetry group \cite{Lipatov94,FK95,Korchemsky95}. This means thatthe high-energy asymptotics of scattering amplitudes in QCD in a certain limit can be
calculated using the Quantum Inverse Scattering Method \cite{QISM}.

\subsection{The BFKL Hamiltonian}

The high-energy asymptotics of the scattering amplitude $\mathcal{A}(s,t)$
is studied most conveniently through the Mellin representation which
corresponds to an expansion in partial waves with complex angular
momentum
\begin{eqnarray}
\label{omega}
\mathcal{A}(s,t)=is \int_{\delta-i\infty}^{\delta+i\infty} \frac{d\omega}{2\pi
i}\, s^\omega \widetilde{\mathcal{A}}(\omega,t)\,.
\end{eqnarray}
Here, the integration contour goes  to the right of all singularities of
$\widetilde{\mathcal{A}}(\omega,t)$ on the complex $\omega-$plane. The
high-energy asymptotics of $\mathcal{A}(s,t)$ is determined by the
singularities of the partial wave amplitudes: If $\widetilde{\mathcal{A}}
(\omega,t) \sim 1/(\omega-\omega_0(t))$ then $\mathcal{A}(s,t) \sim
i s^{1+\omega_0(t)}$. Poles in the $\omega$-plane are called reggeons,
and the position of the pole as a function of
the transfered momentum, $\omega_0(t)$, is called
the reggeon trajectory. From the point of view of QCD perturbation theory,
finding the high-energy asymptotics of the amplitude corresponds to the
resummation of logarithms of the energy $\sim (\alpha_s \ln s)^m$. By
virtue of the Mellin transformation (\ref{omega}), the expansion over
$(\alpha_s\ln s)^m$ is traded for the expansion over $(\alpha_s/\omega)^m$
with coefficients that depend on $t$, and the power-like energy-dependence arises because the resummed expression for
$\widetilde{\mathcal{A}}(\alpha_s/\omega,t)$ develops a nontrivial
singularity for positive $\omega=\omega_0(t)$ in the sum to all orders
in $\alpha_s$.

The partial wave amplitude $\widetilde{\mathcal{A}}(\omega,t)$ can be written
in the so-called impact-parameter representation
\begin{eqnarray}\label{A-Four}
\widetilde{\mathcal{A}}(\omega,t)&=&\int d^2 b_0 \, \textrm{e}^{i(qb_0)} \int
d^2  {b}\, d^2 {b'}\, \Phi_A(\vec{b}-\vec{b}_0)T_\omega(\vec{b},\vec{b'})
\Phi_B(\vec{b'}) \nonumber
\\
&\equiv& \int d^2 b_0 \, \textrm{e}^{i(qb_0)} \langle \Phi(b_0)|
\mathbb{T}_\omega |\Phi(0)\rangle\,,
\end{eqnarray}
where the impact factors $\Phi_A(\vec{b})$ and $\Phi_B(\vec{b'})$ stand
for the parton (gluon) distributions as functions of transverse coordinates
$\vec{b}=\{\vec b_1,\vec b_2,\ldots,\vec b_n\}$ and
$\vec{b}'=\{\vec{b}'_1,\vec{b}'_2,\ldots \vec{b}'_m\}$ in the colliding
hadrons $A$ and $B$, respectively, and $T_\omega(\vec{b},\vec{b'})$ is the
scattering (partial wave) amplitude for a given parton configuration. The
existence of such a representation follows from the space-time picture of
the high-energy collisions: The two Lorentz-contracted hadron discs
approach each other at the speed of light and the interaction between
them takes a finite time in the center-of-mass frame. The
preparation of the initial states is separated in time from the interaction
between hadrons and, therefore, the transverse coordinates of parton
constituents in hadrons can be considered as ``frozen'' during the interaction.
Note that the factorized form in (\ref{A-Four}) implies that the structure
of singularities in the complex $\omega$-plane does not depend on the parton
distribution in hadrons, but rather on general properties of the gluon
interaction in the $t$-channel.

A nontrivial analysis \cite{BFKL} shows that the interaction effects modify the
propagators of the $t-$channel gluons in such a way that they develop their own
Regge trajectory; such $t-$channel gluons ``dressed'' by virtual corrections are
called reggeized gluons. The reggeized gluons with given transverse coordinates
provide one with the relevant degrees of freedom for the study of high energy
interactions and play a r\^ole similar to that of the fundamental fields with
given light-ray coordinates for light-cone dominated processes. In the course of
interaction with each other, the reggeized gluons change their two-dimensional
transverse coordinates and the color charge. The partial waves
$T_\omega(\vec{b},\vec{b'})$ can be classified according to the number of
reggeized gluons propagating in the $t$-channel; the minimum number of two
gluons is required in order to get a colorless exchange. Moreover, it can be
shown that retaining only two $t$-channel reggeized gluons corresponds to the
resummation of energy logarithms to the leading logarithmic accuracy. In this
approximation the amplitude $T_\omega(\vec b_1,\vec b_2;\vec b'_1,\vec b'_2)$
depends on four transverse coordinates and satisfies a Bethe--Salpeter-like
equation -- the so-called BFKL equation -- in the operator form
\begin{equation}\label{BS-coor}
\omega\mathbb{T}_\omega=\mathbb{T}_\omega^{(0)} +\frac{\alpha_s N_c}{\pi}\,
\mathbb{H}_{\rm BFKL}\,\mathbb{T}_\omega\,,
\end{equation}
where $\mathbb{T}_\omega^{(0)}$ corresponds to the free exchange of two gluons.
The solution of this equation can formally be written as
\begin{equation}\label{sol}
\mathbb{T}_\omega=\left(\omega-\frac{\alpha_s N_c}{\pi}\,
\mathbb{H}_{\rm BFKL}\right)^{-1}\mathbb{T}_\omega^{(0)}\, ,
\end{equation}
whence one concludes that the singularities of $\mathbb{T}_\omega$ in the
$\omega$-plane are determined by the eigenvalues of the BFKL operator
\begin{equation}\label{H-bfkl}
\left[\mathbb{H}_{\rm BFKL}\cdot \Psi_\alpha\right](\vec b_1,\vec b_2) =
E_{\alpha}\; \Psi_\alpha(\vec b_1,\vec b_2)\,,
\end{equation}
with $\alpha$ enumerating the solutions. The high-energy behavior of the scattering
amplitude is governed by the right-most singularity of $\mathbb{T}_\omega$,
which corresponds to the maximal eigenvalue ${\rm max}_\alpha E_{\alpha}$.
Equation (\ref{H-bfkl}) has the form of a Schr\"odinger equation for a system of
two interacting particles on the two-dimensional plane. Such particles can be
identified as reggeized gluons and the eigenstates $\Psi_\alpha(\vec b_1,\vec b_2)$
have the meaning of the wave functions of the color-singlet compound states built
from two reggeized gluons.

The BFKL operator $\mathbb{H}_{\rm BFKL}$ has a number of remarkable properties
which allow one to solve the Schr\"odinger equation (\ref{H-bfkl})
exactly~\cite{Lipatov85,Lipatov90}. First of all, $\mathbb{H}_{\rm BFKL}$
splits into the sum of two operators acting on the holomorphic and the
antiholomorphic coordinates
\begin{equation}
\vec{b}_j=\{x_j,y_j\} \quad \Longrightarrow \quad   z_{j} = x_{j}+i
y_{j}\,,\quad
\bar{z}_{j} = x_{j}-i y_{j}\,,
\end{equation}
(with $j=1,2$) on the transverse plane:
\begin{equation}\label{HH}
\mathbb{H}_{\rm BFKL}
=\mathcal{H}_2 +\overline{\mathcal{H}}_2\,
\end{equation}
with
\begin{equation}
\label{BFKLkernel}
\mathcal{H}_2 = \partial_{z_1}^{-1} \ln (z_{12})\, \partial_{z_1} +
\partial_{z_2}^{-1} \ln (z_{12}) \,\partial_{z_2}
+ \ln (\partial_{z_1}\partial_{z_2})-2\psi(1)\,,
\end{equation}
where $z_{12}=z_1-z_2$ and $\overline{\mathcal{H}}_2$ is given by a similar
expression in the $\bar z-$sector.

Another remarkable property of $\mathbb{H}_{\rm BFKL}$  is that it is
invariant under the conformal $SL(2;\mathbb{C})$ transformations of the
reggeon coordinates on the plane
\begin{equation}
z_k \to \frac{a z_k + b_k}{c z_k + d}\,,\qquad (ad-bc=1).
\end{equation}
Consider the generators of the holomorphic $SL(2,\mathbb{C})$ transformations
\begin{equation}\label{c-gen}
L_{k,-}=-\partial_{z_k}\,,\qquad L_{k,0}=z_k\partial_{z_k}\,,\qquad
L_{k,+}=z_k^2\partial_{z_k}\,,
\end{equation}
and the corresponding antiholomorphic  generators $\bar L_{k,-}$, $\bar
L_{k,0}$ and $\bar L_{k,+}$ given by similar expressions with $z_k$ replaced
by $\bar z_k$, with $k=1,2$ enumerating particles. By inspection one finds
that $\mathbb{H}_{\rm BFKL}$ commutes with all two-particle generators
\begin{equation}\label{?-1}
[\mathbb{H}_{\rm BFKL},L_{1,a}+L_{2,a}] =
[\mathbb{H}_{\rm BFKL},\bar L_{1,a}+\bar L_{2,a}] = 0\,
\end{equation}
with $a=+,-,0$. This implies that $\mathbb{H}_{\rm BFKL}$ only depends on the
two-particle Casimir operators of the $SL(2,\mathbb{C})$ group
\begin{equation}\label{SL2C-Casimir}
{L}_{12}^2=-(z_1-z_2)^2\partial_{z_1}\partial_{z_2}\,,\qquad
\bar{{L}}_{12}^2=-(\bar z_1-\bar z_2)^2\partial_{\bar z_1}
\partial_{\bar z_2}\,,
\end{equation}
and leads to $\mathcal{H}_2=\mathcal{H}_2({L}_{12}^2)$ and $\overline{\mathcal{H}}_2
= \overline{\mathcal{H}}_2(\bar{{L}}_{12}^2)$. As a consequence, solutions of the
Schr\"odinger equation (\ref{H-bfkl}) have to be eigenstates of the Casimir
operators
\begin{equation}\label{eig-SL2C}
{L}_{12}^2\Psi_{n,\nu}=h(h-1) \Psi_{n,\nu}\,,\qquad
\bar {L}_{12}^2\Psi_{n,\nu}=\bar h(\bar h-1) \Psi_{n,\nu}\, .
\end{equation}
Here a pair of complex conformal spins is introduced
\begin{equation}\label{h}
h=\frac{1+n}2+i\nu \quad\mbox{and}\quad
\bar h=\frac{1-n}2+i\nu\,
\end{equation}
with a non-negative integer $n$ and real $\nu$ that specify the irreducible
(principal series) representation of the $SL(2,\mathbb{C})$ group to which
$\Psi_{n,\nu}$ belongs. The solutions to Eqs.\ (\ref{eig-SL2C}) read as
\begin{equation}\label{wf-2}
\Psi_{n,\nu}(b_1,b_2) =\left(\frac{z_{12}}{z_{10}z_{20}}\right)^{(1+n)/2+i\nu}
\left(\frac{\bar z_{12}}{\bar z_{10}\bar z_{20}}\right)^{(1-n)/2+i\nu}\,,
\end{equation}
where $z_{jk}=z_j-z_k$ and $b_0=(z_0,\bar z_0)$ is the collective
coordinate, reflecting the invariance of $\mathbb{H}_{\rm BFKL}$ under
translations. We recall that the solution (\ref{wf-2}) defines the wave
function of the color-singlet compound state, which is built from two
reggeized gluons with the coordinates $b_1=(z_1,\bar z_1)$ and
$b_2=(z_2,\bar z_2)$. The integer $n$ fixes the two-dimensional Lorentz
spin of the state, the real valued $\nu$ gives the scaling dimension
$\ell=1+2i\nu$, and the two-dimensional vector $b_0$ sets up the
center-of-mass coordinate of the state.

To find the eigenvalues in Eq.\  (\ref{H-bfkl}), one substitutes
the wave function (\ref{wf-2}) into the Schr\"odinger equation (\ref{H-bfkl})
and uses the explicit form of the BFKL kernel leading to~\cite{BFKL}
\begin{equation}\label{2-en}
E_{n,\nu}=2\psi(1)-\psi\left(\frac{n+1}2+i\nu\right)
-\psi\left(\frac{n+1}2-i\nu\right)\,.
\end{equation}
Its maximal value, $\textrm{max}\,E_{n,\nu}=4\ln 2$, corresponds to $n=\nu=0$,
or equivalently $h=\bar h=1/2$. It defines the position of the right-most
singularity of the partial wave amplitude and is translated into the asymptotic
behavior of the scattering amplitude in the leading logarithmic approximation,
\begin{equation}\label{BFKL}
\mathcal{A}(s,t)\sim i\,s^{1+\frac{\alpha_s N_c}{\pi}4\ln 2}\,,
\end{equation}
known as the BFKL pomeron.%
\footnote{\,Due to accumulation of the energy levels $E_{n,\nu}$ around
$n=\nu=0$, the corresponding Regge singularity is not a pole but a square-root
cut so that (\ref{BFKL}) is modified by additional $(\alpha_s\ln
s)^{-1/2}-$factor.}

Using (\ref{2-en}) one can reconstruct the operator form of the BFKL kernel
$\mathbb{H}_{\rm BFKL}$ on the representation space of the principal series of
the $SL(2,\mathbb{C})$ group
\begin{equation}\label{H-oper}
\mathbb{H}_{\rm BFKL}=\frac12\left[H(J_{12}) + H(\bar J_{12})\right] \,,\qquad
H(j)=2\psi(1)-\psi(j)-\psi(1-j)\,,
\end{equation}
where, as before, the two-particle spins are defined as ${L}_{12}^2 = J_{12}
(J_{12}-1)$ and $\bar{L}_{12}^2=\bar J_{12}(\bar J_{12}-1)$. Notice that we
already encountered the similar Hamiltonian in Sect.~2 and found that it
gives rise to complete integrability for the three-particle evolution equations
for the helicity$-3/2$ baryon distribution amplitudes. It turns out that the
BFKL kernel (\ref{H-oper}) has the same hidden integrability. In order to
appreciate this, we have to consider states containing more than two
particles.

\subsection{Multireggeon compound states}

In the leading logarithmic approximation, the scattering amplitude \re{BFKL}
grows as a power of the energy and violates unitarity constraints. To
restore unitarity, subleading corrections must be included.
Going beyond the leading logarithmic approximation,
 one has to take into account contributions
to the scattering amplitude in Eq.~(\ref{A-Four}) with more than two reggeized
gluons propagating in the $t$-channel. One can argue \cite{Bartels80,CDLO81}
that, in order to restore the unitarity of the scattering amplitude in the
direct ($s-$ and $t-$) channels, it suffices to retain contributions where
the number $N$ of the $t$-channel gluons is conserved, i.e., neglect the
three-reggeon vertices in the Regge theory language. With this assumption, the
amplitude ${\mathcal A}_N(s,t)$ corresponding to the exchange by $N$ reggeized
gluons satisfies the Bartels--Kwiecinski--Praszalowicz (BKP) equation
\cite{Bartels80,KP80}, which is the generalization of the Bethe--Salpeter
equation (\ref{BS-coor}) to the $N-$gluon scattering amplitude. Singularities
of $\mathbb{T}_\omega$ in the complex $\omega$-plane are determined by the
eigenvalues of the BKP operator ${\mathbb H}_N$. They can be thought of as
energies of the $N$-gluon compound states
\begin{equation}
{\mathbb H}_N \Psi_{N,\{q\}}(\vec b_1,\ldots,\vec b_N)= E_{N,\{q\}}
\Psi_{N,\{q\}}(\vec b_1,\ldots,\vec b_N)\,,
\label{BKP}
\end{equation}
with $\{q\}$ being some set of quantum numbers parameterizing the
solutions, and $\vec b_n$ the two-dimensional transverse positions
of $n-$th reggeized gluon.

In the large $N_c$ limit, the relevant Feynman diagrams have the topology of
a cylinder and ${\mathbb H}_N$ reduces to the sum of terms corresponding to
pairwise nearest-neighbor BFKL interactions:
\begin{equation}
{\mathbb H}_N = \frac12 \sum_{k=1}^N {\mathbb H}_{k,k+1}^{\rm BFKL}
\label{H-multi}
\end{equation}
with periodic boundary conditions ${\mathbb H}_{N,N+1}^{\rm BFKL}={\mathbb
H}_{N,1}^{\rm BFKL}$. The holomorphic separability of ${\mathbb H}_{\rm BFKL}$
implies the same property for ${\mathbb H}_N$
\begin{equation}
{\mathbb H}_N = \mathcal{H}_N  + \bar{\mathcal{H}}_N
\end{equation}
with the operators $\mathcal{H}_N$ and $\bar{\mathcal{H}}_N$ acting on $z-$ and
$\bar z-$coordinates, respectively. The conformal symmetry is also inherited by
them, so that each of the two-particle Hamiltonians can be written in terms of
the corresponding two-particle Casimir operators
\begin{equation}
{\mathbb H}_N = \frac12 \sum_{k=1}^N
\Big[
H(J_{k,k+1}) +  H(\bar J_{k,k+1})
\Big],
\end{equation}
where $H(j)$ is given in terms of $\psi$-functions (see Eq.~(\ref{H-oper})).

As a consequence, the (2+1)--dimensional Schr\"odinger equation
(\ref{BKP}) can be replaced by the system of holomorphic and antiholomorphic
(1+1)--dimensional Schr\"odinger equations for the Hamiltonians $\mathcal{H}_N$
and $\bar{\mathcal{H}}_N$. This allows one to expand the wave function of the
$N-$reggeon state over the eigenstates of $H_N$ and $\bar H_N$ as~\cite{JW99}
\begin{equation}
\Psi(\vec b_1,\ldots,\vec b_N) = \sum_{a,b} C_{ab}\, \Psi^{(a)}(z_1,\ldots,z_N)
\bar\Psi^{(b)}(\bar z_1,\ldots,\bar z_N)\,,
\label{CFT-blocks}
\end{equation}
where $C_{ab}$ are the mixing coefficients.

Most remarkably, the Hamiltonian \re{H-multi} is in fact the  Hamiltonian of
the $SL(2,\mathbb{C})$ Heisenberg magnet. The number of sites in the spin chain
equals the number of reggeons and the corresponding spin operators are
identified as six generators, $L_k^\pm, L_k^0$ and $\bar L_k^\pm,\bar L_k^0$,
of the $SL(2,\mathbb{C})$ group. The non-compact $SL(2,\mathbb{C})$ Heisenberg
magnet describes a completely integrable model. It possesses a large-enough set
of mutually commuting conserved charges $q_n$ and $\bar q_n$ $(n=2,...,N)$ such
that $\bar q_n= q_n^\dagger$ and $[\mathbb{H}_N,q_n]=[\mathbb{H}_N,\bar
q_n]=0$. The charges $q_n$ are polynomials of degree $n$ in the holomorphic
spin operators. They have the following form~\cite{Lipatov94,FK95}
\begin{equation}
q_n= \sum_{1\le j_1 < j_2 < ... < j_n \le N} z_{j_1j_2}z_{j_2j_3}... z_{j_nj_1}
p_{j_1}p_{j_2}...p_{j_n}
\label{q_n}
\end{equation}
with $z_{jk}=z_j-z_k$ and $p_j=i\partial_{z_j}$. The ``lowest'' charge $q_2$ is
related to the total spin of the system $h$. For the principal series of the
$SL(2,\mathbb{C})$ it takes the following values
\begin{equation}
q_2=-h(h-1)\,,\qquad h=\frac{1+n_h}2+i\nu_h\,,
\label{h1}
\end{equation}
with $n_h$ integer and $\nu_h$ real. The eigenvalues of the integrals of motion,
$q_2,...,q_N$, form the complete set of quantum numbers parameterizing the
$N-$reggeon states \re{BKP}.

Identification of \re{H-multi} as the Hamiltonian of the $SL(2,\mathbb{C})$
Heisenberg magnet allows one to map the $N-$reggeon states into the eigenstates
of this lattice model. In spite of the fact that the noncompact $SL(2,\mathbb{C})$
Heisenberg magnet represents a generalization of the compact $SU(2)$ spin chain,
very little was known about its energy spectrum until recently. The principal
difficulty is that, in distinction with compact magnets, the quantum space of
the $SL(2,\mathbb{C})$ magnet does not possess a highest weight -- the so-called
``pseudo-vacuum state'' -- and, as a consequence, conventional methods like
the Algebraic Bethe Ansatz method~\cite{QISM} are not applicable. The eigenproblem
\re{H-multi} has been solved exactly in Refs.~\cite{DKM01,KKM02,DKKM02} using the
method of the Baxter $\mathbb{Q}-$operator~\cite{Baxter} which does not rely on
the existence of a= highest weight. In this approach, it becomes possible to
establish the quantization conditions for the integrals of motion $q_3,...,q_N$ and
to obtain an explicit form for the dependence of the energy $E_N$ on the integrals
of motion.

Thanks to complete integrability, the ``conformal blocks'' $\Psi^{(a)}$ and
$\bar \Psi^{(b)}$, Eq.~\re{CFT-blocks}, diagonalize the integrals of motion
in the holomorphic and antiholomorphic sectors, respectively. The function
$\Psi(\vec b_1,\ldots,\vec b_N)$ is a single-valued function on the
two-dimensional plane. It belongs to the principal series of the $SL(2,\mathbb{C})$
group, labelled by the spins $(h,\bar h)$ defined in \re{h} (with $\bar h=1-h^*$),
and is normalizable with respect to the $SL(2,\mathbb{C})$ scalar product. In
contrast with $\Psi(\vec b_1,\ldots,\vec b_N)$, the chiral solutions $\Psi^{(a)}$
and $\bar \Psi^{(b)}$ acquire nontrivial monodromy when $\vec b_k$ encircles other
particles on the plane. The quantization conditions for the integrals of motion
$\Mybf{q}$ ensure that the monodromy cancels in the r.h.s.\ of \re{CFT-blocks}.
The same condition fixes (up to an overall normalization) the mixing coefficients
$C_{ab}$~\cite{JW99,DKKM02}.

In this manner, the spectrum of the $N-$reggeon state has been calculated for $N\ge 3$ particles : For $N=3$ few low-lying states have been found in
\cite{JW99,BLV99} and the complete spectrum of states for $3 \le N \le 8$ was
determined in \cite{KKM02,DKKM02} (see also \cite{VL01}). Close examination
revealed the following properties. The quantized values of the charges $q_k$ (with
$k=3,...,N$) depend on the ``hidden'' set of integers
$\Mybf{\ell}=(\ell_1,\ell_2,...,\ell_{2(N-2)})$
\begin{equation}
q_k = q_k(\nu_h;n_h,\Mybf{\ell})
\, ,
\label{q-ell}
\end{equation}
where the integer $n_h$ and the real number $\nu_h$ define the total $SL(2,\mathbb{C})$ spin of
the state, Eq.~\re{h}. As a function of $\nu_h$, the charges form a
family of trajectories in the moduli space $\Mybf{q}=(q_2,q_3,...,q_N)$ labelled
by integers $n_h$ and $\Mybf{\ell}$ . Each trajectory in the $q-$space induces
a corresponding trajectory for the energy $E_N$
\begin{equation}
E_N=E_N(\nu_h;n_h,\Mybf{\ell})\,.
\end{equation}
A few examples of such trajectories for $N=3$ are shown in
Fig.~\ref{Fig-energy3}.

\psfrag{nu_h}[cc][bc]{$\mbox{$\nu_h$}$}
\psfrag{-E_3/4}[cc][cc]{$E_3/4$}
\psfrag{Im(q3^(1/3))}[cc][cc]{$\Im{\rm m}[q_3^{1/3}]$}
\psfrag{Re(q3^(1/3))}[cc][bc]{$\Re{\rm e}[q_3^{1/3}]$}
\begin{figure}[ht]
\centerline{{\epsfxsize8.0cm \epsfbox{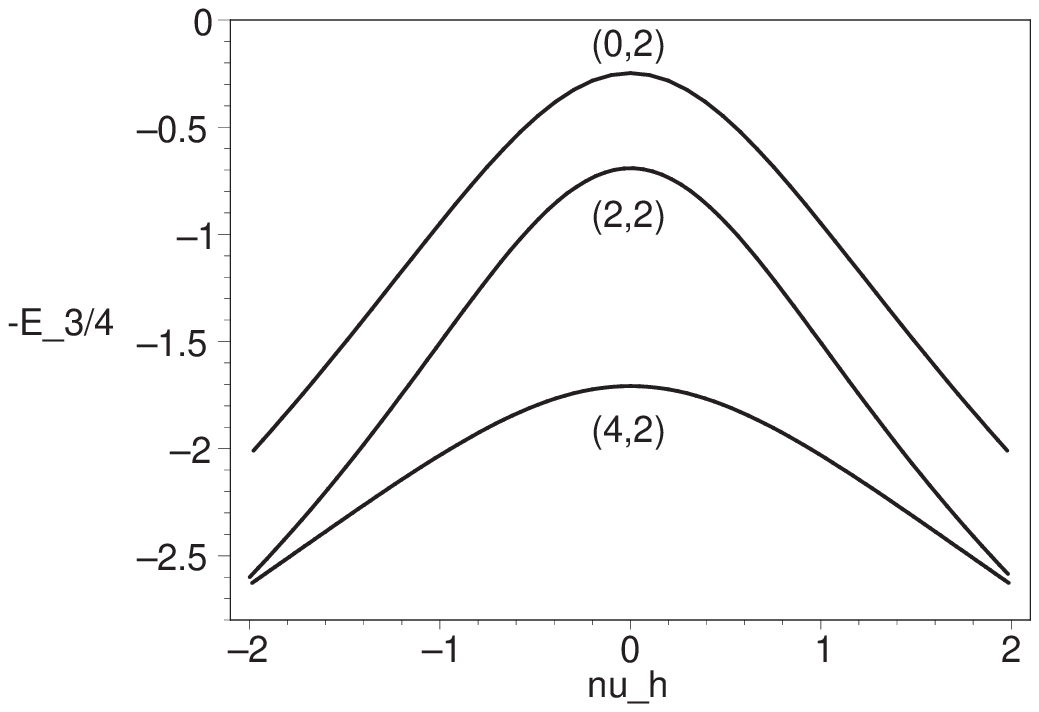}}} \vspace*{10mm}
\centerline{\hspace*{10mm}{\epsfxsize8.0cm \epsfbox{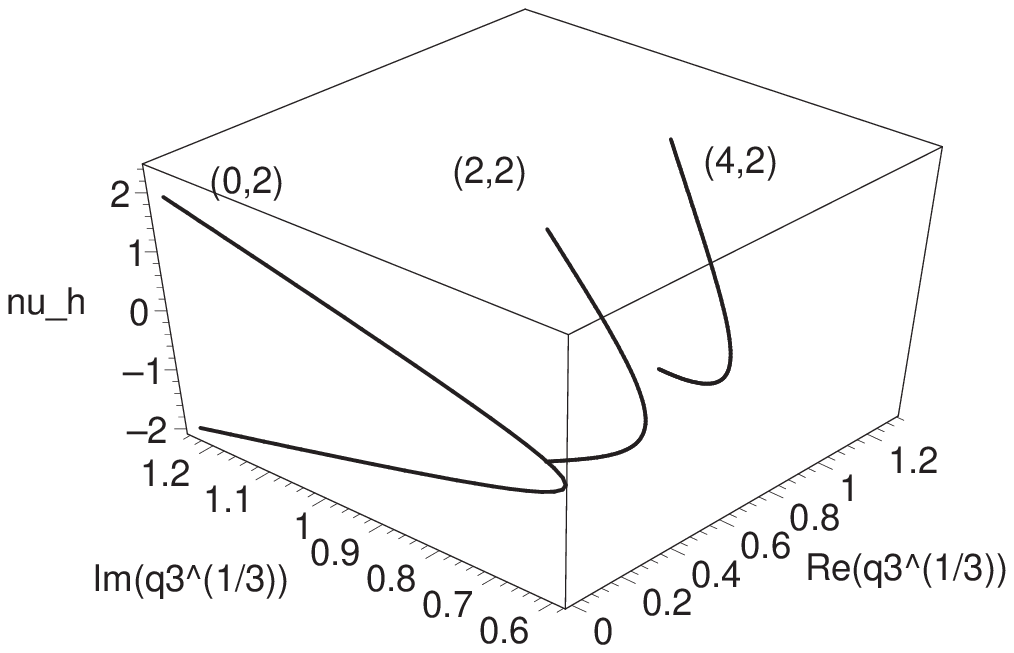}}}
\vspace*{5mm} \caption[]{The dependence of the energy $E_3$ and the conserved
charges $q_3$ on the total spin $h=1/2+i\nu_h$ for $n_h=0$. Three curves
correspond to the trajectories with $(\ell_1,\ell_2)=(0,2)\,, (2,2)$ and
$(4,2)$.}
\label{Fig-energy3}
\end{figure}

Due to the complicated form of the quantization conditions, it remains unclear
what the origin of trajectories is and what the physical interpretation of the integers
$\Mybf{\ell}$ is. Both questions can be answered by solving the Schr\"odinger
equation \re{BKP} within the semiclassical approach~\cite{DKM03}. One might
expect \textsl{a priori} that this approach could be applicable only for
highly-excited states. Nevertheless, as was demonstrated in \cite{DKM03}, the
semi-classical formulae work with good accuracy throughout the whole spectrum.
{}From the viewpoint of classical dynamics, the $N-$reggeon state describes
a chain of $N$ interacting particles on the two-dimensional $\vec b-$plane
\cite{Korchemsky96,KK97,GKK02}. The classical model inherits the complete
integrability of the quantum noncompact spin magnet. Its Hamiltonian and the
integrals of motion are obtained from \re{H-multi}, \re{H-oper} and \re{q_n}
by replacing the momentum operators by the corresponding classical functions.
Since the Hamiltonian \re{H-multi} is given by the sum of holomorphic and
antiholomorphic functions, from point of view of classical dynamics the model
describes two copies of one-dimensional systems ``living'' on the complex $z-$
and $\bar z-$lines. The solutions to the classical equations of motion have a
rich structure and turn out to be intrinsically related to the finite-gap
solutions to the nonlinear equations \cite{NMPZ84,Krichever77}; namely, the
classical trajectories have the form of plane waves propagating in the chain
of $N$ particles. Their explicit form in terms of the Riemann $\theta-$functions
was established in \cite{KK97} by the methods of finite-gap theory
\cite{NMPZ84,Krichever77}. The charges $\Mybf{q}$ define the moduli of the
finite-gap solutions and take arbitrary complex values in the classical model.
Going over to the quantum model, one finds that charges $\Mybf{q}$ are quantized.
In the semiclassical approach, their values satisfy the Bohr--Sommerfeld
quantization conditions imposed on the orbits of classical motion of $N$ particles.

In a standard manner, the WKB ansatz for the eigenfunction of the model
\re{H-multi} involves the ``action'' function, $\Psi_{\rm WKB}(\vec z_1, \ldots,
\vec z_N)\sim \exp(iS_0/\hbar)$. Due to complete integrability of the classical
system, it can be defined as a simultaneous solution of the system of the
Hamilton--Jacobi equations
\begin{equation}
\sum_{k=1}^N \frac{\partial S_0}{\partial z_k}= P\,,\qquad
\mathrm{q}_n\lr{\Mybf{z},\frac{\partial S_0}{\partial \Mybf{z}}}= q_n\,, \qquad
(n=2,...,N)\,,
\label{H-J}
\end{equation}
where $\Mybf{z}=(z_1,...,z_N)$ denotes the set of holomorphic coordinates,
$\mathrm{q}_n(\Mybf{z},\Mybf{p})$ stands for the symbol of the operator
\re{q_n} and $P$ is a holomorphic component of the total momentum of $N$
particles. The $\bar z-$dependence of $S_0$ is constrained by similar relations
in the antiholomorphic sector. To find a general solution to Eq.~\re{H-J}, one
performs a canonical transformation to the classical separated
coordinates~\cite{Sklyanin,NMPZ84}
\begin{equation}
(\vec b_1,\vec b_2,...,\vec b_{N})\ \stackrel{{\rm SoV}}{\mapsto} \ (\vec
b_0,\vec x_1,\vec x_2,...,\vec x_{N-1})\,,
\label{SoV-class}
\end{equation}
with $\vec b_0$ being the center-of-mass coordinate of the system, and $\vec
x_n=(x_n,\bar x_n=x_n^*)$ the new collective (separated) coordinates. Similarly
to the example in Sect.~2.4, the classical dynamics in the separated variables
is determined by the spectral curve (``equal energy'' condition)
\begin{equation}
\Gamma_N: \qquad y^2=t_N^2(x) - 4x^{2N}\,,
\label{curve}
\end{equation}
where $y(x)=2x^N\sinh p_x$ and  $p_x$ are the momenta in the separated
coordinates. Here $t_N(x)$ is a polynomial of degree $N$ with the coefficients
defined by the holomorphic integrals of motion $q_n$
\begin{equation}
t_N(x)=2x^N + q_2 x^{N-2} + ...+ q_{N-1} x + q_N\,.
\label{roots}
\end{equation}
The spectral curve \re{curve} establishes the relation between holomorphic
components of the separated coordinates, $x$ and $p_x$, for a given set of the
charges $q_2,...,q_N$.

In the separated coordinates, the solution to the Hamilton--Jacobi equations
\re{H-J} takes the form $S_0(\vec z_0,\vec x_1,\vec x_2,...,\vec x_{N-1})=(\vec
P\cdot \vec z_0)+ \sum_{k=1}^{N-1}S_0(\vec x_k)$ with~\cite{NMPZ84}
\begin{equation}
S_0(\vec x)
=
\int^x_{x_0} dx\, p_x + \int^{\bar x}_{\bar x_0} d \bar x \, \bar p_{\bar x}
=
2 \Re{\rm e}\int^x_{x_0} dx \, p_x
\, .
\end{equation}
Here $\bar p_{\bar x}=p_x^*$ is the complex momentum defined in \re{curve} and
$\vec x_0$ is arbitrary. The WKB expression for the wave function in the
separated coordinates factorizes into a product of single-particle wave
functions
\begin{equation}
\Psi_{N,q}^{\rm (SoV)} = {\rm e}^{i\vec P \cdot \vec b_0} Q(\vec x_1) \ldots
Q(\vec x_{N-1})
\end{equation}
with $Q(\vec x_k)\sim \exp\lr{{i} S_0(\vec x_k)}$. According to \re{curve}, the
momentum, $p_x$, and, as a consequence, the action function $S_0(\vec x)$ are
multi-valued functions of $x$. Denoting the different branches of the action
function by $S_{0,\,\alpha}(\vec x)$, one writes the WKB expression for the
wave function of the quantum spin magnet as a sum over
branches~\cite{PG92,Korchemsky96}
\begin{equation}
Q(\vec x) = \sum_\alpha A_\alpha(\vec x) \exp\lr{\frac{i}{\hbar}
S_{0,\,\alpha}(\vec x)}\,,\label{WKB-wave}
\end{equation}
with $\hbar =1$. The function $A_k(\vec x)$ takes into account subleading WKB
corrections and is fixed uniquely  by $S_{0,\alpha}(\vec x)$. The quantization
conditions for the charges $\Mybf{q}$ follow from the requirement that
\re{WKB-wave} has to be a single-valued function of $\vec x$. As was shown in
Refs.~\cite{GKK02,DKM03}, these conditions can be expressed in terms of the
periods of the ``action'' differential over the canonical set of the $\alpha-$
and $\beta-$cycles on the Riemann surface corresponding to the complex curve
\re{curve}
\begin{equation}
\Re{\rm e} \oint_{\alpha_k}dx\,p_x =\pi\,\ell_{2k-1}
\, , \qquad
\Re{\rm e} \oint_{\beta_k}dx\,p_x =\pi \, \ell_{2k}
\, ,
\label{WKB-intro}
\end{equation}
with $k=1,..,N-2$ and $\Mybf{\ell}=(\ell_1,\ldots,\ell_{2N-4})$ being the set
of integers. The relations \re{WKB-intro} define the system of $2(N-2)$ real
equations on the $(N-2)$ complex charges $q_3,...,q_N$ (we recall that the
eigenvalues of the ``lowest'' charge $q_2$ are given by \re{h1}). Their
solution leads to the semiclassical expression for the eigenvalues of the
conserved charges. In turn, the energy of the $N-$reggeon states $E_{N,q}$
can be expressed  as a function of $q_3,...,q_N$. In the semiclassical approach,
the corresponding expression is
\begin{eqnarray}\label{E-specc}
E_N^{\rm (as)}
&=& 4 \ln 2
\\
&+& 2 \Re{\rm e} \sum_{k=0}^N
\bigg[
\psi(1 + i \Re{\rm e} \lambda_k + | \Im{\rm m}\lambda_k|)
+
\psi(i \Re{\rm e}\lambda_k + |\Im{\rm m}\lambda_k|) - 2 \psi(1)
\bigg]
\, , \nonumber
\end{eqnarray}
where $\lambda_k$ are roots of the polynomial $t_N(u)$ (see Eq.~\re{roots}).
{The expression in Eq.~\re{E-specc} is similar to the energy of
the $SL(2,\mathbb{R})$ magnet in Eq.~\re{WKB-E} although the properties
of the two models are different.} As was demonstrated in
\cite{DKM03}, the resulting semiclassical expressions for $q_3,...,q_N$ and
$E_N$ are in good agreement with exact results~\cite{KKM02,DKKM02}.
As an example, we show in Fig.~\ref{Fig:q3} a comparison of the exact
spectrum of the conserved charge $q_3$ for the $N=3$ states carrying the
conformal spin $h=1/2$ (or equivalently $n_h=\nu_h=0$)
with the semiclassical expression given by
\begin{equation}
q_3^{1/3}=\frac{\Gamma^3(2/3)}{2\pi}\left[\frac12(\ell_1+\ell_2)
+i\frac{\sqrt 3}2(\ell_1-\ell_2)\right]
\label{q3-N3}
\end{equation}
with integer $\ell_1$ and $\ell_2$ introduced in \re{q-ell}.
\begin{figure}[t]
\psfrag{Im(q3^(1/3))}[cc][cc]{$\Im{\rm m}[q_3^{1/3}]$}
\psfrag{Re(q3^(1/3))}[cc][bc]{$\Re{\rm e}[q_3^{1/3}]$}
\centerline{{\epsfysize6cm \epsfbox{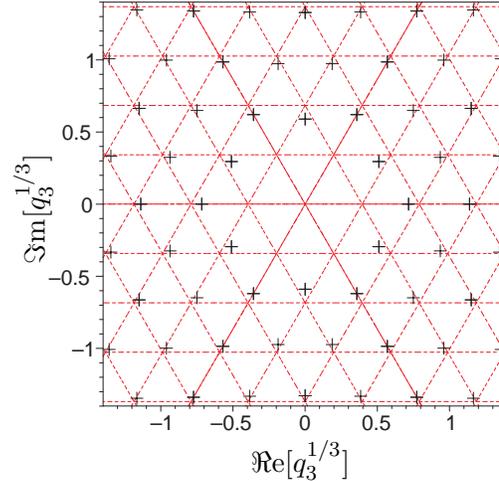}}}
\caption[]{Lattice structure at $N=3$. Crosses denote the exact values of $q_3^{1/3}$
at $h=1/2$. Dotted lines intersect at the points defined in Eq.~\re{q3-N3}.}
\label{Fig:q3}
\end{figure}

A novel feature of the quantization conditions \re{WKB-intro} is that they
involve \textsl{both} the $\alpha-$ and $\beta-$periods on the Riemann surface.
This should be compared with the situation in the $SL(2,\mathbb{R})$ Heisenberg
magnet discussed in Section~2.4. There, the WKB quantization conditions involve
only the $\alpha-$cycles, Eq.~\re{B-S}, since the $\beta-$cycles correspond to
classically forbidden zones. For the $SL(2,\mathbb{C})$  magnet, the classical
trajectories wrap over an arbitrary closed contour on the spectral curve
\re{curve} leading to \re{WKB-intro}. This fact allows one to explore the full
modular group~\cite{Dubrovin81} of the complex curve \re{curve} { and explain
the lattice structure of the spectrum shown in Fig.~\ref{Fig:q3}.} We refer
the interested reader to Ref.~\cite{DKM03} for details.

\section{Spin chains and $\mathcal{N}=2$ super Yang-Mill theories}
\setcounter{equation}{0}

In this section, we shall briefly discuss the integrability phenomenon which
emerged from the calculation of the low-energy effective action in
$\mathcal{N}=2$ SYM theory in Refs.~\cite{SW1,SW2}. Rather than delivering an
extensive review of the subject, we shall outline the similarities and differences
with the cases discussed in the previous sections. Detailed reviews of the
solutions to $\mathcal{N}=2$ SYM theories are given in \cite{Bilal96,Lerche97}
while reviews of the integrable structures in low-energy effective actions
can be found in \cite{GM00,PD99,Marshakov99}. We will argue below that the
low-energy effective action in the $\mathcal{N}=2$ theory with matter in the
fundamental representation is governed by integrable models which belong to the
same universality class as the spin chains that we encountered in the studies of anomalous dimensions of Wilson operators and high-energy scattering in
QCD. The particular form of the model depends on the number of matter
hypermultiplets $N_f$. To preserve the asymptotic freedom in the $\mathcal{N}=2$
SYM theory with the $SU(N_c)$ gauge group, one requires that $N_f$ does not
exceed $2N_{c}$. For $N_f=2N_c$ the beta-function in the $\mathcal{N}=2$ theory
vanishes and the low-energy effective action is described by an inhomogeneous
twisted Heisenberg XXX spin chain while for $N_f< 2N_c$ one encounters its
degenerations \cite{GMMM96,GGM1}.

The classical action of the $\mathcal{N}=2$ SYM theory has the following form
\begin{eqnarray}
\mathcal{S}
&=&
\Im {\rm m}
\int d^4 x \bigg[\tau \int d^2\theta d^2 \bar{\theta}\,
{\rm tr}\,
(
\Phi^\dagger e^{V}\Phi + Q_k^\dagger e^{V}Q^k + \tilde{Q}_k^\dagger e^{V}\tilde{Q}^k
)
\nonumber \\
&&\qquad\qquad
+
\tau \int d^2\theta\,\left({\rm tr}\, W^2 +\tilde{Q}_k\Phi Q^{k}
+
m_{k}^{l}\widetilde{Q}_lQ^k\right)
\bigg]
\label{S-classical}
\end{eqnarray}
where $\tau$ is the complex coupling constant
\begin{equation}
\tau=\frac{4\pi i}{g^{2}}+{\theta\over 2\pi}\,.
\end{equation}
$\Phi$ is the scalar, chiral $\mathcal{N}=1$ superfield involving the scalar fields
and fermions. $V$ is the vector $\mathcal{N}=1$ superfield involving the gauge field
and its superpartner (the gluino), $W$ is the spinor superfield constructed from $V$ in
such a way that ${\rm tr}\,W^2$ gives rise to kinetic terms for the gluons
and gluinos. The superfields $\widetilde{Q}_k$ and $Q_k$ (with $k=1,\dots,N_f$)
represent matter in the fundamental and anti-fundamental representations of the
gauge group $SU(N_c)$,
respectively, while the superfields $\Phi$ and $W$ belong to the adjoint representation.
The $\mathcal{N}=1$ superfields $W$ and $\Phi$ can be combined
together into the $\mathcal{N}=2$ chiral superfield $\Psi_{\mathcal{N}=2}$, while
pairing $\widetilde{Q}_k$ and $Q_k$ yields the $\mathcal{N}=2$ hypermultiplet.

In the $\mathcal{N}=2$ SYM theory, the vacuum states which do not break
supersymmetry are field configurations with vanishing energy. It turns out that
there are vacuum valleys, parameterized by the vacuum expectation values of the
fields $Q_k$ and $\Phi$. We shall be interested in the so-called Coulomb branch
where all matter condensates vanish, $\vev{Q_k}=0$, and the equation defining the
condensate of the scalars reads
\begin{equation}
{\rm tr}\,[\phi,\phi^\dagger]^{2}=0\,.
\end{equation}
Up to an $SU(N_c)$ gauge rotation, the expectation values of the scalar field
can be chosen to lie in the Cartan subalgebra $\phi={\rm
diag}(a_{1},...,a_{N_c})$ with ${\rm tr}\phi=0$. The parameters $a_k$ cannot
serve, however, as good order parameters, since there is still a residual Weyl
symmetry which changes $a_k$ but leaves gauge-invariant quantities intact. An
appropriate set of gauge invariant order parameters that fix the vacuum state
unambiguously is provided by $u_{k}=\vev{{\rm tr}\,\phi^{k}}$ with
$k=2,\ldots,N_c$. The variables $u_2,\ldots,u_{N_c}$ parameterize the space of
vacua in the $\mathcal{N}=2$ theory -- the moduli space, and the choice of a
point on the Coulomb branch is equivalent to a choice of the vacuum state.
Simultaneously, the expectation value of the scalars yields the scale at which
the coupling constant is frozen. At a generic point on the moduli space, for
non-vanishing $\phi$, the $SU(N_c)$ gauge symmetry is broken by the standard
Higgs mechanism and hence massive gauge bosons emerge. Integrating them out, one
arrives at the well-defined problem of the determination of the low-energy
effective action of the $\mathcal{N}=2$ SYM theory.

To calculate the $\mathcal{N}=2$ low-energy effective action, one has to modify
the classical action \re{S-classical} by taking into account one-loop
perturbative corrections and, in addition, include instanton corrections.
The explicit summation over the instantons is a complicated problem which has
been solved only recently \cite{Nekrasov02}. In their original work
\cite{SW1,SW2}, Seiberg and Witten avoided this problem and calculated the
low-energy effective action by making use of three ideas: holomorphicity, duality
and their compatibility with renormalization-group flows. Holomorphicity
implies that the low-energy effective action depends on a single holomorphic
function ${\mathcal F}$ called prepotential
\begin{equation}
\mathcal{S}_{\rm low-energy}
=
\Im {\rm m} \int d^4 x \int d^4 \theta d^4 \bar\theta \,
{\mathcal F}(\Psi_{\mathcal{N}=2})
\, .
\label{low-energy}
\end{equation}
Thus, the problem effectively reduces to the determination of the prepotential.
The holomorphic function ${\mathcal F}(a)$ is defined unambiguously by
its behavior in the vicinity of singular points on the complex $a-$plane. One of
the singularities, corresponding to large values of the condensate, is fixed by the
perturbative contribution to the $\mathcal{N}=2$ low-energy effective action. The behavior at the
remaining singularities is fixed by the use of duality and
non-renormalization theorems for the central charges of the SUSY algebra.

To define the duality transformations relating the weak and strong coupling constant
regimes in the $\mathcal{N}=2$ SYM theory, it is instructive to examine first the
$\mathcal{N}=4$ model. In this model, the coupling constant does not run and the
duality transformations are generated by $\tau \to \tau^{-1}$ and $\tau \to \tau
+1$. The same transformations define the modular group acting on
the moduli space of some Riemann surface parameterized by $\tau$. This suggests
that the duality properties of the SYM theory are encoded in the properties of
Riemann surfaces. Going over from $\mathcal{N}=4$ to $\mathcal{N}=2$ theory, one
finds that naive generalization of the duality meets serious difficulties. The reason for this
is that, in the asymptotically free theory, one has to marry the duality with the
renormalization group flow of the coupling constant $\tau$.
The solution to this problem is that one is still able to connect the duality and
modular transformations if one considers the $\mathcal{N}=2$ theory at different
vacua, connected to each other by the duality transformation. Then, the duality
acts on the moduli space of vacua which is associated with the moduli space
of the auxiliary Riemann surface. At the next step, one has to find out proper
variables whose modular properties fit the field theory interpretation. These
variables are identified with the integrals of a meromorphic 1-form $dS$ over
cycles on some Riemann surface \cite{SW1,SW2}
\begin{equation}
\label{aad}
a_{i}=\oint_{\alpha_{i}}dS\,, \qquad a_{D,i}=\oint_{\beta_{i}}dS\,,
\end{equation}
where $i=1,....,N_{c}-1$ for the $SU(N_{c})$ gauge group.

The integrals \re{aad} play a two-fold role in the Seiberg--Witten approach.
Firstly, they allow one to determine the prepotential ${\mathcal F}(a)$ through
the relation
\begin{equation}
a_{D,i}(a) = \frac{\partial {\mathcal F}(a)} {\partial a_i}
\end{equation}
with $a=(a_1,\ldots,a_{N_c-1})$ and, therefore, evaluate the low energy effective
action \re{low-energy}. Then, using the property of the differential $dS$ that
its variations w.r.t. moduli are holomorphic, one can also calculate the matrix of
coupling constants
\begin{equation}
\label{Tij}
T_{ij}(u)=\frac{\partial^{2}{\mathcal{F}(a)}}{\partial a_{i} \partial a_{j}}\ .
\end{equation}
Secondly, the periods $a$ and $a_D$, Eq.~(\ref{aad}), define the spectrum of stable states in the theory which saturate the Bogomolny--Prasad--Sommerfeld
(BPS) bound. For instance, the dyonic spectrum of stable particles of a $SU(2)$
theory reads
\begin{equation}
M_{n,m}=\big|na(u) +ma_{D}(u)\big|
\end{equation}
with $n$ and $m$ integeral.

The Seiberg--Witten solution admits an elegant interpretation in terms of
\textsl{classical} integrable systems. The Riemann surface entering (\ref{aad})
defines the solution to the classical equations of motion in a classical integrable
system. Moreover, the meromorphic differential $dS$ turns out to be the action
differential in the separated variables in the same system, $dS= p(x)\,d x$, with
$p(x)$ being a single-valued function on the Riemann surface \cite{GKMMM95}. The
dependence of the prepotential on the fundamental scale $\Lambda$ is given by
\cite{Matone:1995rx}
\begin{equation}
\frac{\partial {\mathcal F}(a)}{\partial \ln \Lambda}=\beta u_2(a) \equiv \beta
H\,,
\end{equation}
where $\beta$ is the one-loop beta-function in the
$\mathcal{N}=2$ SYM theory. This equation has another interpretation as the
evolution equation in an integrable system. $H$ coincides with the Hamiltonian of
this system and $\ln\Lambda$ is the evolution time variable. Hence we see, once
again, that the logarithm of the relevant scale plays the r\^ole of time in the
integrable dynamics.

As was already mentioned, the Riemann surface parameterizing the Seiberg--Witten
solution for the low-energy effective action in the $\mathcal{N}=2$ SYM theory
with fundamental matter \cite{GMMM96} coincides with the spectral curve
describing the solution to the classical equation of motion in the periodic
inhomogeneous Heisenberg XXX spin chain of length $N_c$~\cite{APS95,ho}. The
masses of the squarks in the $\mathcal{N}=2$ theory are mapped into the
parameters of the chain while the coordinates on the Coulomb branch of the moduli
space are mapped into the values of the integrals of motion. The spectral curve
for the spin chain system is defined as
\begin{equation}
\det (w - T_{N_c}(x)) =0\,, \label{sp-curve}
\end{equation}
where $T_{N_c}(x)=\tr[L_1(x) \ldots L_{N_c}(x)]$ is the transfer matrix of the
model and $L_i(x)$ is a local $2\times 2$ Lax matrix given by
\begin{equation}\label{LaxXXX}
L_i(x) = (x+\lambda_i) \cdot {\bf 1} + \sum_{a=1}^3 S_{a,i}\cdot\sigma^a\,.
\end{equation}
Here $\sigma^a$ are the Pauli matrices and $\lambda_i$ are the chain
inhomogeneities. The Poisson brackets of the dynamical variables $S_a$,
($a=1,2,3$) are just
\begin{equation}\label{Scomrel}
\{S_{a,j},S_{b,k}\} = -i\epsilon_{abc} S_{c,j} \delta_{jk}\,,
\end{equation}
so that the vector $\{S_a\}$ plays the r\^ole of the angular momentum (``classical
spin''). The spectral curve \re{sp-curve} is now
\begin{equation}\label{scsl2m}
w+{Q(x)\over w}=2P(x),
\end{equation}
with $ P(x) \equiv {\rm tr}\, T(x)/2$ and $Q(x)\equiv \det T(x)$. In the
hyper-elliptic parameterization, $y=(w+Q(x)/ w)/2$, the same curve becomes
\begin{equation}\label{gm8}
y^2=P^2(x)-Q(x)\,.
\end{equation}
The zeroes of $Q(x)$ define the masses of the hypermultiplets in the $\mathcal{N}=2$
SYM theory via
\begin{eqnarray}
Q(x) = \prod_{i=1}^{N_c} \det L_i(x) = \prod_{i=1}^{N_c}(x - m_i^+)(x - m_i^-)
\end{eqnarray}
where $m_i^{\pm} = -\lambda_i \pm s_i$, $S_{a,i}^2 = s_i^2$ and $s_i$
stands for the spins at the $k$th site of the chain. By construction, the
number of matter multiplets is twice the number of colors. Still, Eq.~\re{gm8}
is not the most general form of the spectral curve for $N_f = 2 N_c$.

Equation \re{gm8} can be generalized by taking into account the additional freedom
in the definition of the spin chain and the spectral curve. Namely, one can
multiply the Lax operator of the spin chain by an arbitrary constant matrix
without changing the commutation relations and conservation laws. Moreover,
one can also attach a constant (external magnetic field) matrix $V$ to the
end of the chain, that is $L_{N_c}(x) \to L_{N_c}(x) V$. This amounts to
imposing more general (twisted) boundary conditions along the spin chain.
The spectral curve corresponding to the ``twisted'' spin chain takes the
form \cite{SW2,APS95,ho}
\begin{eqnarray}
\label{AS}
&& w+{Q(x)\over w}=P(x), \\ && P(x)=\prod_{i=1}^{N_c} (x-\phi_i), \nonumber \\ &&
Q(x)=h(h+1)\prod_{j=1}^{2N_c} \left(x-m_j-{2h\over n}\sum_i m_i\right). \nonumber
\end{eqnarray}

Let us now consider these formulae in the special case of a superconformal
$\mathcal{N}$=2 SYM theory with $N_f=2N_c$ massless fundamental hypermultiplets
\cite{SW2,APS95,ho,minahan,argyres,khoze}. The corresponding integrable system is
described by the spectral curve
\begin{equation}
 y^2=P^2(x) -4{x}^{2N_c}(1- \rho^2(\tau_{cl}))\,,
\label{SUSY-curve}
\end{equation}
where $\rho^2(\tau_{cl})$ is some function of the coupling constant in the
$\mathcal{N}=2$ theory and the polynomial $P(x)$ depends on the coordinates on
the moduli space $\vec u=(u_2,...,u_{N_c})$
\begin{equation}
P(x)= \sum_{k=0}^{N_c} q_k(\vec u)\, x^{N_c-k} =2x^{N_c}+ q_2\, x^{N_c-2} + ... +
q_{N_c}\,,
\end{equation}
where $q_0=2$, $q_1=0$ and the other $q_k$ are some known functions of $\vec u$.
Their explicit form is not important for our purposes. The Seiberg--Witten
meromorphic differential on the curve  is given by
\begin{equation}
dS= p\,dx= \ln({\omega}/{{x}^{N_c}})\,dx\,,
\label{lambda-SW}
\end{equation}
where $y=\omega-x^{2N_c}/\omega$ and $\omega={x}^{N_c} {\rm e}^p$.

{}From the point of view of integrable models, the spectral curve \re{SUSY-curve}
corresponds to a classical Heisenberg XXX spin chain of length $N_c$ with
spin-zero at all sites (due to $q_1=0$) and parameter $\rho$ related to the
external magnetic field, or equivalently, to the twisted boundary conditions
\cite{GGM1}. Remarkably enough, the spectral curve \re{SUSY-curve} describing the
low-energy effective action of the superconformal $\mathcal{N}$=2 SYM theory with
$N_f=2N_c$ is closely related to the spectral curve \re{curve} describing the
spectrum of multi-reggeon compound states in multicolor QCD \cite{GKK02}. It is
easy to verify that the two curves coincide if one makes the following
identification:
\begin{itemize}
\item The number of the reggeons $N=N_c$; \item The integrals of motion for the
multi-reggeon state are identified as the above-mentioned functions $q_k(\vec u)$
on the moduli space of the superconformal theory; \item The coupling constant of
the gauge theory should be such that $\rho(\tau_{cl})=0$ that is $\tau_{\rm cl}=
\frac{1}{2}+ \frac{i}{2}\tan \frac{\pi}{2N_c}$
\end{itemize}
Under these conditions the two theories fall into the same universality class.

Let us briefly mention some recent developments concerning the relation between
integrable systems and $\mathcal{N}=1$ SYM theories. In the $\mathcal{N}=2$ SYM
theory, the Riemann surface degenerates at some points on the Coulomb branch of
the moduli space. After soft breaking of $\mathcal{N}=2$ down to
$\mathcal{N}=1$, these points correspond to the vacuum states of the
$\mathcal{N}=1$ SYM theory. Some massless states condense at these points leading
to the formation of a mass gap and to confinement. For instance, in the $SU(2)$
case the $\mathcal{N}=1$ vacua correspond to the points $u_2=\pm\Lambda^2$ where
the monopoles or dyons become massless and condense. It turns out that the
relation to  the integrable systems becomes even more direct after soft
breaking. Namely the tree-level superpotential $W_{\rm tree}=\sum_{k} t_k{\rm tr}
\Phi^k$ amounts to the following effective superpotential
\begin{equation}
W_{\rm eff} =\sum_{k} t_k {\rm tr} L^k\,,
\end{equation}
where $L$ is the Lax operator of the corresponding integrable system governing
the $\mathcal{N}=2$ theory with the same matter content
\cite{Dorey99,BBDW04,Hollowood03}. That is $W_{\rm eff}= \sum_{k} t_k H_k$, and
vacua in the $\mathcal{N}=1$ theory defined by the extremization of $W_{\rm eff}$
are in one-to-one correspondence with the equilibrium states of the
corresponding integrable system with respect to the combination of Hamiltonians
above.

\section{Gauge/string correspondence}
\setcounter{equation}{0}

In previous sections, we described the hidden integrability symmetry of the
one-loop dilatation operator in QCD and its supersymmetric extensions. Its
origin, however, remains obscure on the gauge theory side. In this section we shall
describe attempts to explain the integrability phenomena in Yang--Mills theories
based on the gauge/string correspondence. We will demonstrate that
\textsl{quantum} integrability of the dilatation operator in gauge theory is
ultimately related to \textsl{classical} integrability of the relevant stringy
sigma models. The subject is rapidly developing but a fully satisfactory picture
has not yet emerged.

To start with, let us recall the general features of the gauge/string correspondence
relevant for our discussion. In the case of the $\mathcal{N}=4$ SYM theory it
states that this gauge theory is dual to IIB string theory on an $AdS_5\times S^5$
background with additional flux of a higher-form field \cite{maldacena,obzor}.
The background metric in Poincar\'e coordinates has the form
\begin{equation}
ds^2
=
\frac{r^2}{R^2}(-dt^2 +dx_1^2 +dx_2^2 +dx_3^2 )
+
R^2\frac{dr^2}{r^2}
+
R^2d\Omega^{2}_{5}
\end{equation}
and can be considered as the near-horizon limit of the D3 brane metric. Since the
D3-brane is a source of the Ramond--Ramond four-form field $A_4$, the background
solution is supplemented by the flux of the corresponding field strength
\begin{equation}
F_5 = d A_4\, , \qquad \int_{S^5}*F_5=N_c \, .
\end{equation}
The radii of the $AdS_5$ and $S^5$ are identical and equal to
\begin{equation}
R^4= 4\pi g_s \alpha^{\prime 2} N_c\,.
\end{equation}
The four-dimensional gauge theory is localized on the boundary of the $AdS_5$.
The conformal $SO(2,4)$ group and the $R-$symmetry $SO(6)$ group of the
$\mathcal{N}=4$ SYM theory are identified with the isometry group of the $AdS_5$
and $S^5$ spaces, respectively. According to the gauge/string duality, the
eigenvalues of the dilatation operator in the $\mathcal{N}=4$ SYM theory and the
energy spectrum of the string in radial quantization coincide.

The string tension is related to the gauge theory 't Hooft coupling constant by
\begin{equation}
T = \frac{\sqrt{\lambda}}{2\pi}=\frac{R^2}{2\pi \alpha^\prime} \, , \quad \lambda
= g^2 N_c\,.
\label{string-tension}
\end{equation}
Therefore the strong coupling regime $\lambda\gg 1$ in the gauge theory
corresponds to the semiclassical regime on the stringy side. In this way, the
classical string calculations provide predictions for many interesting quantities
in the $\mathcal{N}=4$ SYM theory (like anomalous dimensions, condensates, Wilson
loops etc.) in the strong coupling regime \cite{obzor}. However, these predictions
can be checked diagrammatically only in a few examples, such as for a circular Wilson loop
\cite{loop,semenoff}. On the other hand, to approach the weak coupling regime in
Yang--Mills theory, one should go beyond the semiclassical approximation and take
into account quantum effects on the string theory side.

At present, an explicit quantum solution of the string theory in the $AdS_5\times
S^5$ background is not available. This makes it impossible to compare the string
spectrum with the complete set of operators in $\mathcal{N}=4$ SYM theory.
Hopefully the explicit solution of the string theory can be found~\cite{metzaev}
in the case of the limiting geometry defined as the Penrose limit of the $AdS_5\times
S^5$ \cite{bmn,Beisert:2002bb,Gross}. The Penrose limit can be described as the
region around the null geodesic in the $AdS_5 \times S^5$. Introducing new
variables
\begin{equation}
x^{+}= \frac{t+\chi}{2\mu},\quad x^{-}=\mu R^2 (t-\chi)
\end{equation}
with $\chi$ being an angular variable in $S^5$ and $\mu$ being some scale, one
takes the limit $R \rightarrow \infty$ and recovers the pp-wave metric
\begin{equation}
ds^2= -4dx^{+}dx^{-} - z^2 dx^{+2} + \sum_{i=1}^{8}dz_i^2\,.
\end{equation}
Here, eight, flat, transverse coordinates $z_i$ come both from the $S^5$ and $AdS_5$
parts of the geometry. In this metric, the string behaves as a particle rotating
with large angular momentum $J$ along the angular coordinate $\chi$ in $S^5$. The
light-cone energy of the string is
\begin{equation}
H=2p^{-}=i(\partial_{t}+\partial_{\chi})=(\Delta -J)\,.
\end{equation}
For $R\to \infty$ it takes finite values provided that the following double
scaling limit is considered:
\begin{equation}
R \rightarrow \infty\,,\qquad \Delta\sim J\rightarrow \infty\,, \qquad
\frac{J^2}{R^4} = \mbox{const}\,.
\end{equation}

Quantization of the string propagating in this background reduces to the
quantization of the oscillators. As a result, the exact spectrum of the type IIB
string in the pp-wave background looks like
\begin{equation}
\label{dim}
\Delta-J=\sum_k N_k\sqrt{1+\frac{\lambda k^2}{J^2}}\,,
\end{equation}
where $k$ labels the Fourier modes, $N_k$ denotes the total occupation number of
oscillatory mode and the condition $P=\sum_{k}kN_k=0$ is imposed.

On the gauge theory side, (\ref{dim}) defines the anomalous dimensions of certain
Wilson operators in the $\mathcal{N}=4$ SYM theory. The length of the string $J$
equals the number of constituents of the composite operator. The ground state
of the string can be identified with the operator built from scalars $Z=\Phi_1
+i\Phi_2$
\begin{equation}
|0,J \rangle \leftrightarrow {\rm tr}\,Z^J\,.
\end{equation}
These operators have charge $J$ with respect to the ``rotation plane" in the pp-wave.
The oscillatory excitations of the string ground state correspond to
insertions of other scalar fields. For example, the so-called BMN operators in
the $\mathcal{N}=4$ SYM theory can be mapped into the stringy modes as follows
\ba
 a_{0}^{i+} |0, J \rangle &\Leftrightarrow& {\rm tr}\,\Phi_i Z^J\,,
\nonumber
\\[2mm]
 a_{n}^{i+}a_{-n}^{j+} |0, J \rangle &\Leftrightarrow& \sum_{l}e^{2\pi i n l/J}
{\rm tr}\,\Phi_iZ^{l}\Phi_{j}Z^{J-l}\,.
\ea
One can deduce from these expressions that calculation of the stringy spectrum
corresponds to diagonalization of the mixing matrix for Wilson operators on the
gauge theory side.

The energy of the string in the pp-wave limit is a function of the ratio of the
coupling constant and angular momentum, ${\lambda}/{J^2}$. It is expected that
expansion of this function in powers of ${\lambda}/{J^2}$ should reproduce a
perturbative series for the anomalous dimension of the corresponding Wilson
operators in the weak-coupling regime. On the other hand, the one-loop
$\mathcal{N}=4$ dilatation operator in the sector of scalar operators
coincides with the Hamiltonian of the Heisenberg $SO(6)$ spin
chain \cite{mz}. This allows one to map stringy states into spin chain
states. The correspondence is very precise for Wilson operators built only from
two complex scalars, in which case the $SO(6)$ spin chain reduces to the
conventional Heisenberg $SU(2)$ spin$-1/2$ chain. Then, the ground state in the
string theory corresponds to all spins aligned in the same direction in the
isotopic space while the stringy excitations correspond to flipping some spins
along the chain.

The BMN operators are the only well-established example in which the exact quantum
answer on the stringy side can be matched into the all-loop anomalous dimension
on the gauge theory side. Later in this section, we shall discuss generalized BMN
operators which can be treated semiclassically on the stringy side and
demonstrate that the corresponding solutions to the classical equations of motion
in the stringy sigma models are ultimately related to classical integrable
models. We shall explain the relation between the string sigma model and quantum
spin chains in the thermodynamical limit and demonstrate that the anomalous
dimensions of the certain operators are in one-to-one correspondence with the
special class of classical solutions to the sigma model for which this model
reduces to finite dimensional integrable systems of the Neumann type. Finally, we
shall comment on the relation between general classical solutions to the sigma
model and the Bethe Ansatz solution to the compact quantum spin chains in the
semiclassical limit.

\subsection{Derivation of the string in the thermodynamical limit}

The sigma model describing the string moving in the appropriate curved background
can be derived from the quantum compact spin chain in the long-wavelength limit.
The corrections to the classical sigma model scale as ${1}/{J}$, where the
angular momentum of the string $J$ corresponds on the gauge theory side to the
number of fields entering the composite operator or equivalently the length of
the spin chain. The transition from the spin chains to the sigma model relies on
the coherent state formalism~\cite{Kruczenski03}.

Let $|ss\rangle$ be the state with total spin $s$ and projection onto the
$z-$axis $S_z=s$. The coherent state for the spin$-s$ representation of the
$SU(2)$ group is defined as
\begin{equation}
|\vec{n}\rangle=\e^{iS_x \phi}\e^{i S_y\theta}|ss\rangle
\end{equation}
where $\vec{n}$ is the unit vector, $\vec n^2=1$,
\begin{equation}
\vec{n}=(\sin\theta \cos\phi,\sin\theta \sin\phi, \cos\theta)
\end{equation}
with $\theta$ and $\phi$ being spherical angles. Expanding the Hamiltonian of the
spin chain $H=\lambda/(4\pi^2) \sum_{k=1}^J (1/4- \vec S_k\cdot\vec S_{k+1})$
over the coherent states, one rewrites the partition function $\tr \e^{-Ht}$ in the standard manner as
a path integral over $\vec S_k=s \vec n_k$ with the following action
\begin{equation}
\mathcal{S}(\vec{n})=s\sum_{k=1}^J\int dt\int _{0}^{1} d\tau\,
\vec{n}_k(\partial_t \vec{n}_k \times
\partial_{\tau}\vec{n}_k) - \frac{\lambda}{8\pi^2} s^2 \int dt
\sum_{k=1}^J(\vec{n}_k-\vec{n}_{k+1})^2\,,
\end{equation}
with $\vec n_{J+1}=\vec n_1$. In the long-wavelength limit, the vectors $\vec n_k(t)$
vary smoothly along the spin chain and, therefore, they can be approximated by a
function $\vec n(\sigma,t)$ with continuous $\sigma$ running between $0$ and the
chain length $J$, leading to
\begin{equation}
\mathcal{S}= -s\int dt d\sigma\, \partial_t \phi \cos\theta  -
\frac{\lambda}{8\pi^2} s^2 \int dt d\sigma \left[(\partial_{\sigma}\theta)^2 +
(\partial_{\sigma}\phi)^2\sin^2\theta \right].
\label{action}
\end{equation}
It turns out~\cite{Kruczenski03} that for $s=1/2$, this expression coincides with
the stringy action
\begin{equation}
\mathcal{S}_{\rm str}=\frac{R^2}{4\pi \alpha'}\int d\sigma d\tau \left[
G_{\mu\nu} \partial_\tau X^\mu \partial_\tau X^\nu-G_{\mu\nu} \partial_\sigma
X^\mu
\partial_\sigma X^\nu\right]
\end{equation}
evaluated for the classical
string propagating in the background $ds^2 =G_{\mu\nu} dX^\mu dX^\nu$
\begin{equation}
ds^2= -dt^2 +d\psi^2 +d\varphi_1^2 +d\varphi_2^2
+2\cos(2\psi)d\varphi_1d\varphi_2\,.
\end{equation}
To see this, one fixes the gauge $t=\chi\tau$, takes the limit $\partial_\tau
X^i\to 0$ and $\chi\to \infty$ with $\chi\partial_\tau X^i= \rm fixed$ and
identifies the variables as
\begin{equation}
\varphi_2=-\frac12\,\phi\,, \qquad \psi=\frac12\,\theta\,.
\end{equation}
Then one eliminates $\varphi_1$ with the help of the classical equations of motion
and arrives at (\ref{action}).

The derivation of the effective action can be also generalized to the $SU(3)$
case~\cite{lopes} and to the string carrying both large Lorentz spin $S$ and
$R-$charge $J$~\cite{ts}. One can improve the effective sigma model action
(\ref{action}) by calculating corrections involving higher derivatives of the
fields. All such terms containing up to four derivatives have been found in
Refs.\ \cite{tr,KruTse04}.

\subsection{Semiclassical string motion and integrable models}

We have argued above that the effective action for long-wavelength excitations in the
compact spin chain coincides with the classical action of the sigma model on the
curved background relevant for calculation of the anomalous dimensions of
BMN-like operators. As the next step, the corresponding solutions to the equations
of motion are compared. To this end, one considers the bosonic part of the
superstring action on the $AdS_5\times S^5$ background. It is given by the sum of
two coset sigma models
\begin{equation}
\mathcal{S}=\frac{\sqrt {\lambda}}{4\pi}\int d\sigma d\tau [G_{mn}^{AdS}\partial
y_{m}
\partial y_{n} + G_{kl}^{S^5}
\partial x^{k}\partial x^{l}]\,,
\end{equation}
where the string tension is proportional to the 't Hooft coupling,
Eq.~\re{string-tension}. It is convenient to rewrite the action with the
constraint imposed by the Lagrangian multiplier
\begin{equation}
\mathcal{S}=\frac{\sqrt {\lambda}}{4\pi}\int d\sigma d\tau [\partial X_{m}
\partial X_{m}+ \Lambda_{x}(X^2 -1)+
\partial Y^{k}\partial Y^{k}+ \Lambda_{y}(Y^2 +1)]
\label{S-Polyakov}
\end{equation}
where $X_n$ $(n=1,\dots6)$ and $Y_k$ $(k=0,\dots,5)$ are the two sets of the
embedded coordinates in the flat $R^6$ space with signatures $(6,0)$ and
$(4,2)$, respectively. The action has to be supplemented by the Virasoro
constraint for the vanishing of the two-dimensional energy momentum tensor
\begin{equation}
 \dot{Y_k}\dot{Y_l} + Y_{k}' Y_{l}' + \dot{X_n}\dot{X_n}+ X_{n}'X_{n}'
=\dot{Y_k}Y_{k}' +\dot{X_n}X_{n}'=0
\label{Virasoro-const}
\end{equation}
and by the periodic boundary conditions 
\begin{equation}
Y_k(\sigma +2\pi)=Y_k(\sigma), \qquad X_n(\sigma +2\pi)=X_n(\sigma)\,.
\label{boundary-cond}
\end{equation}
Due to the $SO(2,4)$ and $SO(6)$ symmetries, the classical action possesses the
set of conserved charges
\begin{eqnarray}
S_{kl} &=& \sqrt{\lambda}\int d\sigma(Y_k\dot{Y_l}- Y_l\dot{Y_k})
\, ,
\nonumber \\
J_{nm} &=& \sqrt{\lambda}\int d\sigma(X_n\dot{X_m}- X_m\dot{X_l})\,.
\end{eqnarray}
Among them, one distinguishes 6 Cartan generators: the energy $E=S_{05}$, the
Lorentz spins $S_{12},S_{34}$ and the $S^5$ angular momenta
$J_{12},J_{34},J_{56}$. These conserved charges parameterize general solutions to
the classical equations of motion~\cite{ft1}.

To describe a particular operator on the gauge theory side, we have to identify
the corresponding solution to the classical equations of motion in the sigma
model~\re{S-Polyakov}, subject to the constraints \re{Virasoro-const} and
\re{boundary-cond}. The simplest ansatz, corresponding to a string located at the
center of the $AdS_5$ and rotating in the $S^5$, is
\begin{equation}
Y_5+i Y_0=\e^{it}
\, , \qquad
X_{2i-1}+iX_{2i}=r_{i}(\sigma)\,\e^{i\omega_{i}\tau+i\alpha_i(\sigma)}
\,.
\end{equation}
with $i=1,2,3$ and the remaining $Y-$coordinates set to zero. Substituting
into the sigma model action \re{S-Polyakov} yields the Lagrangian
\cite{Arutyunov:2003za}
\begin{equation}
\mathcal{L} = \sum_{i=1}^{3} \left( r_{i}^{\prime 2} + r_{i}^2\alpha_{i}^{\prime
2} - \omega_{i}^2 r_{i}^2 \right) - \Lambda_x \sum_{i=1}^{3} \left( r_{i}^{2} - 1
\right) \, .
\label{system}
\end{equation}
Solving the equations of motion for $\alpha_i$ one gets $\alpha_{i}^\prime =
{v_i}/{r_{i}^2}$ with $v_i$ being the integration constants. The resulting
Lagrangian describes the integrable Neumann--Rosochatius system.%
\footnote{\,For
vanishing $v_i$ the system \re{system} reduces to the Neumann model with three
degrees of freedom.} It admits five independent integrals of motion: $v_1, v_2,
v_3$ plus two additional integrals of the form
\begin{equation}
I_{i} = r_{i}^2 + \sum_{j\neq i}^{3} \frac{1}{\omega_{i}^2 - \omega_{j}^2} \left[
\left( r_i r_{j}^\prime - r_i r_{j}^\prime \right)^2 + \frac{v_i^2
r_{j}^2}{r_i^2} + \frac{v_j^2 r_{i}^2}{r_j^2} \right]
\end{equation}
subject to $\sum_{i=1}^3 I_i=0$. The periodicity condition trades $v_i$ for three
integers $m_i$ and $I_i$ for two integers $n_i$. As a result, the energy depends
on the frequencies $\omega_i$ and five integers. These variables are not independent
since the Virasoro constraint imposes a relation between them. For this type of
string motion, the infinite set of conserved charges in the sigma model is parameterized
by a finite set of integrals of the motion \cite{am1,am2}. The energy corresponding to
classical solutions of the string sigma model defines the anomalous dimension of the
dual composite scalar operators in the $\mathcal{N}=4$ SYM theory. There are many
examples of such correspondences discussed in the literature, initiated in
Ref.~\cite{ft1} and further developed in
Refs.~\cite{Minahan:2002rc,Stefanski:2003qr,Engquist:2003rn,Engquist:2004bx,Kristjansen:2004ei,Tseytlin:2003ii}.

It is interesting to note that the Neumann system is isomorphic to the stationary
solutions of the Landau--Lifshitz equation. We recall that the time variable in
the Neumann system is identified with the coordinate along the string. This
identification survives time discretization in which case the Neumann system is
equivalent to the discretized version of the Landau--Lifshitz equation describing the
XYZ spin chain \cite{veselov} and it can be used for the formulation of the
integrability in the context of the string bit model \cite{gorsky}.

In the above example, the string was rotating in the $S^5$. In general, it moves
both in the $AdS_5$ and $S^5$ and could have large angular momenta in both
spaces. Contrary to the situation with pure scalar operators, when the comparison with the loop
expansion on the gauge side can be performed for operators with large $R$
charge, the situation with operators carrying large Lorentz spin $S$ is more
subtle. The folded, closed string rotating in the $AdS_5$ yields the dependence
for the anomalous dimensions of twist-two operators $F_{+ \perp } (D_+)^S
F_{+\perp}$ at large coupling of the form \cite{gkp}
\begin{equation}
\gamma^{\rm (tw=2)}_S =  \frac{\sqrt{\lambda}}{2\pi} \ln S^2\,.
\label{twist-2-open}
\end{equation}
This result can be generalized to higher twist operators of the form $F_{+\perp}
D_+^{S_1} F_{+\perp} \dots D_+^{S_{L-1}} F_{+\perp}$. The energy of the
corresponding revolving string coincides with the energy of the classical
Heisenberg spin chain of length $L$ and leads to \cite{bgk}
\begin{equation}
\gamma^{{\rm (tw=}L)}_{S_1, S_2, \dots , S_{L-1}} = \frac{\sqrt{\lambda}}{2\pi}
\ln q_L(S_1,S_2,\dots , S_{L-1}) \,.
\label{twist-L-open}
\end{equation}
Here $q_L$ is the highest integral of motion of the spin chain. For $S_k\sim S
\gg 1$ with $k=1,\ldots, L-1$ one has $q_L \sim S^L$.

Notice that the logarithmic scaling of the anomalous dimensions is a universal
feature of Wilson operators with large Lorentz spin in gauge theories, unrelated
to the presence of supersymmetry~\cite{Korchemsky88,km}. However stringy
description of this scaling at weak coupling remains unknown and it is doubtful
whether such classical string sigma model solutions exist~\cite{Tseytlin:2003ii}.
One should expect instead that $\sim \ln S$ behavior at weak coupling is driven
by the quantum sigma model.

The integrability phenomenon offers the possibility of extending the gauge/string
duality beyond the special class of classical string solutions described above.
Namely, instead of comparing particular solutions one can identify the integrable
structures corresponding to \textsl{quantum} spin chains describing the dilatation
operator on the gauge theory side and  to classical equations of motion on the
string theory side~\cite{kmmz}. It turns out that, in both cases, integrability is
encoded in the properties of Riemann surfaces.

For the quantum spin chains, the appearance of Riemann surfaces within the
framework of the Bethe Ansatz is not surprising. As we already explained in
Sect.~2.4, semiclassical solutions to the Baxter equation are determined by the
properties of the spectral curve whose genus is proportional to the number of
sites in the chain. In Sect.~2 we discussed baryon operators built from three
quarks and the corresponding Riemann surface, Eq.~\re{GamN}, had genus equal
to 1. In the case of the BMN-like operators, the number of constituent scalar
fields goes to infinity in the thermodynamic limit and, therefore, the
corresponding Riemann surface would have an infinite genus. However, by choosing the
appropriate values of the integrals of motion, it is possible to pinch almost all
handles and obtain a finite genus surface. It is this degenerate surface which
parameterizes general solutions to the classical equations of motion of the
string~\cite{krichever,kmmz}. The agreement between semiclassical solutions to
the Bethe Ansatz equations and solutions to the classical string equations of
motion has been carefully checked up to the two-loop level~\cite{kmmz}.

Although the correspondence between one- and two-loop\,\footnote{\,In a closed subsector
only.} dilatation operators in the $\mathcal{N}=4$ SYM theory Yang--Mills theory,
stringy states and integrable quantum spin chains is well established, the situation
with higher loops in perturbation theory is unclear. Several proposals have been
made concerning integrable structures behind a higher loop dilatation
operator~\cite{SS,bdm}. At the same time, starting at three-loop order the
discrepancy seems to arise between expressions for the anomalous dimensions of
composite scalar operators with large-$R$ charge and the energy spectrum of the
string~\cite{callan}. More work is needed to clarify this issue.

\subsection{Open string picture for anomalous dimensions}

There exists an alternative description of the logarithmic growth of the anomalous
dimensions of Wilson operators with large Lorentz spin, Eq.~\re{twist-2-open} and
\re{twist-L-open}, in terms of Wilson loops in the gauge theory and open strings
on the $AdS_5$ background. This picture relies on the correspondence between
the anomalous dimensions of the composite operators with large number of light-cone
derivatives and the so-called cusp anomaly of Wilson
loops~\cite{Korchemsky88,km}. It was shown a long time ago \cite{polyakov2} that the
Wilson loop $W[C]=\tr\{P\exp(i g \int_C dx^\mu A_\mu(x))\}$ acquires a nontrivial
anomalous dimension $\Gamma_{\rm cusp}(\lambda, \theta)$ if the integration
contour $C$ has a cusp
\begin{equation}
\vev{W[C]} \sim \mu^{\Gamma_{\rm cusp}(\lambda,\theta)}\,,
\end{equation}
with $\mu$ being a UV cut-off. The cusp angle $\theta$ is restricted to the
interval $[0,2\pi[$ in Euclidean space but is unrestricted in Minkowski space.
The correspondence between the anomalous dimension of twist-2 spin operators
with large Lorentz spin $S$ and the cusp anomaly is
follows~\cite{Korchemsky88,km}
\begin{equation}
\gamma^{\rm (tw=2)}_S(\lambda)=2\Gamma_{\rm cusp}(\lambda, \theta=\ln S)
\label{gamma=cusp}
\end{equation}
and is valid for an arbitrary coupling constant $\lambda$. At weak coupling and
$\theta\gg 1$, one has
\begin{equation}
\Gamma_{\rm cusp}(\lambda, \theta)=\theta \left[\frac{\lambda}{4\pi^2} +
\mathcal{O}( \lambda^2)  \right]
\end{equation}
with perturbative coefficients known up to three-loop order~\cite{MVV04}. The
calculation of $\Gamma_{\rm cusp}(\lambda, \theta)$ at the strong coupling can be
effectively done via the open string picture. In this limit, the Wilson loop is
proportional to the area of the minimal surface swept out by an open string which
penetrates into the fifth AdS dimension and whose ends trace the integration
contour $C$ in Minkowski space~\cite{loop}. This leads to~\cite{kru2,mak}
\begin{equation}
\Gamma_{\rm cusp}(\lambda, \theta)=\theta
\left[\left(\frac{\lambda}{4\pi^2}\right)^{1/2} + \mathcal{O}(\lambda^0) \right]
\label{cusp-strong}
\end{equation}
for $\theta\gg 1$. Being combined together, Eqs.~\re{gamma=cusp} and
\re{cusp-strong} reproduce the strong coupling result \re{twist-2-open} based on
the folded closed string picture~\cite{gkp}.

The correspondence \re{gamma=cusp} can be generalized to higher twist operators.
In that case, the anomalous dimension of the Wilson operator built from $L$
constituent fields and a total number of derivatives $S$, such that $S\gg L$,
can be mapped into the anomalous dimension of the product of $L$ Wilson loops in the
fundamental representation of the $SU(N_c)$ group; the total number of cusps
varies between $4$ and $2L$~\cite{bgk}. At large $N_c$, the expectation value of
the product of Wilson loops factorizes into the product of their expectation
values. This implies that, at strong coupling the area of the minimal surface
corresponding to the product of $k=2,\ldots,L$ Wilson loops with cusps is given
by the sum of $k$ elementary areas leading to
\begin{equation}
2 \,\Gamma_{\rm cusp}(\lambda, \theta=\ln S) \le \gamma^{{\rm (tw=}L)}_S(\lambda)
\le L\, \Gamma_{\rm cusp}(\lambda, \theta=\ln S)\,.
\label{band1}
\end{equation}
We are reminded that the anomalous dimensions of higher twist operators are not solely
determined by the total number of derivatives $S$. They form instead a band whose
internal structure at weak coupling is governed by integrals of the motion of the
quantum Heisenberg $SL(2)$ magnet. Equation \re{band1} defines the boundaries of the
band both at strong and weak coupling.

\section{Conclusion}
\setcounter{equation}{0}

The main objective of the present work was to demonstrate how the phenomenon of
integrability arises in certain limiting cases of Yang--Mills dynamics and exhibit
their similarities or ultimate relation whenever it was obvious. In all
cases we have studied, integrability appears as a hidden symmetry of an
underlying effective theory. It either describes elementary fields living on the
light cone, in the case of renormalization group evolution in QCD and its
supersymmetric extensions, or reggeons as new degrees of freedom in high-energy
QCD, or branes as effective degrees of freedom for the Seiberg--Witten solution to
the $\mathcal{N} = 2$ SYM theory. A number of nontrivial questions still have to
be answered. The most obvious and at the same time, the most profound and most
difficult question is ``What is the origin of integrability?'' or, in other words,
``What is the symmetry, if any, of the gauge theory which leads to it?'' More
specific questions that have recently kept theoretical physicists busy concern
issues like the fate of the integrability of the dilatation operator in SYM at
higher orders of perturbation theory, the integrability of the SYM dual sigma
models on curved backgrounds and the matching of integrable structures on both sides of
the correspondence, to name but a few. Ultimately, if the integrability is
indeed a property of the full quantum Yang--Mills theory as well as the dual
string theory, it will provide the most sophisticated test of duality and endow
us with powerful machinery to tackle the strong-coupling regime of field theories.
On this quest, unfortunately, we will be missing Ian Kogan.

\section*{Acknowledgements}

Three of us (A.B., V.B and G.K.) are most grateful to Sergei Derkachov
and Alexander Manashov for enjoyful and fruitful collaboration on the topics
covered in the present work. A.G. would also like to thank A.~Marshakov, A.~Mironov,
A.~Morozov, A.~Tseytlin and K.~Zarembo for the collaboration and discussions of
various aspects of integrability and related issues. This work supported (A.B.) in
part by U.S.\ Department of Energy under grant no.\ DE-FG02-93ER-40762.

\end{document}